\documentclass[%
twocolumn,
superscriptaddress,
showpacs,preprintnumbers,
 amsmath,amssymb,
 aps,
 prd,
]{revtex4-1}

\usepackage{braket}            %for braket notation
\usepackage{graphicx}          %Include figure files
\usepackage[utf8]{inputenc}    %For Bali paper references
\usepackage{hyperref}          %Hyperlinks for references, cites, etc
\usepackage{dcolumn}           % Align table columns on decimal point
\usepackage{color}             %for \textcolor
\usepackage{adjustbox}         %For making the chiEFT coeffs table fit better
\usepackage[caption=false]{subfig}        %Subfigures
\usepackage{csquotes}        %enquote
\usepackage{physics}
\usepackage{mathtools}
\usepackage{amsmath}
\usepackage{enumerate}
\usepackage{placeins}
\usepackage{url}

\newcolumntype{d}[1]{D{.}{.}{#1}}

\makeatletter
\let\oldtheequation\theequation
\renewcommand\tagform@[1]{\maketag@@@{\ignorespaces#1\unskip\@@italiccorr}}
\renewcommand\theequation{(\oldtheequation)}
\makeatother

\newcommand{\mpi}{m_{\pi}}
\newcommand{\mpipacs}{m_{\pi}^{\rm PACS-CS}}
\newcommand{\mpis}{m_{\pi}^2}
\newcommand{\vx}{\vec{x}}
\newcommand{\vB}{\vec{B}}
\newcommand{\pivB}{\psi_{i,\vB}}
\newcommand{\toG}{\Braket{\Omega | T\left\{\chi\left(\vx,t\right)\,\bar{\chi}\left(\vec{0},t\right)\right\}|\Omega}}
\newcommand{\toGr}{\Braket{\Omega | T\left\{\chi\left(\vec r,t\right)\,\bar{\chi}\left(\vec{0},t\right)\right\}|\Omega}}
\newcommand{\pvB}{\psi_{\vB}}
\newcommand{\rb}[1]{\left(#1\right)}

\newcommand{\sigmap}{$\Sigma^+\,$}
\newcommand{\sigmam}{$\Sigma^-\,$}
\newcommand{\cascadez}{$\Xi^0\,$}
\newcommand{\cascadem}{$\Xi^-\,$}
\newcommand{\etal}{{\it et al.}}

\newcommand{\Kp}{{K^{+}}}

\newcommand{\Kz}{K^{0}}
\newcommand{\Kzb}{\overline{K}^0}

\newcommand{\Km}{K^{-}}

\newcommand{\pim}{\pi^{-}}
\newcommand{\pip}{{\pi^{+}}}

\newcommand{\mcC}{\mathcal{C}}

\newcommand{\CSSM}{Special Research Centre for the Subatomic Structure
  of Matter (CSSM),\\Department of Physics, University of
  Adelaide, Adelaide, South Australia 5005, Australia}

\newcommand{\Swansea}{Department of Physics, Swansea University, Swansea, SA2 8PP, United Kingdom}

\newcommand{\Trinity}{School of Mathematics, Trinity College, Dublin, Ireland}

\begin{document}

\preprint{ADP-24-13/T1252}

\title{Magnetic polarizability of octet baryons via lattice QCD}
\author{Thomas Kabelitz}
\affiliation{\CSSM}
\author{Ryan Bignell}
\affiliation{\CSSM}
\affiliation{\Swansea}
\affiliation{\Trinity}
\author{Waseem Kamleh}
\affiliation{\CSSM}
\author{Derek Leinweber}
\affiliation{\CSSM}

\date{\today}

\begin{abstract}
    Drawing on recent advances in lattice-QCD background-field techniques, the magnetic polarizability
    of octet baryons is calculated from the first principles of QCD. The results are presented in the
    context of new constituent quark-model calculations providing a framework for understanding the
    lattice results and a direct comparison with simulation results at unphysical quark masses.  Using
    smeared quark sources, low-lying Laplacian eigenmode projection and final-state Landau-mode
    projection, considerable attention is devoted to ensuring single-state isolation in the lattice
    correlation functions.  We also introduce new weighting methods to reduce the sensitivity to
    correlation-function fits, averaging over many fits based on merit drawn from the full correlated
    $\chi^2$ of the fits.  The techniques are implemented on the $32^3 \times 64$, $2+1$-flavor
    dynamical-fermion lattices provided by the PACS-CS Collaboration following the introduction of
    uniform magnetic fields quantized to the lowest nontrivial values available.  After some scaling
    of the constituent quark model parameters, we find the model captures the patterns observed
    in the lattice QCD results very well, providing important insights into the physics underpinning the
    magnetic polarizabilities.  Finally, comparison with the most recent results from experiment
    proceeds through an effective field theory formalism which incorporates estimates of finite-volume
    corrections and small electroquenching corrections as the results are brought to the physical
    point.  We find excellent agreement with experiment where available, including the proton and
    neutron polarizabilities.
\end{abstract}

% insert suggested PACS numbers in braces on next line
\pacs{13.40.-f, 12.38.Gc, 12.39.Jh}
%13.40-f  = Electromagnetic processes and properties
%12.38.Gc = Lattice QCD calculations
%12.39.Jh = Nonrelativistic quark model
%12.39.Fe = Chiral Lagrangians
%
% insert suggested keywords - APS authors don't need to do this
%\keywords{}

%\maketitle must follow title, authors, abstract, \pacs, and \keywords
\maketitle

\section{Introduction}

The magnetic polarizability is the second-order response of an electrically charged composite
particle to an external magnetic field. It encapsulates the manner in which the internal structure
of the particle is changed by the field. From a perturbative point of view, the second-order
process induces virtual electromagnetic transitions to nearby excitations in the hadron spectrum
and probes the distribution of quarks within the hadron. In a nonperturbative sense, these virtual
transitions act to change the structure of the hadron and thus the energy of the particle as it
resides in the magnetic field.

Our focus is on lattice QCD calculations \cite{Gross:2022hyw} of the magnetic polarizability of
octet baryons.  We draw on recent advances in lattice-QCD background-field techniques, established
in an analysis of nucleon polarizabilities \cite{bignell2020nucleon}.  In particular, smeared quark
sources are utilized to capture the QCD aspects of the hadron structure. Then low-lying QCD+QED
SU(3)$\times$U(1) Laplacian eigenmode projection is considered at the quark level to capture the
low-energy response of individual electrically charged quarks to QCD and the external magnetic
field.  Electrically charged baryons will be in a superposition of Landau-level states and
therefore final-state color-singlet U(1) Landau-mode projection is used at the sink to ensure the
baryon is in the lowest-lying Landau
level~\cite{bignell2018neutron,bignell2020nucleon,tiburzi:2012:projection}.  We extend the previous
analysis by devoting considerable attention to ensuring single-state isolation in the lattice
correlation functions.  We further introduce new weighting methods to reduce the sensitivity to
correlation-function fits, averaging over many fits based on merit determined by the full
covariance-matrix-based $\chi^2$ of the fits.

To help in understanding the pattern of octet-baryon magnetic polarizabilities observed, we
consider the simple nonrelativistic constituent quark model and extend calculations of the nucleon
polarizabilities \cite{Capstick:1992:qmpol,Lipkin:1992:qmpol,bhaduri1998quarkmodel} to include the
hyperons of the baryon octet.  As we will demonstrate, chiral nonanalytic behavior is subtle at
the quark masses we consider on the lattice.  Therefore, we anticipate this simple model will be of
utility in understanding the competing effects that generate some complexity in the pattern of
magnetic polarizabilities observed in lattice QCD.  Qualitatively, it explains the large difference
between negatively charged baryons and the remainder of the octet.  After some scaling of the
constituent quark model parameters, we find the model captures the patterns observed in the lattice
QCD results very well, providing important insights into the physics underpinning the magnetic
polarizabilities.

Our approach to the magnetic polarizability in lattice QCD centers on the uniform background
magnetic-field method.  Historically, this approach has been the method of choice for determining
baryon magnetic polarizabilities
\cite{Zhou:2002:polarisability,Lee:2006:polarisability,primer2014magnetic,Chang:2015:polarisability,bignell2018neutron,bignell2020nucleon}.
As a leading effect in the expansion of the baryon energy in terms of the magnetic-field strength,
the background field must be weak to ensure higher ${\cal O}(B^3)$ terms are small.
Thus, subtle shifts in the energy of the baryon induced by the magnetic field need
to be related to the polarizability.  To reveal the effects of the magnetic field, correlated
ratios are constructed to enable QCD-based fluctuations to cancel.

Our techniques are implemented on the $32^3 \times 64$, $2+1$-flavor dynamical-fermion lattices
provided by the PACS-CS Collaboration. QCD correlations are maintained by applying the external
magnetic field to these existing gauge field-configurations.  A consequence of this is that our
gauge fields are "electroquenched" in that the sea-quarks are blind to the magnetic field.
At the quark masses considered herein, effective field theory indicates the approximation is
reasonable, and we use effective field theory to estimate the small electroquenching corrections
\cite{Deshmukh:2018:octet,hall2014chiral,bignell2020nucleon}.  We also draw on effective field
theory to estimate the small finite-volume corrections to our lattice QCD calculations
\cite{hall2014chiral,bignell2020nucleon}.

In constructing the required correlated ratios, several two-point correlation functions are
combined.  These include the zero-field correlator and spin-field aligned and antialigned baryon
correlators at finite magnetic-field strength. These all display different behavior as a function
of Euclidean time and their combination can hide excited-state contamination effects. To address
this, we examine the Euclidean time behavior of each of these underlying correlators to ensure
that the effective energy of each two-point correlator has reached single-state isolation through
Euclidean time evolution. We find that in many cases plateaulike behavior in the correlated
ratios precedes the single-state isolation of the individual underlying correlation functions.
Thus the results presented herein provide an important step forward in suppressing contamination
from excited states.

We also carefully examine the dependence of the results on the selection of the fit window where a
significant dependence can arise in some cases.  To address this
we introduce a weighting method to reduce the sensitivity to fit window selection.
Following the ideas presented in Ref.~\cite{NPLQCD:2020multihadron}, we
average over many fit windows with their weight based on merit calculated with  the full
covariance-matrix-based $\chi^2$ of the fits.

We find the processes described above bring a new level of rigor to the polarizability results,
producing the highest precision lattice QCD calculations of baryon magnetic polarizabilities to
date.

The presence of a magnetic field and the difference in the $u$ and $d$ quark charges breaks the
isospin symmetry of equal mass quarks.  The $\Lambda$ and $\Sigma^0$ baryons will mix in the
magnetic field complicating the interpretation of the mass shifts in the background-field approach
considered herein.  Correlation matrix techniques can be introduced to address the state isolation
issue. However, together these issues add additional layers of complexity that are not required for
the other six members of the baryon octet. These are the so-called outer members of the octet,
composed with a doubly represented quark flavor and a singly represented flavor.  It is these
baryons that hold the focus of the present investigation.

Recently four-point function methods have been considered in the calculation of the magnetic
polarizability~\cite{Wilcox:2021:towards,Lee:2023:pion}.  This time the baryons are probed
perturbatively by two electromagnetic-current insertions.  As such, the baryon states are
eigenstates of QCD alone and the $\Lambda$ and $\Sigma^0$ baryons do not mix.  Thus, this alternative
approach presents a significant advantage to understanding the magnetic polarizabilities of the
$\Lambda$ and $\Sigma^0$ baryons and we anticipate future calculations for these baryons.

The presentation of this work is as follows. In \autoref{sec:quarkmodel} we derive the generalized
expression for the magnetic polarizability of an octet baryon in a simple constituent quark model.
Highlighting the competing effects of magnetic transitions and the distribution of charge within
the baryon, we derive predictions for the magnetic polarizabilities of the outer octet baryons and
use these to provide context for the results of the lattice QCD calculations.
In \autoref{sec:backgroundfieldmethod} we describe our implementation of the background
field for the lattice QCD calculation, emphasizing the colocation of the field and baryon
correlator origin.
In \autoref{sec:magneticpolarisability} we discuss the extraction of the magnetic polarizability
from two-point baryon correlation functions and ratios.
In \autoref{sec:simulationdetails} we discuss relevant details of the lattice simulation including
the ensembles used, the quark level SU(3)$\times$U(1) projection at the quark propagator sink, and
the U(1) hadronic projection which allows for proper handling of Landau-level physics.
In \autoref{sec:fitting} we discuss our advances in the correlation-function fitting methods,
including the examination of the excited-state contamination in the underlying correlation
functions and the weighted averaging of fit windows.
In \autoref{sec:latticeresults} we present the
results of the lattice calculation and utilize them to improve the quark model such that it can
capture the essential physics governing baryon magnetic polarizabilities.
In \autoref{sec:chiralextrapolation} we use chiral effective field theory to incorporate
electroquenching corrections and estimate finite-volume corrections before extrapolating the
lattice results to the physical regime.
In \autoref{sec:comparison} we compare our findings with experimental measurements and previous
determinations of octet-baryon magnetic polarizabilities.
In \autoref{sec:proton-neutron} we briefly examine the proton-neutron magnetic polarizability
difference.
In \autoref{sec:conclusion} we summarize our findings.

\section{Constituent Quark Model}\label{sec:quarkmodel}

The constituent quark model is renowned for its capacity to provide a simple explanation for the
pattern of baryon magnetic moments observed in experiment. Founded on SU(6) spin-flavor symmetry,
symmetric ground-state wave functions with zero angular momentum, and antisymmetry via SU(3)
color-singlet states \cite{Close:1979bt}, the model captures the essence of the physics with
subtle corrections arising from the environment sensitivity of the constituent quark moments
\cite{Leinweber:1990dv,Boinepalli:2006xd}. We consider the constituent quark model predictions for
octet-baryon magnetic polarizabilities in a similar spirit, hoping to capture the predominant
features in a simple model and providing a framework for the description of more subtle
corrections.

\subsection{Magnetic polarizabilities from the quark model}

Bhaduri {\it et al.}~\cite{bhaduri1998quarkmodel} present an expression for the
magnetic polarizability of the proton in the constituent quark model. Here we review the derivation
of this expression and generalize it for an arbitrary octet baryon.  In doing so, we aim to provide
insight into the salient features of QCD that give rise to the magnetic polarizabilities of
ground-state baryons, at quark masses where pion dressings of the baryons are of secondary
importance \cite{hall2014chiral,bignell2020nucleon}.

A quark model expression for the magnetic polarizability may be obtained using
Rayleigh-Schr\"{o}dinger perturbation theory, reviewed in Ref.~\cite{sakuraimodern} for example.
We consider an unperturbed Hamiltonian $\mathcal{H}_0$, and label the energy eigenstates of this Hamiltonian $\ket{n_i}$
\begin{equation}
    \mathcal{H}_0\ket{n_i}=E_i\ket{n_i}, \quad i=1,2,3,\ldots \, .
\end{equation}
%
%% In our case these energy eigenstates correspond to the tower of states including and above the
%% relevant octet baryon. 
%
This Hamiltonian is perturbed with an interacting Hamiltonian $\mathcal{H}_{\rm int}$. In our case this will be a Hamiltonian relevant to the inclusion of a magnetic field. Substituting this perturbation into the Schr\"{o}dinger equation we define
\begin{equation}
    (\mathcal{H}_0 + \lambda \mathcal{H}_{\rm int})\ket{k_i}
    = E^\prime_i\ket{k_i},\quad i=1,2,3,\ldots \, ,
\end{equation}
where $\ket{k_i}$ are the energy eigenstates of the full Hamiltonian and $\lambda$ controls the strength of the perturbation.
%
%% We wish to write the energy levels and eigenstates as a power series, which requires the
%% interaction to be weak. $\lambda$ is used to ensure this condition holds, though it may be set
%% to one for a sufficiently weak interaction. 
%
%% These power series expressions are then substituted into the Schr\"{o}dinger equation and solved
%% iteratively to obtain the various energy levels and eigenstates in terms of the unperturbed
%% energy levels and states. Two iterations of the process results in the following expression for
%% the perturbed energy of the $i$th energy level
%
At second order in $\lambda$, the perturbed energy of the $i$th energy level is
\begin{eqnarray}
    E^\prime_i(\lambda) &=& E_i
    + \lambda\bra{n_i}\mathcal{H}_{\rm int}\ket{n_i} \nonumber \\
    &&+ \lambda^2\sum_{j\neq i}\frac{|\bra{n_i}\mathcal{H}_{\rm int}\ket{n_j}|^2}{E_i-E_j}
    + \order{\lambda^3} \, , \label{eqn:QCD:PerturbativeExpansion}
\end{eqnarray}
and the shift in energy
\begin{equation}
    \Delta E = E^\prime_i(\lambda) - E_i \, , \label{eqn:QCD:PerturbationShift}
\end{equation}
includes terms both linear and quadratic in $\lambda$.  At this point we set $\lambda = 1$ and draw
on the electric charge, $e$, in $\mathcal{H}_{\rm int}$ to ensure the interaction is perturbative.

The magnetic polarizability is associated with terms quadratic in the magnetic-field strength and
thus $e^2$ and we will see both correction terms of Eq.~\ref{eqn:QCD:PerturbativeExpansion} provide
a contribution.  Focusing on terms proportional to $e^2$, we define the magnetic polarizability,
$\beta$ via
\begin{equation}
    \Delta E_2=-\frac{1}{2}\, 4\pi \,\beta \, B^2 \, , \label{eqn:QCD:PolarisabilityShift}
\end{equation}
where $\Delta E_2$ contains the terms second order in $e$ from Eq.~\ref{eqn:QCD:PerturbativeExpansion}.

In deriving the interacting Hamiltonian, we first consider the case of a free particle, where we
can write the Hamiltonian in terms of the momentum operator and the mass
\begin{equation}
    \mathcal{H}=\frac{\vec{p}^{\,2}}{2m} \, .
\end{equation}
Minimal substitution provides~\cite{anderson1971textbook}
\begin{equation}
    \vec{p}\rightarrow\vec{p}+q\, e\,\vec{A} \, ,
\end{equation}
where $q\, e$ is the particle charge and $\vec{A}$ is the electromagnetic vector potential. Making
this substitution, we obtain
\begin{equation}
    \mathcal{H}
    =\frac{\vec{p}^{\,2}}{2m}
    + \frac{q\,e}{m}(\vec{p}\vdot\vec{A})
    + \frac{q^2\,e^2\,\vec{A}^{\,2}}{2m}\, , \label{eqn:QuarkModel:Polarisability:CoupledHamiltonian:General}
\end{equation}
where we note that $\vec{p}$ and $\vec{A}$ commute when the background field is uniform.
For $\vec{B}$ in the $z$-direction, $\vec{B}=B\,\hat{k}$, the following relation equitably
distributes the field strength among the components of $\vec{A}$
\begin{align}
    \vec{A}&=\frac{1}{2}\,\vec{B}\cp\vec{r}\, , \label{eqn:QuarkModel:Polarisability:VectorPotential:Uniform:General}\\
    &=\frac{1}{2}(-B\,y, B\,x, 0)\, , \label{eqn:QuarkModel:Polarisability:VectorPotential:Uniform:Specific}
\end{align}
This relation allows us to rewrite the dot product in terms of angular momentum
\begin{align}
    \vec{p}\cdot\vec{A}
    &= \frac{1}{2}\,\vec{p}\cdot\left(\vec{B}\times\vec{r}\right)\, , \\
    &= \frac{1}{2}\,\vec{L}\cdot\vec{B} \, .
\end{align}

Here we are purely interested in ground-state baryons. As such, the quarks are in a relative $S$-wave state and have zero angular momentum causing that term to vanish, leaving
\begin{equation}
    \mathcal{H}=\frac{\vec{p}^{\,2}}{2m}+\frac{q^2\,e^2\,\vec{A}^{\,2}}{2m}\, .
\end{equation}

We must also account for the particle's spin. Inclusion of a vector potential in the Dirac
equation for a charged spinor produces a term~\cite{schwartz2014textbook}
\begin{equation}
    \frac{q\,e}{2 m}\,\vec{B} \cdot \hat{\sigma}\, ,
\end{equation}
where $\hat{\sigma}$ is the Pauli spin operator. As such our Hamiltonian becomes
\begin{equation}
    \mathcal{H}=
    \frac{\vec{p}^{\,2}}{2m}
    + \hat{\mu}\cdot\vec{B}
    + \frac{q^2\,e^2\,\vec{A}^{\,2}}{2m}\, ,
\end{equation}
where the magnetic moment operator is given by
\begin{equation}
    \hat{\mu}=\frac{q\,e}{2 m}\,\hat{\sigma} \, .
\end{equation}
The sign of the magnetic moment term comes from the fact that an addition of a magnetic field
will cause the particle's spin to align opposite to the field, thereby decreasing its energy via
the dot product.

Drawing on Eq.~\ref{eqn:QuarkModel:Polarisability:VectorPotential:Uniform:Specific},
\begin{equation}
    \vec{A}^{\,2} = \frac{1}{4}(x^2+y^2)B^2\, ,
\end{equation}
and
\begin{align}
    \mathcal{H} &=
    \frac{\hat{\vec{p}}^{\,2}}{2m}
    + \hat{\mu}\cdot\vec{B}
    + \frac{q^2\,e^2}{8m}( \hat{x}^2+\hat{y}^2 )B^2\, ,
\end{align}
where we now see that the $e^2$ term contains the position operators. Thus, the magnetic
polarizability contains contributions probing the distribution of quarks within the hadron.

Recalling $\vec{B}=B\,\hat{k}$, we reduce the dot product to simply the magnetic moment operator acting purely in the $z$-direction. Further, as we are working within the framework of the quark model we include a sum over the quarks in our baryon. As such, we write our interaction Hamiltonian as
\begin{equation}
    \mathcal{H}_{\rm int}=
    \hat{\mu}_{z}\,B
    + \sum_{f=1}^3 \frac{q_f^2\,e^2}{8 m_f}( \hat{x}^2+\hat{y}^2 )B^2\, ,
\end{equation}
where our magnetic moment operator is now
\begin{equation}
    \hat{\mu}_{z}=\sum_{f=1}^3\frac{q_f\,e}{2 m_f}\,\hat{\sigma}_z\, .
\end{equation}

This interaction Hamiltonian is then substituted into Eq.~\ref{eqn:QCD:PerturbativeExpansion}. Such a substitution requires the determination of two matrix elements. First,
\begin{align}
    \bra{\mathcal{B}}\mathcal{H}_{\rm int}\ket{\mathcal{B}}
    &= \mu_{\mathcal{B}}\,B
    + \sum_{f=1}^3\frac{q_f^2\,e^2}{8 m_f}
    \left(\expval{x^2}_f +\expval{y^2}_f\right) \, , \nonumber\\
    &=  \left(\frac{4}{3}\mu_D-\frac{1}{3}\mu_S\right)B
    + \sum_{f=1}^3\frac{q_f^2\,e^2}{12 m_f}\expval{r^2}_f \, ,
    \label{eqn:magmom}
\end{align}
where $\mu_{\mathcal{B}}$ is the magnetic moment of the baryon, and $\mu_D$ and $\mu_S$ are the
intrinsic magnetic moments of the doubly represented and singly represented quark flavors
respectively.  In the weak-field limit, spherical symmetry of the $S$-wave states provides
$\expval{x^2}=\expval{y^2}=\frac{1}{3}\expval{r^2}$.

The second matrix element of interest in Eq.~\ref{eqn:QCD:PerturbativeExpansion} contains an excited
state of the octet baryon in the ket. The sum is estimated by considering dominance of the first
nearby excitation, the corresponding decuplet baryon state. This matrix element generates terms
proportional to $e^2$ via
\begin{equation}
    \abs{\bra{\mathcal{B}}\mathcal{H}_{\rm int}\ket{\mathcal{B}^*}}^2
    = \abs{\bra{\mathcal{B}}\hat{\mu}_f\ket{\mathcal{B}^*}}^2B^2 \, .
\end{equation}
This contribution is the transition magnetic moment for an octet baryon to its complementary
decuplet baryon. For a baryon on the outer ring of the octet, the quark
model provides
\begin{equation}
    \bra{\mathcal{B}}\hat{\mu}_f\ket{\mathcal{B}^*} = \mu_{\mathcal{B}\mathcal{B}^*}
    = \frac{2\sqrt{2}}{3}\left(\mu_D - \mu_S\right)\, , \label{eqn:QCD:transitionmoment}
\end{equation}
where $D$ and $S$ once again label the constituent quark flavors for the doubly and
singly represented sectors; for example for the proton, $\mu_D-\mu_S=\mu_u-\mu_d$.

Substituting these matrix elements into Eq.~\ref{eqn:QCD:PerturbativeExpansion}, the energy shift
quadratic in $e^2$ and thus $B^2$ is
\begin{eqnarray}
    \Delta E_2 &=&
    \sum_{f=1}^3\frac{q_f^2\,e^2}{12 m_f}\expval{r^2}_f\, B^2
    + \sum_{\mathcal{B}^*}
    \frac{\abs{\bra{\mathcal{B}}\hat{\mu}_z\ket{\mathcal{B}^*}}^2 \, B^2}{E_{\mathcal{B}}-E_{\mathcal{B}^*}}\, ,
    \nonumber\\
    &=& -\frac{1}{2}\, 4\pi\, \beta \, B^2
\end{eqnarray}
where the definition of Eq.~\ref{eqn:QCD:PolarisabilityShift} has been used.  Thus, the constituent
quark model prediction for the magnetic polarizability of octet baryon $\mathcal{B}$ composed of
three quark flavors denoted by $f$ is
\begin{align}\label{eqn:quarkmodel:polarisability}
    \beta &=
    \frac{1}{2\pi}\sum_{\mathcal{B}^*}\frac{\abs{\bra{\mathcal{B}}\hat{\mu}_z\ket{\mathcal{B}^*}}^2}{E_{\mathcal{B}^*}-E_{\mathcal{B}}}
    - \sum_{f=1}^3\frac{q_f^2\,\alpha}{6 m_f}\expval{r^2}_f \, , \nonumber \\
    &\equiv \beta_1 - \beta_2 \, ,
\end{align}
where $\alpha={e^2}/{4\pi}$ is the fine structure constant.  The magnetic polarizability has its
origin in two competing terms which we have defined as $\beta_1$, a magnetic transition term, and
$\beta_2$ probing the distribution of quarks within the baryon.  The two terms and their opposing
signs in this expression highlight the complex nature of the baryon polarizability.  This is the
constituent quark model description for the magnetic polarizability of an octet baryon with two
quarks of one flavor and another quark with a different flavor.

Using Eqs.~\ref{eqn:magmom} and \ref{eqn:QCD:transitionmoment} it can be shown that
$\bra{p}\hat{\mu}_z\ket{\Delta}=\frac{2\sqrt{2}}{3}\mu_p$ for constituent quark masses $m_u = m_d$.
Upon substitution into Eq.~\ref{eqn:quarkmodel:polarisability}, one obtains the original result of
Bhaduri {\it et al.}~\cite{bhaduri1998quarkmodel}
\begin{equation}
    \beta_p =
    \frac{16}{9}\frac{1}{4\pi}\frac{\mu_p^2}{m_{\Delta} - m_p}
    -\frac{\alpha}{6\,m_l}\left(2\,q_u^2\expval{r^2}_u+q_d^2\expval{r^2}_d\right),
\end{equation}
where $m_l = m_u = m_d$.

To compare the predictions of the quark model to the results of our lattice QCD calculations, we
require a relationship between the baryon masses calculated in lattice QCD and the associated
constituent quark mass to be used in the model.  We turn our attention to this in the following
section.

\subsection{Implementing the quark model}

To present the predictions of the quark model for the magnetic polarizability as a function of
quark mass, we require a relationship between the lattice baryon masses and the constituent quark
masses.  This relationship has been long established for the baryon octet at the experimental point
and we need only extend this relationship away from the physical point.  In addition, the
octet-decuplet mass splitting in Eq.~\ref{eqn:quarkmodel:polarisability} demands knowledge of both
octet and decuplet baryons.  Similarly, $\expval{r^2}_f$ describing the distribution of quark
flavors within the octet baryons is also required. We seek smooth, interpolated expressions for
each of these quantities.

Our lattice calculations are performed on the PACS-CS gauge-field
ensembles~\cite{PACS-CS2008ensembles} and therefore we draw on their published baryon masses in
the Sommer scheme \cite{Sommer:1993ce}.  Reference~\cite{Walker-Loud:2014:ruler} demonstrated that the nucleon masses
observed in finite-volume lattice QCD simulations display a ruler-style linear behavior when
plotted as a function of $m_{\pi}$. This provides a simple characterization of lattice baryon
masses and in the spirit of the simple quark model, we take this approach to interpolate the
PACS-CS baryon masses for the octet and decuplet.

To estimate the constituent quark masses away from the physical point, we utilize a simple linear
model
\begin{align}\label{eqn:quarkmodelimplementation:octetmassmodel}
    m^0_{\rm oct} + \alpha_{\rm oct}\,(3\,m_l) &= m_N, \nonumber\\
    m^0_{\rm oct} + \alpha_{\rm oct}\,(m_l+2\,m_h) &= m_{\Xi},
\end{align}
which leverages the maximum difference in strange quarks in the baryon octet.  Here $m^0_{\rm oct}$ and
$\alpha_{\rm oct}$ are fit parameters, and $m_l$ and $m_h$ are the light and heavy constituent quark masses
already determined in the simple constituent quark model summarized in Ref.~\cite{pdg2020}. The
parameter $m^0_{\rm oct}$ allows some of the nucleon mass to have its origin in the confining potential and
$\alpha_{\rm oct}$ allows for spin-dependent effects to change the slope from 1.  Using experimental baryon
masses and the constituent quark masses $m_l=338$ and $m_h=510\,$MeV from the PDG~\cite{pdg2020},
we solve for the fit parameters $m^0_{\rm oct}$ and $\alpha_{\rm oct}$.

For decuplet baryons, we repeat the analysis with two new fit parameters
\begin{align}\label{eqn:quarkmodelimplementation:decupletmassmodel}
    m^0_{\rm dec} + \alpha_{\rm dec}\,(3\,m_l) &= m_{\Delta}, \nonumber\\
    m^0_{\rm dec} + \alpha_{\rm dec}\,(3\,m_h) &= m_{\Omega}.
\end{align}

The fit parameters corresponding to the two models are shown in
\autoref{tab:quarkmodelimplementation:massfitparameters}. We see that in the octet case the slope
parameter $\alpha_{\rm oct} = 1.09$ is larger than unity. This is in accord with anticipated hyperfine
effects.  Because the strength of the attractive hyperfine interaction is inversely proportional to
the product of the constituent quark masses, the attraction diminishes with increasing quark
mass. Thus the baryon mass grows at a rate exceeding the rate of the quark mass increase.

On the other hand, the hyperfine interaction is repulsive in decuplet baryons. This time the
repulsion diminishes with increasing mass such that the decuplet slope is expected to be less than
unity. The value of $\alpha_{\rm dec}=0.85$ is in accord with these expectations.

\begin{table}[tb]
    \centering
    \caption{Fit parameters for the simple models of
      Eqs.~\ref{eqn:quarkmodelimplementation:octetmassmodel}
      and~\ref{eqn:quarkmodelimplementation:decupletmassmodel}.}
    \label{tab:quarkmodelimplementation:massfitparameters}
    \begin{ruledtabular}
        \begin{tabular}{lcc}
            \noalign{\smallskip}
            Baryon   & $\alpha$ & $m_0$ (MeV)      \\
            \noalign{\smallskip}\hline\noalign{\smallskip}
            Octet    & 1.09     & $-169$           \\
            % Updated decuplet value
            Decuplet & 0.85     & $\phantom{-}367$ \\
            \noalign{\smallskip}
        \end{tabular}
    \end{ruledtabular}
\end{table}

We use the octet fit parameters and the ruler-style interpolated $N$ and $\Xi$ masses to obtain the
constituent quark masses as a function of $m_{\pi}$.  These masses are given in
\autoref{tab:quarkmodelimplementation:latticequarkmasses}. We see the expected approach to the
physical constituent quark mass of 338 MeV~\cite{PDG2022} while the strange-quark mass remains stable
as expected.

\begin{table}[tb]
    \centering
    \caption{Constituent quark masses for the PACS-CS gauge-field
    ensembles~\cite{PACS-CS2008ensembles}. Constituent masses are obtained from the model
    Eq.~\ref{eqn:quarkmodelimplementation:octetmassmodel} with $(\alpha_{\rm oct},m^0_{\rm oct})=$($1.09,-169\,$MeV). All
    masses are in MeV.} \label{tab:quarkmodelimplementation:latticequarkmasses}
    \begin{ruledtabular}
        \begin{tabular}{cccc}
            \noalign{\smallskip}
            $\kappa$ & $\mpipacs$ & $m_l$ & $m_h$ \\
            \noalign{\smallskip}\hline\noalign{\smallskip}
            0.13700  & 701        & 480   & 528   \\
            0.13727  & 570        & 447   & 528   \\
            0.13754  & 411        & 402   & 529   \\
            0.13770  & 296        & 365   & 529   \\
            0.13781  & 156        & 328   & 529   \\
            \noalign{\smallskip}
        \end{tabular}
    \end{ruledtabular}
\end{table}

With the constituent quark masses $m_l$ and $m_h$ determined as a function of $m_{\pi}$, one can
use the form of Eqs.~\ref{eqn:quarkmodelimplementation:octetmassmodel} to obtain the $\Sigma$-baryon
masses by counting two light quarks and one strange quark
\begin{equation}
    m_\Sigma = m^0_{\rm oct} + \alpha_{\rm oct}\,(2\, m_l + m_h)\, .
\end{equation}
A similar approach is used to get the interpolated decuplet baryon masses. Drawing on
Eqs.~\ref{eqn:quarkmodelimplementation:decupletmassmodel} and modifying them to count the number of
light and strange quarks, the decuplet baryon masses are obtained within the constituent quark
model framework in a consistent manner.

Referring back to Eq.~\ref{eqn:quarkmodel:polarisability} for the magnetic polarizability, the
intrinsic quark magnetic moments contained within the octet-decuplet baryon magnetic transition
matrix element are simply
\begin{equation}
    \mu_{f} = \frac{q_f\,e}{2 m_f}.
\end{equation}

Quark distribution radii on the PACS-CS ensembles have been determined by Stokes {\it et
        al.}~\cite{stokes2020formfactor} where they examined the proton and neutron. We interpolate the
proton and neutron squared radii as linear in $\log(m_{\pi})$ and this provides an excellent
description of those lattice results. Equation.~\ref{eqn:quarkmodel:polarisability} requires quark
distribution radii for single quark flavors of unit charge, as the charge factors appear elsewhere
in the expression.  We define radii for the doubly $D$ and singly $S$ represented quark sectors
for single quark flavors of unit charge by
\begin{align}
    \expval{r^2}_p &= \phantom{-} 2\, \frac{2}{3}\, \expval{r^2}_D - \frac{1}{3} \expval{r^2}_S\, , \\
    \expval{r^2}_n &=            -2\, \frac{1}{3}\, \expval{r^2}_D + \frac{2}{3} \expval{r^2}_S\, ,
\end{align}
where charge and quark-counting factors are explicit.  Then
\begin{align}
    \expval{r^2}_D &= \frac{1}{2}\left(2\expval{r^2}_p + \expval{r^2}_n\right), \\
    \expval{r^2}_S &= \expval{r^2}_p + 2\expval{r^2}_n.
\end{align}
We approximate the doubly and singly represented strange-quark distribution radii to be that of the
light-quark at the heaviest PACS-CS quark mass where the light quark mass is similar to that
of the strange quark as seen in ~\autoref{tab:quarkmodelimplementation:latticequarkmasses}.

\begin{figure}[t]
    \includegraphics[width=\columnwidth]{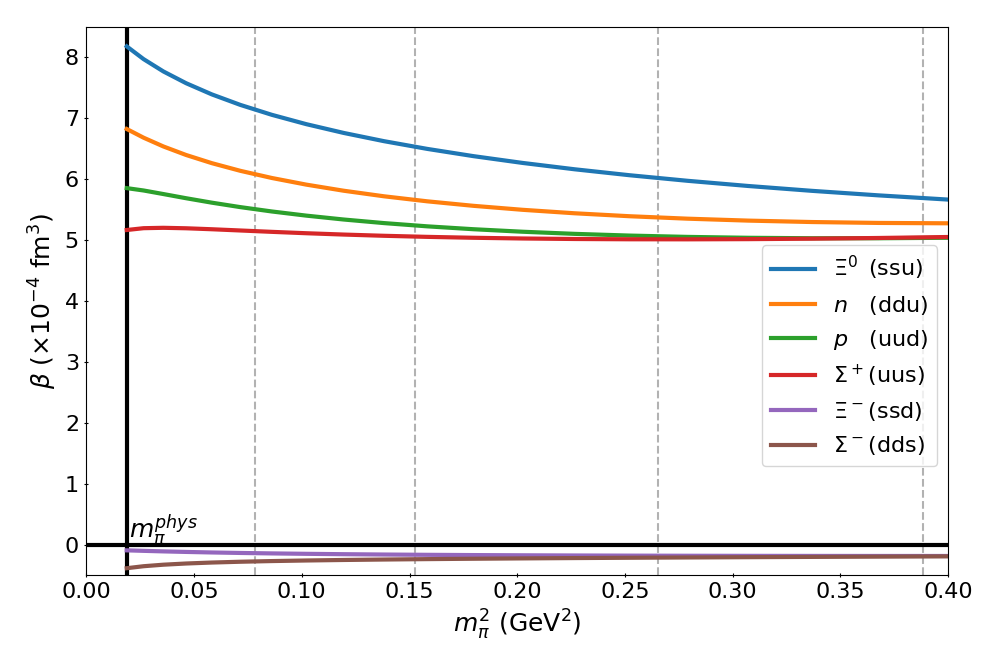}
    \caption{Quark model predictions for the magnetic polarizabilities of octet baryons are plotted
        as a function of the squared pion mass as a proxy for the quark mass. Dashed lines represent
        the pion masses of the PACS-CS ensembles. The legend is ordered to match the vertical
        ordering at the physical point. } \label{fig:quarkmodel:polarisabilitypredictions}
\end{figure}

With the octet and decuplet baryon masses, constituent quark masses, magnetic moments, and quark
flavor distribution radii parameterized as a function of $m_\pi^2$, we are now able to present
values for the magnetic polarizability as predicted by the constituent quark model as illustrated
in \autoref{fig:quarkmodel:polarisabilitypredictions}. We note that the predicted magnitudes of the
proton and neutron magnetic polarizabilities are somewhat larger than expected from experimental
measurements. While we will address this issue in the following, the trends of the model are worthy
of discussion.

First, we note the quark model predicts qualitatively different values for the negatively charged
baryons.  Their polarizabilities are predicted to be opposite in sign and much smaller in
magnitude.
Recalling Eq.~\ref{eqn:quarkmodel:polarisability} with the transition magnetic moment written
explicitly in terms of quark moments using equation Eq.~\ref{eqn:QCD:transitionmoment}
\begin{align}
    \beta &=
    \frac{4}{9\pi}\sum_{\mathcal{B}^*}\frac{\left(\mu_D - \mu_S\right)^2}{E_{\mathcal{B}^*}-E_{\mathcal{B}}}
    - \sum_{f=1}^3\frac{q_f^2\,\alpha}{6 m_f}\expval{r^2}_f, \nonumber \\
    &\equiv \beta_1 - \beta_2,
\end{align}
we first highlight the squared difference between the magnetic moments of each sector
\begin{equation}
    \left(\mu_D - \mu_S\right)^2 = \left(\frac{q_D\, e}{2 m_D} - \frac{q_S \,e}{2 m_S}\right)^2.
\end{equation}
When the quark charges of the sectors are opposed, such as in the proton, these quark moments sum
to produce a large contribution. Conversely, in the case of the negative baryons \cascadem and
\sigmam, the $d$ and $s$ quarks both carry a charge of $-1/3$ providing cancellation between the
terms. However, the splitting of the $d$ and $s$ quark masses admits a small contribution.  It is
this cancellation that gives rise to the very small nature of the magnetic polarizabilities of the
negative baryons.  Further cancellation comes from the opposite sign contribution of $\beta_1$ and $\beta_2$.
It is clear the $\beta_2$ term describing quark distributions
dominates the sum.

Conversely, it is the opposition of quark sector electric charges that is the main generator of
large magnitudes for the magnetic polarizability.  Moreover, this is achieved through the inclusion
of a $u$ quark whose charge magnitude is double that of the $d$ or $s$ quarks.  When added and
squared the contribution of opposing charge pairs, $u - d$ or $u - s$, is an order of magnitude
larger than the contribution of a single $d$ or $s$ quark flavor.

It is also insightful to consider the explicit constituent quark mass dependence of both terms. We have
\begin{equation}
    \beta_1\propto \frac{1}{m^2_f},\quad \beta_2\propto\frac{1}{m_f} \, .
\end{equation}
Given that the quark distribution radius $\expval{r^2}_f$ is also smaller for the strange quark,
together these effects suppress the magnitude of the contributions of the strange quark.

As a result, the polarizability is very sensitive to the up quark.  The up quark generates a large
$\beta_1$, but having two up quarks causes a large opposite contribution from $\beta_2$.  This
explains why the \cascadez and $n$ have greater magnetic polarizabilities than the \sigmap and
proton respectively.

Another observation is that the mass splitting between the octet and decuplet baryons in $\beta_1$
benefits the hyperon magnetic polarizabilities over the nucleon polarizabilities.  In the nucleon,
the scalar diquark is composed of two light quarks. The strength of this hyperfine interaction is
inversely proportional to the product of the constituent quark masses; hence the hyperfine
attraction is strong in the nucleon and the hyperfine repulsion is strong in the $\Delta$.
Together, these hyperfine interactions produce a large mass splitting in the magnetic transition
term, $\beta_1$.  In the hyperons considered herein, the scalar diquark is composed of a light $u$
or $d$ quark and a strange quark.  This time the magnitude of the hyperfine interaction is weaker
and the octet-decuplet mass splitting is smaller. Thus the magnitude of the magnetic transition
term, $\beta_1$, is enhanced.  This effect explains why the magnitude of the \cascadez
polarizability exceeds that of the neutron and would suggest the \sigmap exceeds the proton.

The alternate ordering of the magnetic polarizability of the proton and \sigmap highlights the
complexity of the magnetic polarizability. Here the larger mass of the strange quark causes a
reduction in the transition magnetic moment which outweighs that of the octet-decuplet mass
splitting and the reduced magnitude of $\beta_2$.

The \cascadez and \sigmap both have a $u$-$s$ scalar diquark and therefore $\beta_1$ is
similar. The difference in the total polarizability is associated with $\beta_2$ where the second
$u$ quark in \sigmap makes a larger negative contribution to the polarizability of the \sigmap
compared to the second $s$ quark in the \cascadez whose magnitude is small due to both the charge
and the mass of the strange quark.

Finally, we note that the magnetic polarizability of the \cascadez rises quite steeply as the pion
mass approaches the physical point. For all baryons, both $\beta_1$ and $\beta_2$ increase in
magnitude as the physical pion mass is approached.  However, for the \cascadez the increase in
$\beta_1$ significantly outpaces the increase in $\beta_2$ near the physical point.  It is a
combination of all aspects discussed thus far that make this possible.  The magnetic transition
term, $\beta_1$ benefits from both the presence of a $u$ quark enhancing the numerator, but also
the presence of a $u$-$s$ scalar diquark suppressing the mass splitting and thus further enhancing
$\beta_1$.  Moreover, the presence of two $s$ quarks ensures $\beta_2$ is as small as possible,
through the small electric charge, the suppression factor of the strange-quark mass and the reduced
distribution radius.

We now progress to the lattice QCD calculation to evaluate the veracity of these predictions.

\section{Background Field Implementation}\label{sec:backgroundfieldmethod}

The background-field method is a well-established approach to calculating the magnetic
polarizability in a lattice QCD calculation. The discussion here is based on
Ref.~\cite{primer2014magnetic} founded on Ref.~\cite{Smit:1986fn}.

We commence by considering the continuum formulation, where the covariant derivative is modified by
the addition of a minimal coupling
\begin{equation}
    D_{\mu} = \partial_{\mu}+i\, g\, G_{\mu} + i\, q_f\,e\, A_{\mu}\, ,
\end{equation}
where $G_{\mu}$ is the gluonic four-potential with coupling $g$, and $A_{\mu}$ is the
electromagnetic four-potential with quark-charge coupling $q_f\, e$.  On the lattice this addition
corresponds to a multiplication of the QCD gauge links by an exponential phase factor
\begin{equation}
    U_{\mu}(x) \rightarrow U_{\mu}(x)\,e^{i\, a\,q_f\,e\,A_{\mu}(x)}\, ,
\end{equation}
where $a$ is the lattice spacing.
To obtain a uniform magnetic field along the $z$-axis we use $\vec{B} = \vec{\nabla} \times
    \vec{A}$, such that
\begin{equation}
    B_z = \partial_x A_y - \partial_y A_x \, .
    \label{eqn:Bz}
\end{equation}
We exploit the second term with $A_x = -B\, y$ to produce a constant magnetic field of magnitude $B$
in the $z$-direction.

The resulting field can be calculated by examining a single plaquette in the
$(\mu,\nu) = (x,y)$ plane as
\begin{equation}
    \Box_{\mu\nu}(x) = \exp\left(i\, q_f\, e\, a^2\, F_{\mu\nu}(x) \right) \, .
\end{equation}
This relation is exact for a constant background field as all higher-order terms involve a second-
or higher-order derivative. The plaquette at coordinates $x,y$,
\begin{eqnarray}
    \Box_{\mu\nu}(x,y) &=& \exp(-i\, a\, q_f\, B\, y)\exp(i\, a\, q_f\, B(y+a)) \nonumber\\
    &=& \exp(i\, a^2\, q_f\, e\, B)\, ,
\end{eqnarray}
giving the desired field away from the $y$-direction boundary.  However, on a finite lattice with
sites labeled $(0 \le x/a \le N_x-1)$ and $(0 \le y/a \le N_y-1)$, there is a discontinuity at the
$y$ boundary due to the periodic boundary conditions used for the QCD fields. To address this, we make
use of the first term from Eq.~\ref{eqn:Bz}, $\partial_x A_y$ and assign
\begin{equation}
    A_y(x,y) = \begin{cases} 0\, ,           & \mathrm{for\ } y/a < N_y-1\, , \\
              N_y\, B\, x\, , & \mathrm{for\ } y/a = N_y-1\, .\end{cases}.
\end{equation}
This ensures that the discontinuity at the boundary in the $y$-direction at $y/a=N_y-1$ is
compensated via the $y$ boundary contribution from $A_y$.

Having used both terms available in Eq.~\ref{eqn:Bz}, the double boundary, $x/a = N_x-1$ and
$y/a=N_y-1$, gives rise to a quantization condition for the field strength.  Here the plaquette
takes the value
\begin{equation}
    \Box_{\mu\nu}(x,y) = \exp\left(i\, a^2\, q_f\, e\, B \right) \, \exp\left(-i\, a^2\, N_x\, N_y\, q_f\, e\, B \right) \, ,
\end{equation}
which provides the required value when the second exponential $\exp(-i\, a^2\, N_x\, N_y\, q_f\, e\,
    B) = 1$. Setting $a^2\, N_x\, N_y\, q_f\, e\, B$ to be an integer multiple of $2\pi$ provides the
field-strength quantization
\begin{equation}
    q_f\, e\, B = \frac{2\pi\, n}{N_x\, N_y\, a^2} \, ,
    \label{eq:qc}
\end{equation}
where $n$ is an integer specifying the field strength in multiples of the minimum field strength
quantum.

Throughout this work, we will specify the field quanta in terms of the charge of the down
quark. The quantization condition becomes
\begin{equation}\label{eqn:backgroundfield:quantizationcondition}
    e\,B=\frac{2\,\pi}{N_x\,N_y\,a^2}\frac{1}{q_d}\,k_d,
\end{equation}
such that a field corresponding to integer $k_d = 1$ is oriented in the negative $\hat{z}$-direction.

While in principle the fermion propagator source can be placed anywhere within the spatial volume,
one obtains the optimal signal-to-noise ratio in the baryon correlation functions when the source is
placed at the origin of the electromagnetic potential \cite{primer2014magnetic}. Hence, as one
increases statistics by considering many spatial source positions on a single gauge field, we cycle
the selected fermion source to the (0,0,0) position by periodic circular shifts of the gauge field.
Then the magnetic field is introduced as a phase on the gauge-field links as described above.

\section{Lattice QCD Formalism}\label{sec:magneticpolarisability}

In the presence of a uniform background magnetic field, the energy of a baryon changes as a
function of magnetic-field strength \cite{Martinelli:1982:expansion,primer2014magnetic}
\begin{equation}\label{eqn:backgroundfield:energyexpansion}
    E(B) = m + \vec{\mu}\cdot\vec{B} + \frac{|q_B\,e\,B|}{2\,m}\left(n+1\right) -
    \frac{1}{2}\, 4 \pi\,\beta\,B^2 + \order{B^3} \, .
\end{equation}
Here the mass of the baryon, $m$, is complemented by contributions from the magnetic moment
$\vec{\mu}$, the Landau term proportional to $|q_B\, e\, B|$ where $q_B$ is the charge of the
baryon, and the magnetic polarizability $\beta$. For neutral baryons, the Landau term does not
contribute, and for charged baryons, we will use a U(1) Landau-mode projection to select
$n=0$. This aspect of the calculation is detailed in
\autoref{sec:simulationdetails:landauprojection}.

This expression arises as a nonrelativistic Taylor expansion of the relativistic
energy~\cite{itzykson:QFT}. The large mass scale of the octet baryons enables the use of the
nonrelativistic approximation with systematic errors of $\order{1\%}$ at the field strengths
considered in this work. See for example, Chap. 3.6.2 of Ref.~\cite{bignellthesis}.  This
approach simplifies the required calculations and improves the signal relevant to the
polarizabilities.

The magnetic polarizability may be extracted by isolating the quadratic $B^2$ term in
Eq.~\ref{eqn:backgroundfield:energyexpansion}. The sign of the magnetic moment term is dependent on
the alignment of the baryon's spin, quantized in the $z$-direction, and the magnetic field. Hence, it may
be eliminated by summing energies associated with different spin orientations
\begin{equation}\label{eqn:polarisability:energyalignedsum}
    E_{\uparrow\uparrow}(B) + E_{\uparrow\downarrow}(B) - 2\,m = 2\,\frac{|q_B\,e\,B|}{2\,m} -
    2\,\frac{4\pi}{2}\,\beta\,|e\,B|^2 + \order{B^3} \, ,
\end{equation}
where $\uparrow\uparrow$ indicates spin-field alignment and $\uparrow\downarrow$ indicates spin-field
anti-alignment.

The magnetic polarizability and Landau terms are now isolated as the only remaining field-strength-dependent
terms. This process could be mirrored in a lattice QCD calculation, determining the
effective energies in the aligned and antialigned case, however this neglects
the opportunity to cancel the highly correlated QCD fluctuations contained within the correlation
functions of different field strengths and spin alignments.

To optimize this cancellation, we define a "spin-field aligned" correlator
\begin{equation}\label{eqn:polarisability:alignedcorrelator}
    G_{\uparrow\uparrow}(B) = G(+s,+B) + G(-s, -B) \, ,
\end{equation}
where the baryon's spin is aligned with the magnetic field and a "spin-field antialigned"
correlator
\begin{equation}\label{eqn:polarisability:antialignedcorrelator}
    G_{\uparrow\downarrow}(B) = G(+s,-B) + G(-s, +B) \, ,
\end{equation}
where the spin and field are opposed. To enable the cancellation of QCD fluctuations, we form the
correlator ratio
\begin{equation} \label{eqn:polarisability:ratio}
    R(B,t) = \frac{G_{\uparrow\uparrow}(B,t)\,G_{\uparrow\downarrow}(B,t)}{G(0,t)^2} \, .
\end{equation}

When taking the $\log$ of $R(B,t)$ in determination of the effective energy, the zero-field
correlator acts to subtract the mass term, while the numerator product results in the subtraction
of the magnetic moment, in an analogous manner to the energy sum of
Eq.~\ref{eqn:polarisability:energyalignedsum}.

As such, we define the magnetic polarizability energy shift $\delta E_{\beta}(B,t)$
\begin{align}\label{eqn:polarisability:energyshift}
    \delta E_{\beta}(B,t) &= \frac{1}{2}\frac{1}{\delta t}\lim_{t\rightarrow \infty}\log\Bigl(\frac{R(B,t)}{R(B,t+\delta t)}\Bigr),\nonumber\\
    &= \frac{1}{2}\left[
        \delta E_{\uparrow\uparrow}(B) + \delta E_{\uparrow\downarrow}(B)
        \right] - \delta E(0)\nonumber \\
    &= \frac{\abs{q_B\, e\, B}}{2m} - \frac{4\pi}{2}\beta|B|^2 +\order*{B^3}.
\end{align}

We note that this is the analog to Eq.~\ref{eqn:polarisability:energyalignedsum}. The first term,
the Landau term, vanishes for neutral baryons allowing for direct access to the magnetic
polarizability. For charged baryons, the Landau term must be carefully considered in the fitting
process.

Fitting a single-parameter quadratic fit to Eq.~\ref{eqn:polarisability:energyshift} with the
Landau term fully specified as discussed in \autoref{sec:fitting} allows the extraction of the
magnetic polarizability.

\section{Simulation Details}\label{sec:simulationdetails}
\subsection{Gauge ensembles}\label{sec:simulationdetails:gaugeensembles}

\begin{table}[tb]
    \centering
    \caption{Details of the PACS-CS ensembles used in this work. The lattice spacing of each
        ensemble is set using the Sommer scale with $r_0=0.4921(64)(+74)(-2)\,$fm. In all cases
        $\kappa_s^{\rm sea}=0.13640$ and $\kappa_s^{\rm val}=0.13665$~\cite{menaduethesis}. $N_{\rm
                    con}$ describes the number of
        configurations.} \label{tab:simulationdetails:pacsensembles}
    \begin{ruledtabular}
        \begin{tabular}{cccc}
            \noalign{\smallskip}
            $\mpipacs$(MeV) & $\kappa_{u\, d}$ & $a\,$(fm)  & $N_{\rm con}$ \\
            \noalign{\smallskip}\hline\noalign{\smallskip}
            701             & 0.13700          & 0.1022(15) & 399           \\
            570             & 0.13727          & 0.1009(15) & 397           \\
            411             & 0.13754          & 0.0961(13) & 449           \\
            296             & 0.13770          & 0.0951(13) & 399           \\
            \noalign{\smallskip}
        \end{tabular}
    \end{ruledtabular}
\end{table}

The four gauge ensembles used in this work are the four heaviest of five $2+1$-flavor dynamical
gauge configurations provided by the PACS-CS collaboration~\cite{PACS-CS2008ensembles} through
the International Lattice Data Grid (ILDG)~\cite{ILDG}. The configurations have a range of
degenerate up and down quark masses while the strange quark mass is fixed. 

The ensembles were generated in the absence of a background magnetic field.  As a result, the sea
quarks are blind to the magnetic field and the ensembles may be regarded as electroquenched.  On
the fifth and lightest ensemble, we encounter uncertainties which do not respond to increased
statistics, hinting at an exceptional configuration problem associated with the electroquenching
of the light sea-quark sector. For this reason, it has been omitted.

The strange-quark mass of the ensembles which corresponds to $\kappa_s=0.13640$ does not
extrapolate to the physical kaon mass~\cite{menadue2012lambda}. Use of $\kappa_s=0.13665$ for the
valence strange quark mass produces the correct value for the kaon mass extrapolated to the
physical point~\cite{menaduethesis}. We note that the mass of the strange quarks in the sea remain
at the heavier mass.

Each of the ensembles considered are a $32^3\times 64$ lattice. The gauge action is the Iwasaki gauge action
and the clover fermion action with $C_{\rm SW}=1.715$ is the background-field-corrected clover fermion
action~\cite{bignell:2019:cloverpion} which is tuned to remove the unphysical magnetic-field-induced
additive mass renormalization. Details of the ensembles are summarized in
\autoref{tab:simulationdetails:pacsensembles}.

As only the valence quarks interact with the background magnetic field the ensembles are
electroquenched. While it is possible to include the background field in the process of generating
each gauge-field configuration~\cite{fiebig1989polarisability}, this would require a separate
Monte Carlo simulation at each field strength. Such separate simulations would remove the
correlated QCD fluctuations which are otherwise efficiently removed through the ratio in
Eq.~\ref{eqn:polarisability:ratio}. Without this correlation, a very significant increase in
statistics would be required. Instead we preserve the correlations and estimate the small
corrections for electroquenching through chiral effective field theory in the process of
extrapolation to the physical point.

The pion masses obtained by the PACS-CS Collaboration in their renormalisation scheme are used as a
label for their ensembles. We denote these masses $\mpipacs$ as in
\autoref{tab:simulationdetails:pacsensembles}. Our analysis uses our pion masses calculated in the
Sommer scheme \cite{Sommer:1993ce} with PACS-CS $r_0$ values from Table XII in
Ref.~\cite{PACS-CS2008ensembles}.  We also use their determination of the physical value
$r_0=0.4921(64)(+74)(-2)\,$fm to set the lattice spacing for each ensemble. The pion masses
obtained in this scheme are denoted $\mpi$ and are given in
\autoref{tab:latticeresults:allpolarisability}.

\subsection{Baryon interpolating fields}

The commonly used proton interpolating field in lattice QCD is given by~\cite{Leinweber:1990dv}
\begin{equation}
    \chi_p(x) = \epsilon^{abc} \, \left[ {u^a}^T \, C \, \gamma_5 \, d^{\,b}(x)\right]\,u^c(x) \, ,
\end{equation}
where $C$ is the charge-conjugation matrix.  The interpolating fields of the other outer
octet baryons may be easily obtained through appropriate substitution of doubly and singly
represented quark flavors. This is the form of the interpolating field used throughout
this work.

Such interpolating fields, utilized purely with traditional gauge-covariant Gaussian
smearing are ineffective at isolating the baryon ground state in a uniform background
field
\cite{primer2014magnetic,Deshmukh:2018:octet,bignell2020nucleon,Bruckmann:2017pft,Chang:2015:polarisability}. The
uniform background field breaks the spatial symmetry and Landau-mode physics presents
at both the quark and hadronic levels, and this must be accommodated.

\subsection{Quark operators}

Asymmetric source and sink operators have been shown to improve the overlap of the lowest-energy
eigenstates of baryons in a magnetic field \cite{bignell2020nucleon}. Following
this work we utilize standard Gaussian smearing at the source and a low-lying eigenmode
projection at the sink. The low-mode sink projection employs QED+QCD eigenmodes such that
the quark propagators are sensitive to the dynamics of nontrivial electric charges in the
magnetic field. As the QCD+QED eigenmodes contain the effects of the breaking of spatial
symmetry by the uniform field in the $z$-direction, these modes further aid in the
isolation of the lowest-energy baryon in the correlation functions.  Finally, we include a
U(1) projection of the color-singlet baryon state for charged baryons rather than a
traditional Fourier projection.  This ensures the isolation of the
lowest Landau-level. Each of these steps is described in detail below.

\subsubsection{Link smearing}

In constructing the smeared source and sink projections described below, stout link
smearing~\cite{morningstar:2003linksmearing} is utilized on the spatially oriented gauge links. Ten
smearing sweeps are applied with an isotropic smearing parameter of $\alpha_{stout}=0.1$. These
gauge links are used in the process of $\delta$-function source smearing and in the calculation of the
sink projection via low-lying eigenmodes of the lattice Laplacian.

\subsubsection{Quark propagator source}

The quark source is constructed using three-dimensional, gauge-invariant Gaussian
smearing~\cite{gusken:1990:smearing}. In the process of smearing, we use
stout-links~\cite{morningstar:2003linksmearing} as described above.

At all quark masses, $\alpha=0.7$ is used for the Gaussian smearing. The number of
gauge-invariant Gaussian smearing sweeps considered is quark mass dependent, with smaller
numbers of sweeps associated with heavier quark masses. In tuning the smearing to optimize
the onset of early effective-mass plateaus~\cite{bignell2020nucleon}, 150-350
sweeps are utilized.

\subsubsection{Boundary conditions}

For the calculation of the quark propagators, periodic boundary conditions are used in the
spatial dimensions. To avoid signal contamination from the backward propagating states, we
use fixed boundary conditions in the temporal direction. The source is then placed at
$t=16$, one quarter of the total time-dimension length such that one is always away from
the fixed boundary by using the middle part of the lattice time dimension.

\subsubsection{Quark propagator sink}\label{sec:simulationdetails:sink}

The source construction is designed to provide a representation of the QCD interactions
with the intent of isolating the QCD ground state. The sink operators are then constructed
in such a manner as to encapsulate the quark-level physics of the electromagnetic and QCD
interaction. This is done through the eigenmode projection techniques demonstrated in
Ref.~\cite{bignell2020nucleon}, where a comprehensive explanation of the mechanisms may be
found.

In particular, the basis of eigenmodes of a fermion operator describing a quark or charged baryon
in a constant magnetic field depends only on the lattice Laplacian \cite{bignell2018neutron}.  In
other words, the Landau modes for a charged Dirac particle in a constant magnetic field $\vec B = B
    \, \hat z$ correspond to the eigenmodes of the two-dimensional U(1) gauge-covariant lattice
Laplacian.  Thus, the fully gauge-covariant sink operator is constructed by first calculating the
low-lying eigenmodes of the two-dimensional lattice Laplacian
\begin{equation}\label{eqn:sink:latticelaplacian}
    \Delta_{\vec{x},\vec{x}'} = 4\,\delta_{\vec{x},\vec{x}'} - \sum_{\mu=1,2}U_{\mu}(\vec{x})\,
    \delta_{\vec{x}+\hat{\mu},\vec{x}'}
    +U^{\dagger}_{\mu}(\vec{x}-\hat{\mu})\,\delta_{\vec{x}-\hat{\mu},\vec{x}'}\, ,
\end{equation}
where $U_{\mu}(\vec{x})$ are the full SU(3)$\times$U(1) gauge links described in
\autoref{sec:backgroundfieldmethod}.  Again, stout links are used in constructing the eigenmodes.

Due to the two-dimensional nature of the Laplacian, the modes are calculated on each $(z,t)$-slice
of the lattice independently. Considering the four-dimensional coordinate space representation of
the eigenmode with $\vec r = (x,y,z)$
\begin{equation}\label{eqn:sink:coordpsi}
    \expval{\vec r,t\,\Big|\,\pivB} = \pivB(x,y\,|\,z,t) \, ,
\end{equation}
this may be interpreted as the selection of the two-dimensional eigenmode $\pivB(x,y)$ in the
coordinate space representation on slice $(z,t)$ thus forming the full four-dimensional eigenmode
$\pivB(x,y\,|\,z,t)$.

Through the completeness relation
\begin{equation}\label{eqn:sink:completenessrelation}
    1 = \sum_{i=1}\,\ket{\psi_i}\bra{\psi_i}\, ,
\end{equation}
we form a coordinate space projection operator
\begin{eqnarray}\label{eqn:sink:projectionoperator}
    P_n\left(\vec r,t;\vec r^\prime,t'\right)
    &=& \sum_{i=1}^{n}\,\expval{\vec r, t \,\Big|\, \pivB}\expval{\pivB \,\Big|\, \vec r^\prime, t'}\, \\
    &=& \sum_{i=1}^{n}\,\expval{x, y \,\Big|\, \pivB}\expval{\pivB \,\Big|\, x',y'}\,
    \delta_{z z'}\,\delta_{t t'} \, , \nonumber
\end{eqnarray}
which may then be applied to the quark propagator at the sink.
As we are looking for low-energy states, we utilize only the $n$ lowest-lying eigenmodes of the
Laplacian. It is shown in Ref.~\cite{bignell2020nucleon}, that including too few modes results in a
noisy hadron correlation function in much the same manner as applying too many sweeps of traditional
sink smearing. As such,
the number of modes is chosen to be large enough to minimize the noise of the correlation function,
but small enough to retain the focus on the aforementioned low-energy physics. The work of
Ref.~\cite{bignell2020nucleon} found that $n=96$ modes provides balance to these two effects and is
what we use here.

\subsection{Hadronic projection}\label{sec:simulationdetails:landauprojection}

The inclusion of the background magnetic field induces a change to the wave function of a charged
baryon~\cite{roberts:2010:protonwf}. The quark level electromagnetic physics is highlighted by the
eigenmode projection at the sink. However, we must also ensure that our operator has the
appropriate electromagnetic characteristics on the hadronic level. By projecting final-state
charged baryons to the Landau state corresponding to the lowest-lying Landau level, we ensure $n=0$
in Eq.~\ref{eqn:backgroundfield:energyexpansion}.

Due to the color-singlet nature of the baryon, we need only project the eigenmodes of the U(1)
Laplacian rather than the full lattice Laplacian used for the sink. Again we follow the formalism
of Ref.~\cite{bignell2020nucleon}.

In the zero-field case, correlators are momentum projected
\begin{align}
    G\left(\vec{p},t\right) = \sum_{\vec{x}}\,e^{-i\,\vec{p}\cdot\vec{x}}\,\toG,
\end{align}
to $\vec p = 0$. As neutral baryons do not feel the effects of the background field, these
correlation functions are also momentum projected to $\vec p = 0$.

However, the standard approach of a three-dimensional Fourier projection is not
appropriate for a charged baryon when the uniform background magnetic field is
present. With a background field present, the baryon's energy eigenstates are no longer be
eigenstates of the $p_x, p_y$ momentum components. Instead, the $x,y$ dependence of the
two-point correlator is projected onto the baryon's lowest Landau level,
$\psi_{\vB}\rb{x,y}$.  A Fourier transform of the $z$-coordinate selects a specific value
for the $z$-component of momentum
\begin{align}
    G\left(p_z, \vB,t\right) = \sum_{x,y,z}\,&\pvB\left(x,y\right)\,e^{-i\,p_z\,z}\nonumber \\
    &\times \toGr.
\end{align}

In the infinite-volume continuum limit, the lowest Landau mode has a Gaussian form, $\pvB \sim
    e^{-\abs{q_Be\,B}\,\left(x^2+y^2\right)/4}$. However, in a finite volume the periodicity of the
lattice causes the wave function's form to be
altered~\cite{tiburzi:2012:projection,bignell2020nucleon}. As such, we instead calculate the
lattice Landau eigenmodes using the two-dimensional U(1) lattice Laplacian in an analogous way to
Eq.~\ref{eqn:sink:latticelaplacian}~\cite{bignell2018neutron}. Here $U_\mu$ contains only the
U(1) phases appropriate to the background magnetic field quantized on the lattice.

The correlator projection is then onto the space spanned by the degenerate modes $\pivB$
associated with the lowest lattice Landau level available to the proton
\begin{align}
    G\left(p_z, \vB,t\right) = \sum_{x,y,z}\,\sum_{i=1}^{n}\,&\pivB\left(x,y\right)\,e^{-i\,p_z\,z}
    \nonumber \\
    &\times \toGr.
    \label{eqn:19Draft}
\end{align}
The degeneracy of the lowest-lying Landau mode is governed by the magnetic-field quanta
$\abs{k_d}$.  Each mode in the finite volume has a degeneracy equal to the magnetic flux quanta,
$\abs{k_d}$. Thus, for a baryon with charge $q_B$ the degeneracy for the first and second field
strengths considered, $k_d=\pm 1,\ \pm 2$, are
\begin{equation}
    n= \left | k_d \, \frac{q_B}{q_d} \right | \, .
\end{equation}
For example, the proton's degeneracy for the first and second field strengths are 3 and 6
respectively.

In evaluating Eq.~\ref{eqn:19Draft}, one can also consider the case of fixing $n=1$ and
considering only one of the degenerate eigenmodes through an optimization procedure.  To ensure
excellent overlap with the source, we rotate the $U(1)$ eigenmode basis to maximize overlap with
the baryon source.  Recalling the baryon delta-function source in the $x$-$y$ plane
$\rho(x,y)=\delta_{x0}\, \delta_{y0}$, we proceed to optimize the overlap of the source with the
first mode $i=1$ by maximizing the value of $\left | \braket{\rho\,}{\,\psi_{i=1,\vec{B}}} \right|^2$.
An optional phase is then applied such that $\psi_{i=1,\vec{B}}(0,0)$ is purely real at the
source point.  Illustrations of these eigenmodes are provided in Ref.~\cite{bignell2020nucleon}.

\begin{figure}
    \includegraphics[width=0.99\columnwidth]{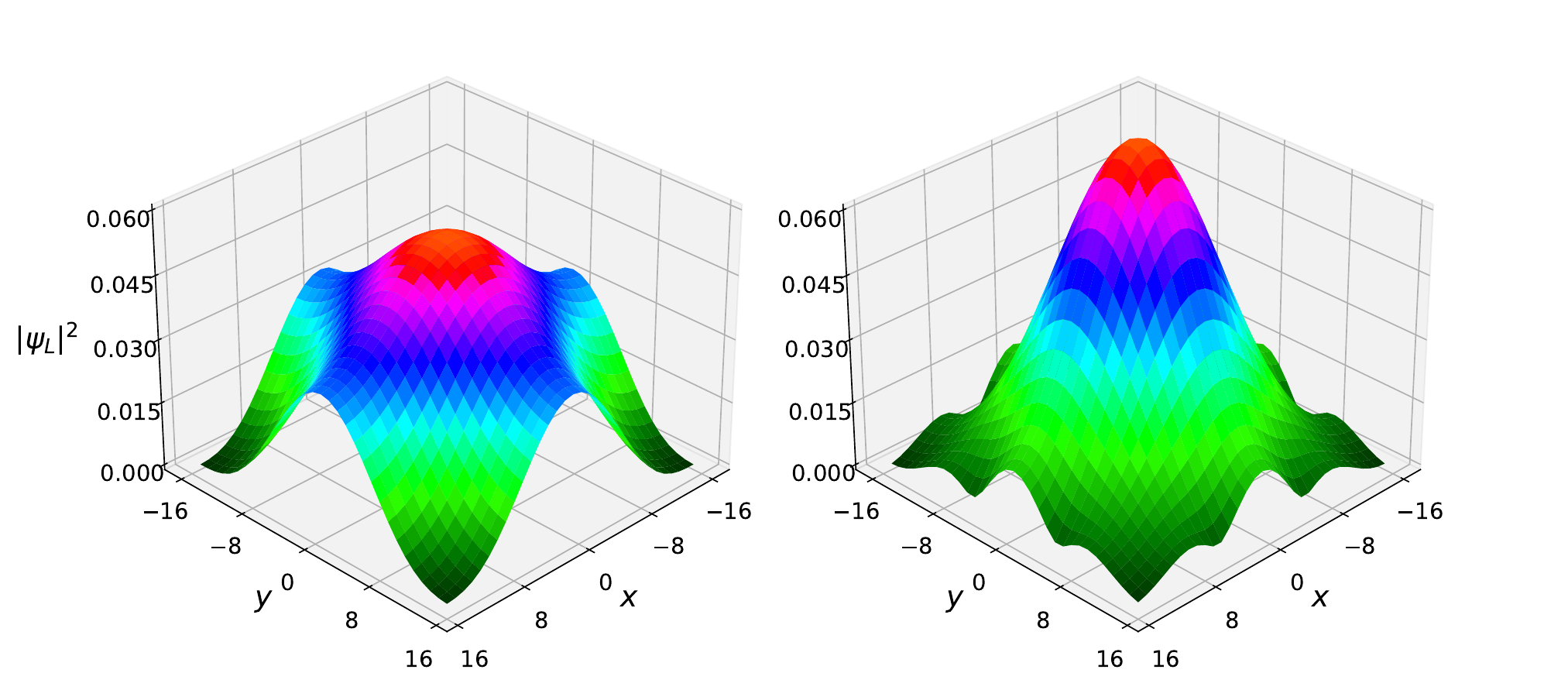}
    \caption{Landau-mode probability densities for charge $|q_B|=1$ baryons at our two lowest field
    strengths.  Recalling $n=| k_d \,{q_B}/{q_d}|$, the $n=3$ (left) and $n=6$ (right)
    probability densities are illustrated for field strengths $|k_d|=1$ and $2$ respectively. The
    degeneracy of each mode is $n$. }
    \label{fig:LandauDensities}
\end{figure}

For large field
strengths, the probability density of the projection mode has a Gaussian shape.  As the field
strength decreases the width of the Gaussian increases, making the probability distribution flatter
as zero-field strength is approached. Landau-mode probability densities for charge $|q_B|=1$
baryons in fields strengths $|k_d|=1$ and $2$ are illustrated in Fig.~\ref{fig:LandauDensities}.

This hadronic eigenmode-projected correlator offers superior isolation of the ground state
as shown in Ref.~\cite{tiburzi:2012:projection} and is crucial for the fitting of constant
plateaus in the energy shift of Eq.~\ref{eqn:polarisability:energyshift} herein.

\subsection{Statistics}

As periodic boundary conditions are used in all four dimensions for the gauge-field
generation, one can exploit the associated translational invariance of the gauge fields. A
quark source can be placed at any position on the lattice and then circularly
cycled to the standard source position of $(x,y,z,t)=(1,1,1,16)$. This enables additional
sampling of the full gauge field.

Further, the two-dimensional nature of the lattice Laplacian operator allows the
eigenmodes for the sink projection to be reused when the gauge field is cycled solely in the
time direction.
Hence, we increase our statistics on the PACS-CS ensembles by considering four random
spatial sources at $t=16$. The gauge field is then circularly cycled in the temporal
direction by an eighth of the lattice time extent (eight slices in our case) for each
random source. This results in a further increase in statistics by a factor of 8.
Together random sources and time-direction cycles increase our statistics by a factor of
32.  These multiple samples are binned and averaged as a single configuration estimate in
the error analysis.

\subsection{Magnetic field}

Baryon correlation functions are calculated for five magnetic fields corresponding to
$k_d=-2,\, -1,\, 0,\, 1,\, 2$. In doing so, quark propagators and eigenmodes are
calculated at $k_d=0,\, \pm 1,\, \pm 2,$ and $\pm 4$ in accounting for the up quark. The
nonzero field strengths correspond to magnetic fields in the $z$-direction of $e\, B=\pm
    0.087$, $\pm 0.174$, and $\pm0.348\,$GeV$^2$.

\section{Fitting}\label{sec:fitting}

As discussed in \autoref{sec:magneticpolarisability}, we construct the spin-field aligned and antialigned correlation functions
\begin{align}
    G_{\uparrow\uparrow}(B) &= G(+s,+B) + G(-s, -B) \, , \label{eqn:fitting:alignedcorrelator}\\
    G_{\uparrow\downarrow}(B) &= G(+s,-B) + G(-s, +B) \, , \label{eqn:fitting:antialignedcorrelator}
\end{align}
which are combined in
\begin{equation}\label{eqn:fitting:ratio}
    R(B,t) = \frac{G_{\uparrow\uparrow}(B,t)\,G_{\uparrow\downarrow}(B,t)}{G(0,t)^2} \, ,
\end{equation}
which aggregates the positive and negative field strengths together to remove the magnetic moment
term from the energy expansion of Eq.~\ref{eqn:backgroundfield:energyexpansion}. Due to this
aggregation, any reference to the magnetic-field strength from this point refers to the aggregated
positive field strength.

Recalling Eq.~\ref{eqn:polarisability:energyshift}, we then construct the magnetic polarizability energy shift
\begin{align}\label{eqn:fitting:energyshift}
    \delta E_{\beta}(B, t) &= \frac{1}{2}\frac{1}{\delta t}\lim_{t\rightarrow \infty}\log\Bigl(\frac{R(B,t)}{R(B,t+\delta t)}\Bigr), \nonumber \\
    &= \frac{\abs{q_B\, e\, B}}{2m} - \frac{4\pi}{2}\beta|B|^2 +\order*{B^3}.
\end{align}
Fitting to the above ratio in the long time limit and repeating this as a function of our two
positive field strengths provides access to the magnetic polarizability.

In practice, it is simple to fit in terms of the field strength quanta $k_d$. Using the
quantization condition of Eq.~\ref{eqn:backgroundfield:quantizationcondition}
\begin{equation}
    e\,B=\frac{2\,\pi}{N_x\,N_y\,a^2}\frac{1}{q_d}\,k_d,
\end{equation}
we substitute for $e\, B$ in Eq.~\ref{eqn:fitting:energyshift} to rewrite $\delta E_{\beta}$ in
terms of the field strength quanta
\begin{align}
  \delta E_{\beta}(k_d) &= \frac{q_B}{2m}\abs{\frac{2\pi}{N_xN_ya^2}\frac{1}{q_d}k_d}
  - \frac{4\pi}{2}\beta\left(\frac{2\pi}{e N_xN_ya^2}\frac{1}{q_d}k_d\right)^2, \nonumber\\
  &= \frac{2\pi}{N_x N_y a^2}\frac{1}{2  m}\abs{\frac{q_B}{q_d}  k_d} 
  - \frac{1}{2  \alpha}\left[\frac{2\pi}{N_x  N_y  a^2}\right]^2\frac{1}{q_d^2}\beta  k_d^2, \nonumber \\
  \label{eqn:fitting:fieldstrengthfitverbose} \\
  &\equiv L(k_d, m) + C  \beta  k_d^2  . \label{eqn:fitting:fieldstrengthfit}
\end{align}
It is important to note the distinction between $q_B$ which is the charge of the baryon in question
and $q_d$, the down quark charge introduced in the magnetic-field quantization condition. Here
$\alpha=\frac{e^2}{4\pi}$ is the fine structure constant.  For convenient future reference, we have
defined $L(k_d,m)$ the Landau term and $C$ the remaining coefficient to the magnetic polarizability
term.

Previous fitting procedures \cite{bignell2018neutron,bignell2020nucleon} for the magnetic
polarizability in the background-field method followed the following steps
\begin{enumerate}[(i)]
    \item Consider all field strengths under investigation and select a common fit window for the
          magnetic polarizability energy shift.  Fit the energy shift to determine $\delta E_{\beta}(k_d)$.
    \item Construct $\delta E_{\beta}(k_d) - L(k_d,m)$ by subtracting the known Landau term at each
          field strength using the best estimate for the baryon mass on that ensemble.
    \item Fit $\left[\delta E_{\beta}(k_d) - L(k_d,m)\right]/C$ as a function of the integer $k_d^2$
          to obtain the magnetic polarizability.
\end{enumerate}

Specifically, fits for $\delta E_\beta(k_d)$ were judged by the full covariance matrix
$\chi^2_{\rm d.o.f.}$ coefficient, estimated by the jackknife method~\cite{efron1979computers}. We
required $\chi^2_{\rm d.o.f.}<1.2$ at both field strengths with the start and end points of the fit window
being common for each field strength. These criteria are designed to maximise the cancellation of
correlated QCD fluctuations.
\begin{figure}
    \includegraphics[width=0.45\textwidth]{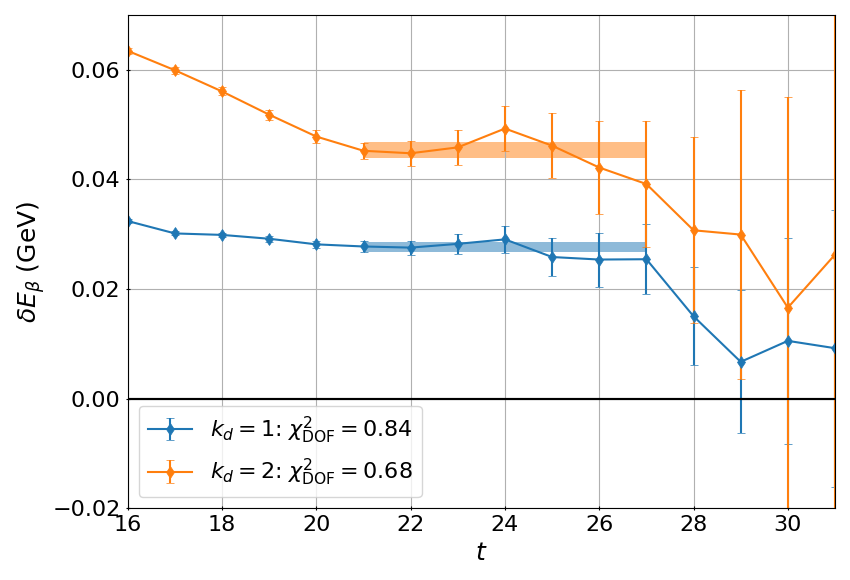}
    \caption{A plateau fit for a proton at $m_\pi=296\,$MeV. This effective energy plateau fit to
        $\delta E_{\beta}(B,t)$ is an example of a fit that looks acceptable, but is in fact fit too
        early due to underlying excited-state contamination.}
    \label{fig:fitting:poorprotonfit}
\end{figure}
An example of such a fit is shown in \autoref{fig:fitting:poorprotonfit}.

\subsection{Fitting issues}

This method presents a couple of issues. It makes the assumption that once the $\delta
    E_\beta(k_d,t)$ energy shift plateaus, the underlying correlation functions used to construct the
ratio are also independently exhibiting plateaulike behavior. That is an assumption that a
plateau in $\delta E_\beta(k_d,t)$ implies single-state isolation of all underlying correlation
functions. We have seen that this is not necessarily the case.

Consider the plateau plot for the proton in \autoref{fig:fitting:poorprotonfit} at
$\kappa=0.13770$, $\mpi=296\,$MeV. The selected fit window \mbox{[21, 27]} appears very
reasonable. The $\chi^2_{\rm d.o.f.}$ values are both acceptable, the energy shift appears plateaulike at
each field strength, and we do not appear to be fitting significant noise.

However, we need to check that the underlying correlators that form the ratio $R(k_d,t)$ of
Eq.~\ref{eqn:polarisability:ratio} have all reached single-state isolation.

\begin{figure*}
    \includegraphics[width=0.75\linewidth]{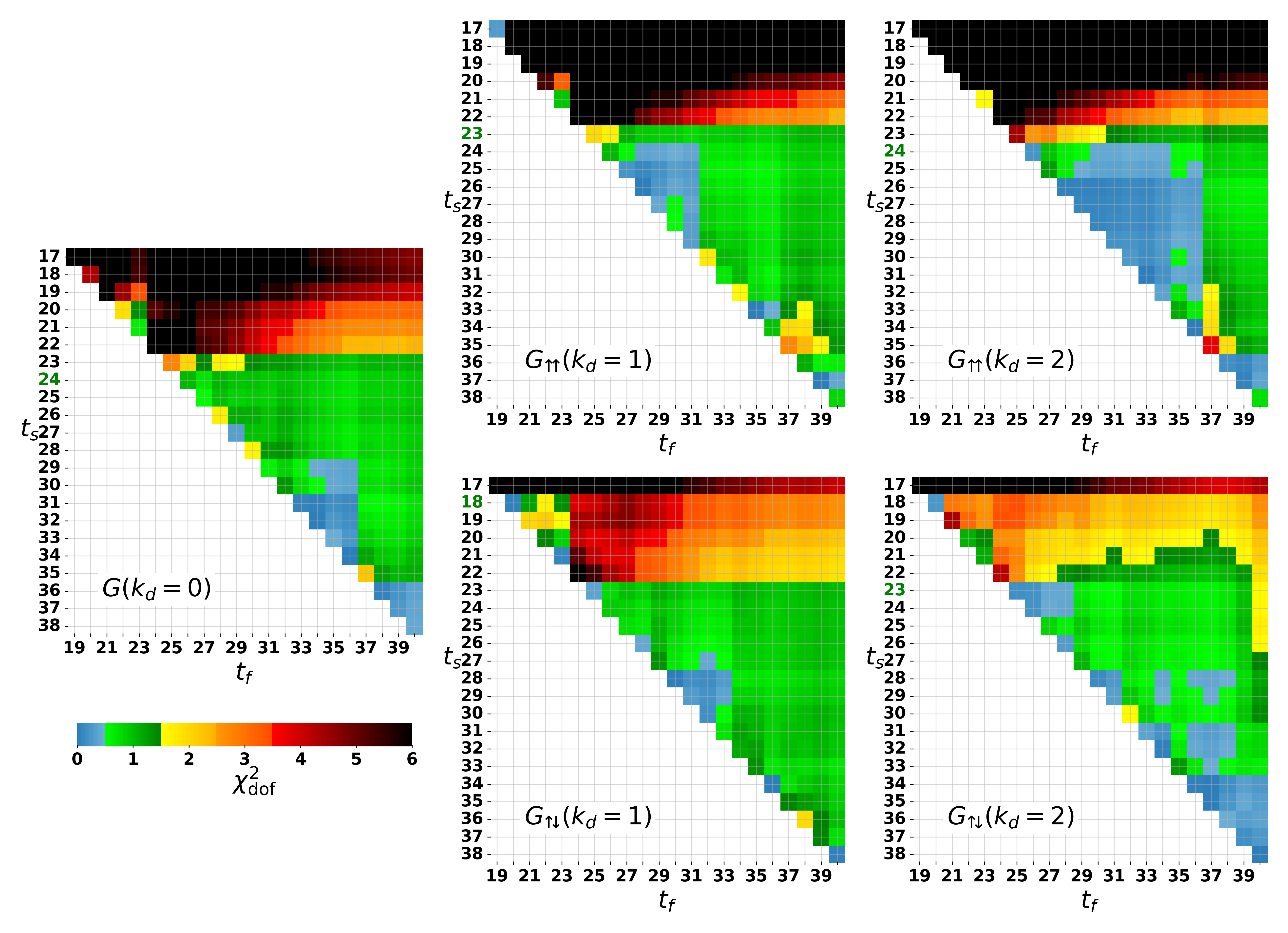}
    \caption{$\chi^2_{\rm d.o.f.}$ heat maps for fits to $\log G$ for all fit windows from $t_s \to t_f$
    of the aligned $(\uparrow\uparrow)$ and antialigned $(\uparrow\downarrow)$ correlators at
    each field strength of interest. The numbers in parentheses denote the field strength quanta
    governed by $k_d$. This example is a proton at $\mpi=296\,$MeV as in
    \autoref{fig:fitting:poorprotonfit}.} \label{fig:fitting:protonheatmaps}
\end{figure*}

The general form of these correlators is
\begin{equation}
    G(k_d,t) = \sum_{\alpha}e^{-E_{\alpha}t}\lambda^{\alpha}\bar{\lambda}^{\alpha}\, ,
\end{equation}
where $E_{\alpha}$ are the energies of this constructed state and
$\lambda^{\alpha},\bar{\lambda}^{\alpha}$ are the couplings of the baryon interpolating fields to
the baryon states at the sink and source respectively. Due to the asymmetric source and sink
operators used here, the couplings are not adjoint.
Taking the log
\begin{equation}
    \log (G_A(B,t)) \stackrel{t \rightarrow\infty}{=} \log(\lambda\bar{\lambda^\prime}) - E_0\, t \, ,
\end{equation}
which reaches a linear form in the long time limit once single-state isolation has been reached. As
such, the $\chi^2_{\rm d.o.f.}$ of a linear fit to $\log G$ is an excellent metric for single-state
isolation of the underlying correlator.

To provide an easy interface to the vast number of $\chi^2_{\rm d.o.f.}$ values which need to be checked,
we produce a series of heat maps. The heat maps corresponding to the underlying correlators of the
proton magnetic polarizability are shown in \autoref{fig:fitting:protonheatmaps}.

Each plot corresponds to a different underlying correlator. We must consider the zero-field correlator
in addition to the aligned $G_{\uparrow\uparrow}$ and antialigned
$G_{\uparrow\downarrow}$ correlators at each field strength. In each case we consider all windows
which may be of interest. The heat maps clearly show that $t_s=21$, the starting time slice chosen
in \autoref{fig:fitting:poorprotonfit}, is too early.

Noting that the $\chi^2_{\rm d.o.f.}$ has a distribution, our absolute criteria for the choice of a
starting time slice is the first three fit windows must satisfy the criteria $\chi^2_{\rm d.o.f.} <
    2.5$. We regard this criteria as a cut to ensure the fit windows considered in the averaging
procedure described below are relevant and do not adversely affect the final result. At the same
time, it is desirable to consider several meritorious fit windows such that an average weighted by
the $\chi^2_{\rm d.o.f.}$ may be considered. The details of the weighted averaging are provided in
\autoref{sec:weightedaveraging}.

The criteria $\chi^2_{\rm d.o.f.} < 2.5$ corresponds to the three leftmost boxes in a row being blue,
green, or yellow and this corresponds to plateaulike behavior across five time slices. Earliest
possible starting time slices in each case according to the criteria have been highlighted with
green text in the column denoting $t_s$.

Based on this, we see that we cannot begin fitting until time slice $t=24$, much later than $t=21$
which we selected based only on the plateau behavior of the polarizability energy shift. This
highlights the importance of checking the single-state isolation of the underlying
correlators. Pushing the fit later here changes the result significantly. We do note that, in many
cases, the polarizability energy shift does not plateau until the underlying correlators have
reached single-state isolation, however it is important to check due to cases such as the
proton presented here.

\begin{figure}
    \includegraphics[width=0.85\linewidth]{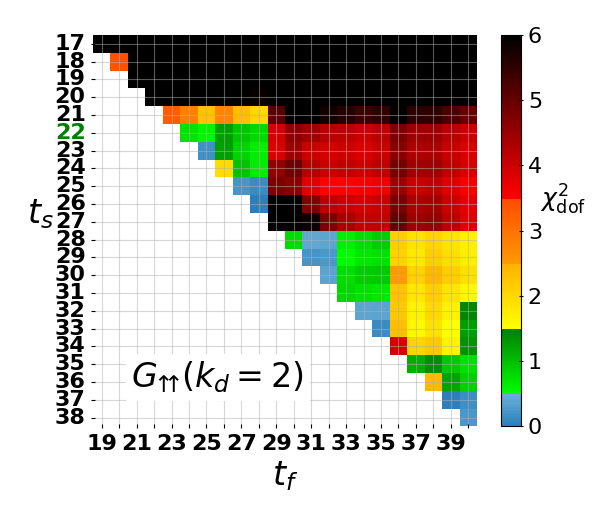}
    \caption{$\chi^2_{\rm d.o.f.}$ heat map for fits to $\log G$ for all fit windows $t_s \to t_f$ of the
    spin-field aligned correlator at the second field strength. Here we see the behavior
    exhibited in some cases in which the quality of $\log G$ fit is good in two places and
    unsatisfactory in between. This example is a neutron at $\kappa=0.13700$, $\mpi=701$
    MeV. Further inspection reveals the red quadrant is associated with a loss of signal in the
    correlator.} \label{fig:fitting:singleneutronheatmap}
\end{figure}

Finally, we do not require the fit for $\log G$ to maintain a good quality of fit indefinitely as
the correlation function can suffer a loss of signal. Instead, we use the measure as a metric for
excited state contamination. We need only determine when $\log G$ becomes linear. As an example,
consider the spin-field aligned, $k_d=2$, $\chi^2_{\rm d.o.f.}$ heat map for the neutron at
$\kappa=0.13700$, $\mpi=701\,$MeV shown in \autoref{fig:fitting:singleneutronheatmap}. Here we see
two clear regions. The criteria is satisfied at starting time slices $t_s=22,23,$ and $24$, then
again at $t_s=28$ onward. In this case, we take the earliest start point to be
$t_s=22$, as the $\chi^2_{\rm d.o.f.}$ suggests clear single-state isolation over seven time slices. This
scale is long enough to eliminate the possibility of a short-lived false plateau associated with
an excited state. Further inspection reveals the source of the red and black blocks in the heat
map commencing at $t_f=29$ is noise due to a loss of signal.

\subsection{Fit window dependence}

With confidence in hand that we are choosing fit windows with reasonable suppression of
excited-state contamination, we still need to select a fit window. This time we consider
the proton at $\kappa=0.13727$, $\mpi=570\,$MeV.

The start point criteria derived from the heat maps as described in the previous section requires
$t_s=29$ in this case. Based purely on the $\chi^2_{\rm d.o.f.}$ values for the two field strengths, any
fit window between $t=29$ and $t=35$ is reasonable. In many cases, such a flexibility in the choice
of window produces minimal variation in the resulting value. However, this is not the case here. A
summary of the polarizability obtained from various acceptable fit windows are summarized in
\autoref{tab:fitting:windowdependence}. Here we see some variation in the resulting magnetic
polarizability. This presents a problem as we desire a systematic and consistent scheme through
which we can extract the magnetic polarizability. To resolve this issue, we employ a weighted
averaging method~\cite{NPLQCD:2020multihadron} to systematically extract a consistent result from a
given set of fitting regions.

\begin{table}[tb]
    \centering
    \caption{Dependence of the magnetic polarizability of the proton $\beta_p$ on the choice
        of fit window at $\mpi=570\,$MeV. Magnetic polarizability values in are units of $\times
            10^{-4}\,$fm$^3$.} \label{tab:fitting:windowdependence}
    \begin{ruledtabular}
        \begin{tabular}{ccccc}
            \noalign{\smallskip}
            $t_s$ & $t_f$ & $\beta_p$ & $\chi^2_{{\rm d.o.f.},k_d=1}$ & $\chi^2_{{\rm d.o.f.},k_d=2}$ \\
            \noalign{\smallskip}\hline\noalign{\smallskip}
            29    & 32    & 2.41(19)  & 0.87                       & 0.42                       \\
            29    & 34    & 2.35(19)  & 0.58                       & 0.56                       \\
            30    & 34    & 2.22(25)  & 0.68                       & 0.70                       \\
            31    & 34    & 2.01(33)  & 0.41                       & 0.46                       \\
            32    & 34    & 1.78(46)  & 0.21                       & 0.47                       \\
            \noalign{\smallskip}
        \end{tabular}
    \end{ruledtabular}
\end{table}

\subsection{Weighted averaging}\label{sec:weightedaveraging}

Our weighted averaging method is based on that described in the Appendix of
Ref.~\cite{NPLQCD:2020multihadron}. Within a region $t\in [t_{\rm min}, t_{\rm max}]$, all eligible fits
are weighted based on their uncertainty, $\chi^2$, and number of degrees of freedom. We choose
$t_{\rm min}$ based on the heat maps as discussed above. $t_{\rm max}$ is chosen simply as the final time
slice before signal is clearly lost to noise. This corresponds to $t_{\rm max} = 27$ in
\autoref{fig:fitting:poorprotonfit}.

We consider a specific fit window $[t_s,t_f]$ to be eligible if the window has minimum length three
($t_f \ge t_s+2$) and $t_f = t_{\rm max}$. Fixing $t_f = t_{\rm max}$ ensures the collection of possible
fit windows is sampled equitably. Consideration of every possible end point for a fit starting at a
given time slice would favour fit windows commencing at early time slices; they would contribute on
more occasions to the weighted average than a fit that starts later. The criteria $t_f =
    t_{\rm max}$ ensures each starting time slice is considered once.

For example in a fitting range $t\in[20,24]$ where all starting points are equally favorable, the
possible windows are $[20,22],[20,23],[20,24],[21,23],[21,24],[22,24]$. We clearly see that windows
with $t_s=20$ would be favored in the calculation due simply to occurring earlier. Fixing
$t_f=t_{\rm max}$ provides $[20,24],[21,24]$ and $[22,24]$ which equitably samples possible $t_s$
values. This is important as the value of a fit is mostly determined by the first few data points
contained in a fit where the uncertainties are small.  Over counting early fits is likely to lead
to systematic errors.

With the candidate fit windows determined, the $i$th window of $N$ candidates is assigned a weight
according to
\begin{equation}\label{eq:weights}
    w_i = \frac{1}{\mathcal{Z}}\frac{p_i}{(\delta E_i)^2}\, ,
\end{equation}
where
\begin{equation}
    \mathcal{Z} = \sum_{i=1}^N\frac{p_i}{(\delta E_i)^2}\, .
\end{equation}
Here, $\delta E_i$ is the uncertainty of the $i$th fit, and $p_i$ is the $p$-value of the fit. The
$p$-value is the probability of a given fit occurring in a $\chi^2$ distribution. It is most easily
calculated using the $\gamma$ distribution and $\gamma$ function in the following manner
\begin{equation}
    p_i = \frac{\Gamma(N_{\rm d.o.f.}/2, \chi^2/2)}{\Gamma(N_{\rm d.o.f.}/2)}\, .
\end{equation}

Dividing the degrees of freedom and $\chi^2$ of the fit by two and normalising the $\gamma$
distribution in the numerator with the $\gamma$ function in the denominator causes the gamma
distribution to rescale precisely to the $\chi^2$ distribution, hence producing the probability of
any given fit occurring based on its $\chi^2$ and number of degrees of freedom.

With these weights, the average effective energy is
\begin{equation}
    E = \sum_i^n w_i \, E_i \, ,
\end{equation}
and the statistical error
\begin{equation}
    (\delta E)^2 = \sum_i^n w_i(\delta E_i)^2\, .
\end{equation}

We now apply this method to the proton correlators considered in
\autoref{tab:fitting:windowdependence}. We assign a start point $t_{\rm min} = 29$ based on the heat
map of the underlying correlators and choose $t_{\rm max} = 34$. Beyond that, all signal has been
lost to noise.

The weights for each window are shown in \autoref{tab:fitting:protonweights}. We can see that the
later windows, which are more susceptible to noise due to their larger uncertainties are
suppressed.

\begin{table}[tb]
    \centering
    \caption{Fit weights of Eq.~\ref{eq:weights} for the proton at $\mpi=570\,$MeV. Weight
        is distributed among the candidate fit windows with later fit windows
        suppressed. Magnetic polarizability values are in units of $\times
            10^{-4}\,$fm$^3$.} \label{tab:fitting:protonweights}
    \begin{ruledtabular}
        \begin{tabular}{ccccccc}
            \noalign{\smallskip}
            $t_s$ & $t_f$ & $\beta_p$ & $\chi^2_{{\rm d.o.f.},k_d=1}$ & $\chi^2_{{\rm d.o.f.},k_d=2}$ & $w_{k_d=1}$ & $w_{k_d=2}$ \\
            \noalign{\smallskip}\hline\noalign{\smallskip}
            28    & 34    & 2.35(15)  & 0.51                       & 0.77                       & 0.23        & 0.33        \\
            29    & 34    & 2.35(19)  & 0.58                       & 0.56                       & 0.28        & 0.23        \\
            30    & 34    & 2.22(25)  & 0.68                       & 0.70                       & 0.31        & 0.27        \\
            31    & 34    & 2.01(33)  & 0.41                       & 0.46                       & 0.13        & 0.11        \\
            32    & 34    & 1.78(46)  & 0.21                       & 0.47                       & 0.04        & 0.06        \\
            \noalign{\smallskip}
        \end{tabular}
    \end{ruledtabular}
\end{table}

While our procedure assigns the same $t_{\rm min}$ and $t_{\rm max}$ to both $k_d$ values governing
the field strength, we fit and weight the two field strengths independently. It is not uncommon for
their correlators to have significantly different $\chi^2_{\rm d.o.f.}$ values in spite of their
correlated nature.
We also explored the possibility of forcing common weights for the field strengths to maximize the
opportunity for cancellation of correlated fluctuations.  However, we observed this approach to
have minimal effect on the final uncertainty.  As a result, we fit and weight the two field
strengths independently.

Having defined values for $\delta E_{\beta}(k_d)$ at $k_d=1,2$, we turn our attention to the Landau
term $L(k_d,m)$ of Eq.~\ref{eqn:fitting:fieldstrengthfit}, which depends on the baryon mass. This
mass is also determined through the weighted average approach.

\begin{figure*}[p]
    {\includegraphics[width=0.9\columnwidth]{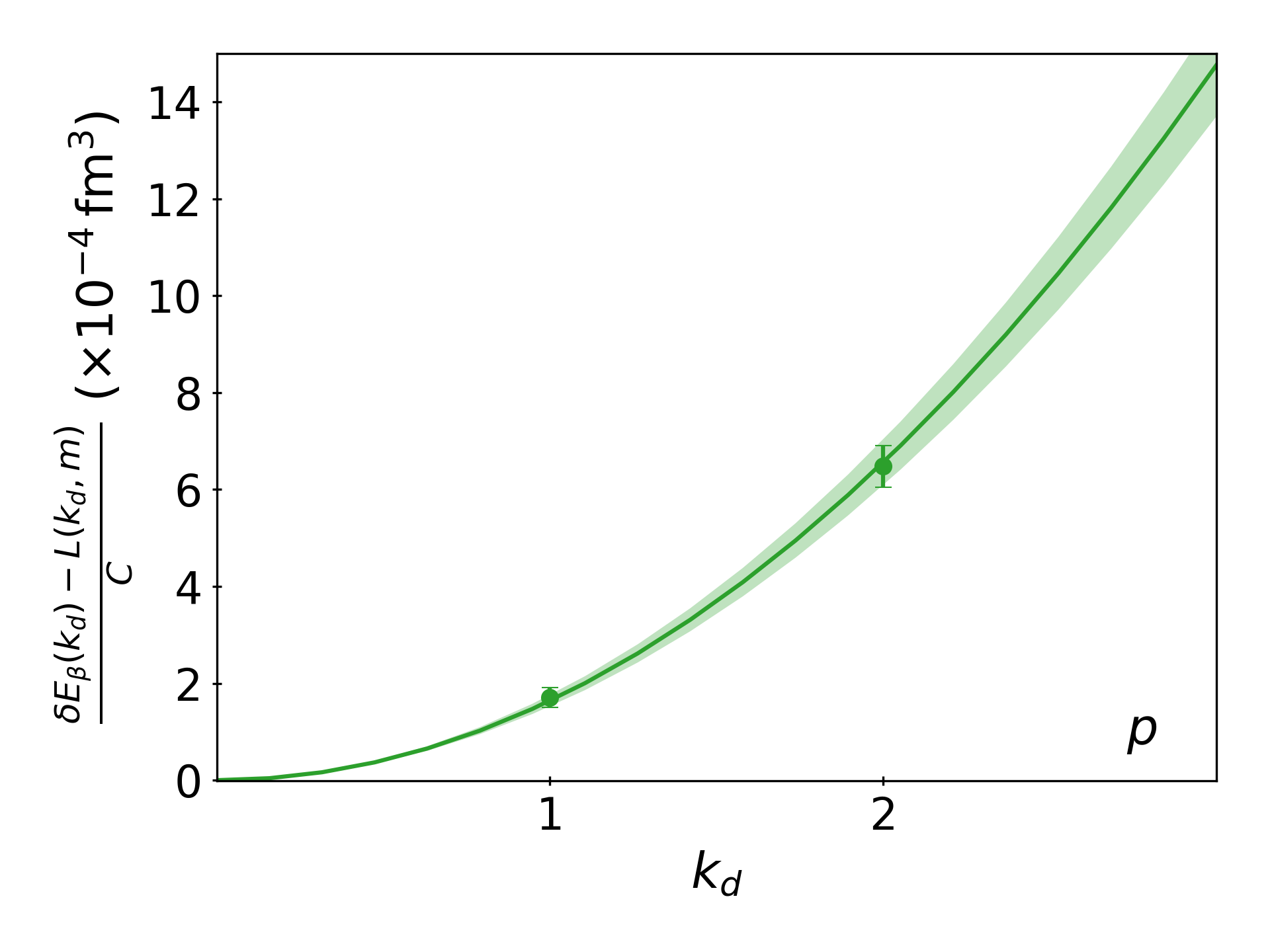}}\quad
    {\includegraphics[width=0.9\columnwidth]{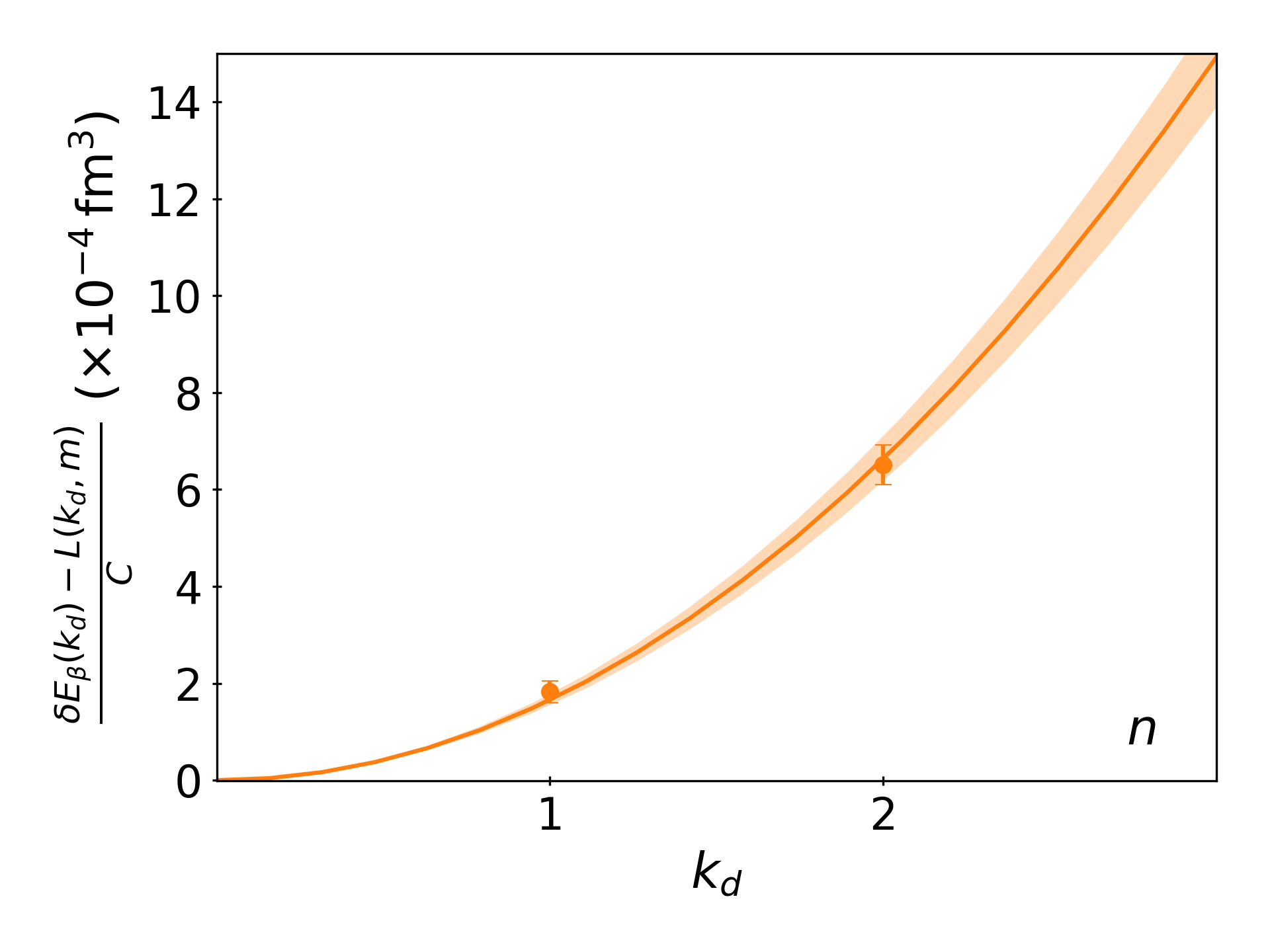}}
    {\includegraphics[width=0.9\columnwidth]{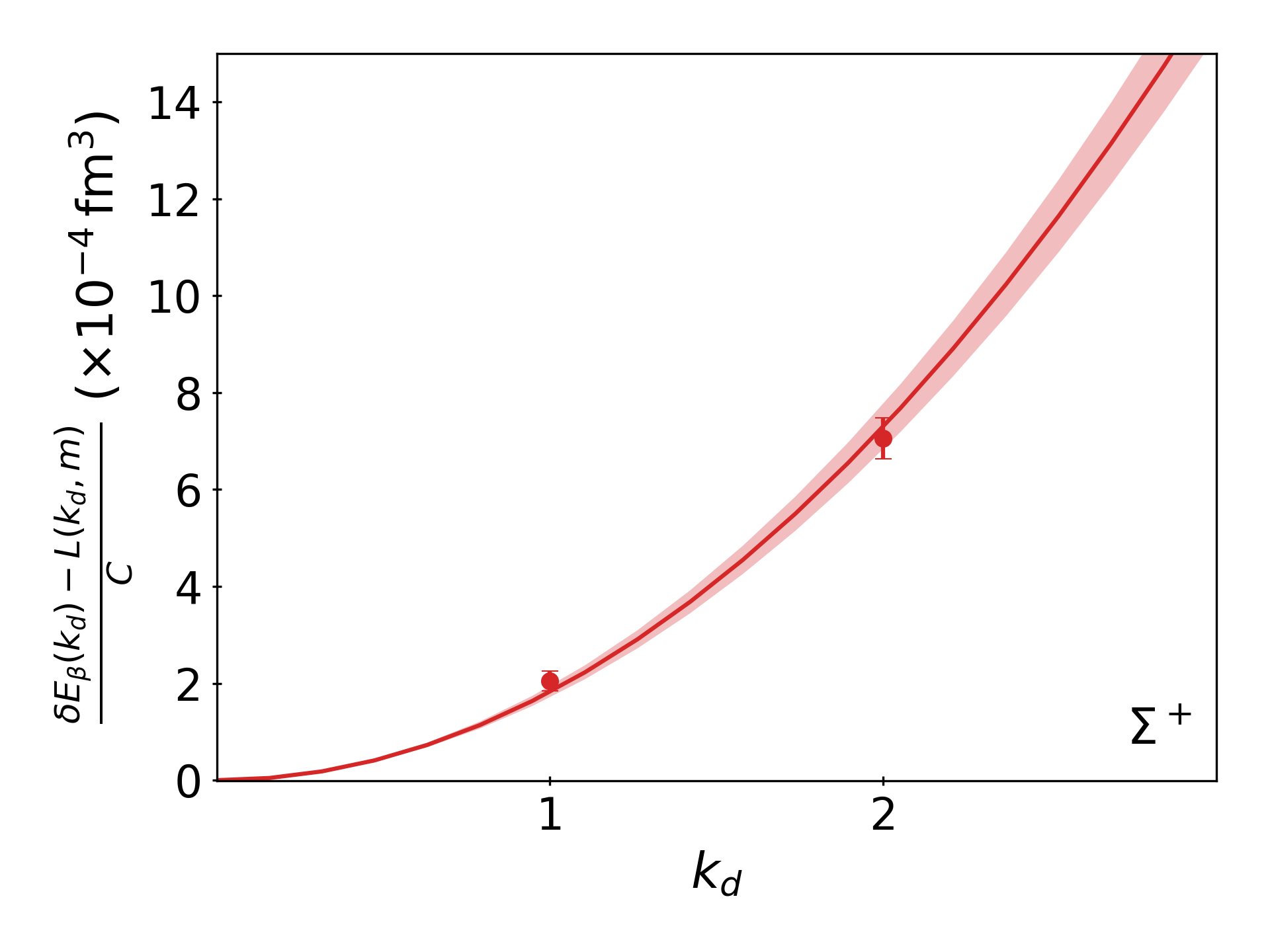}}\quad
    {\includegraphics[width=0.9\columnwidth]{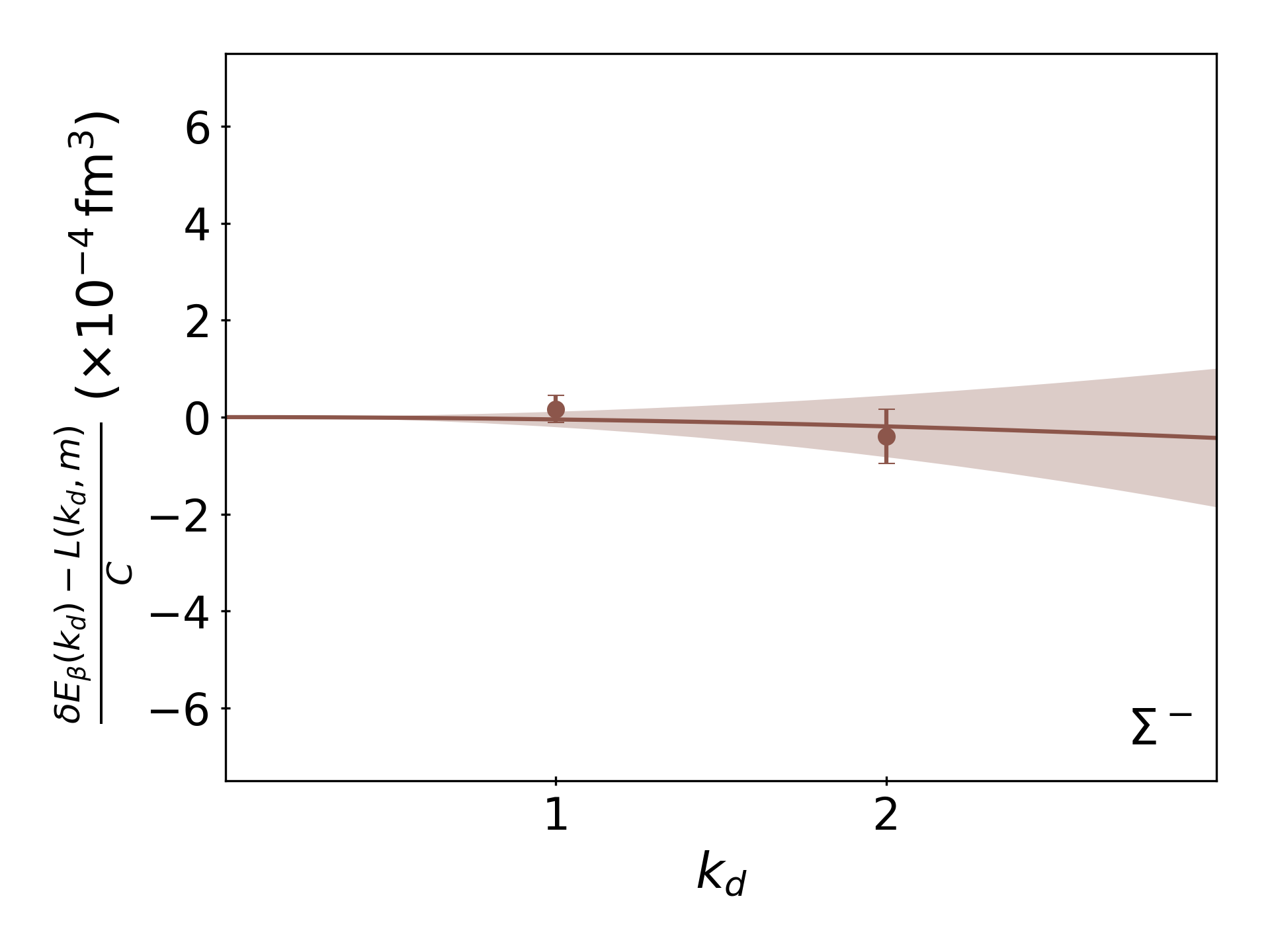}}
    {\includegraphics[width=0.9\columnwidth]{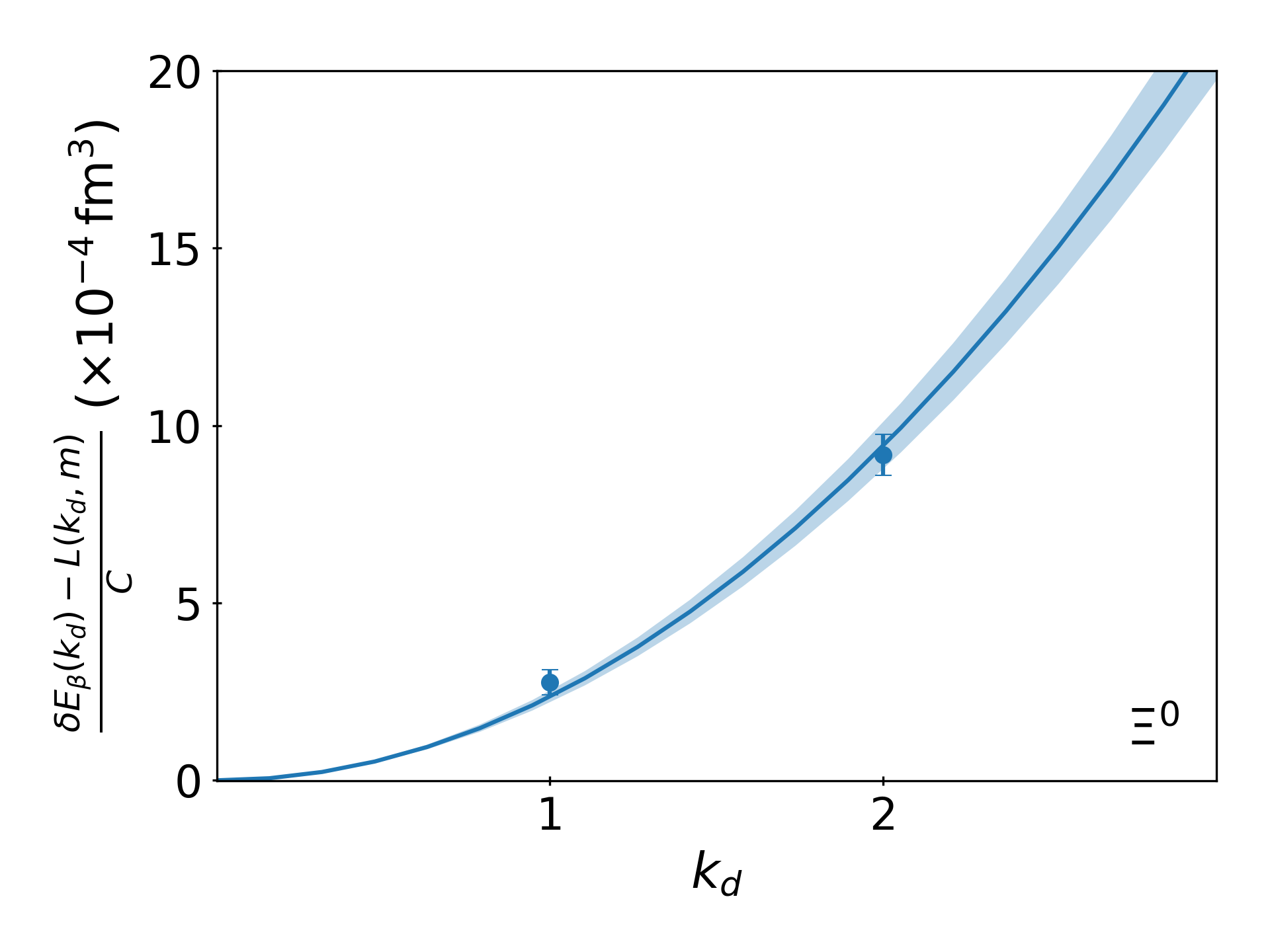}}\quad
    {\includegraphics[width=0.9\columnwidth]{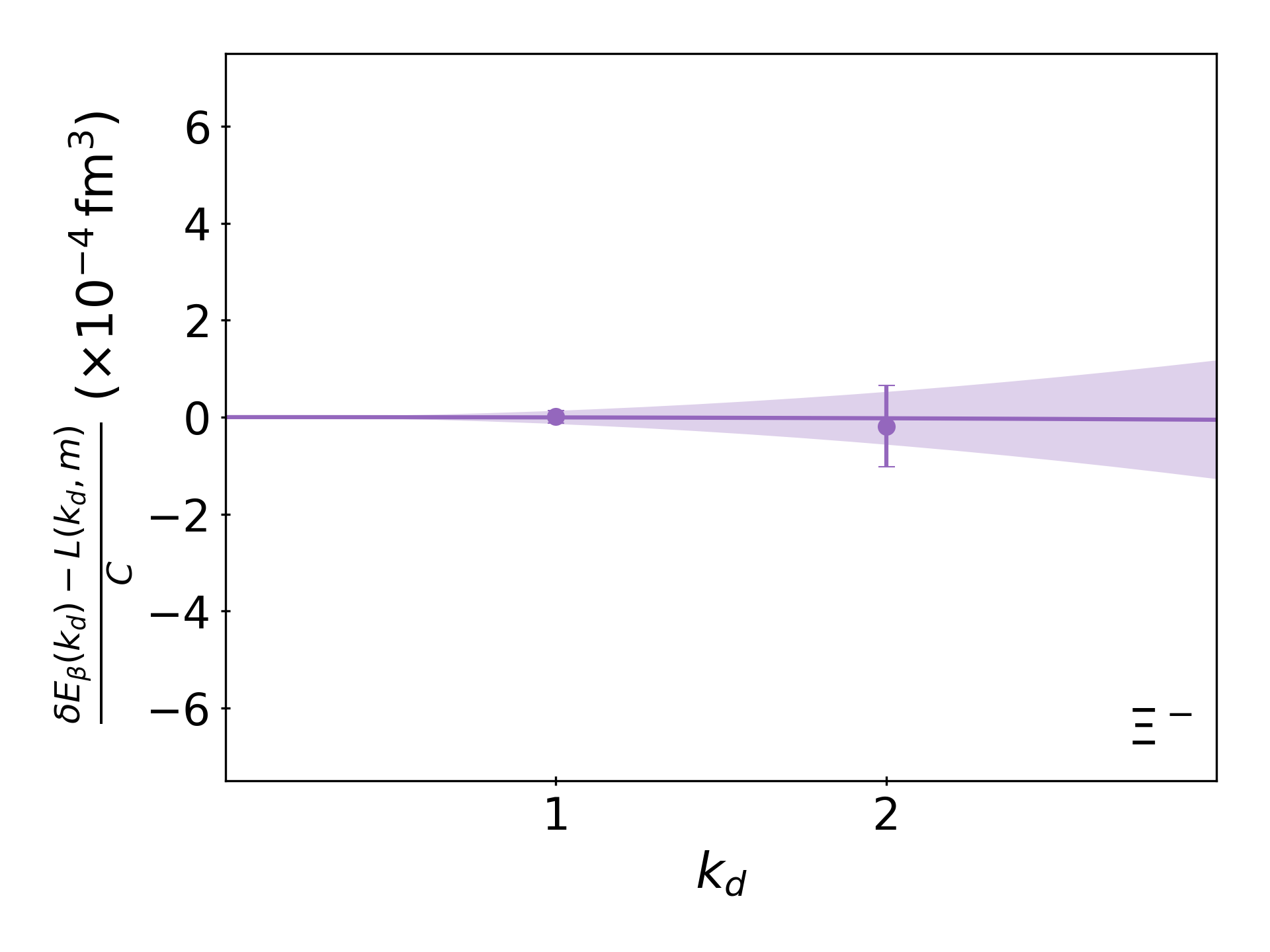}}
    \caption{Fits to Eq.~\ref{eqn:fitting:kdsqfit} as a function of $k_d$. These representative
        examples are at $\kappa=13754$, $\mpi=411\,$MeV.}
    \label{fig:fitting:kdsqfit}
\end{figure*}

The magnetic polarizability is determined in a single-parameter fit of
\begin{equation}\label{eqn:fitting:kdsqfit}
    \frac{ \delta E_{\beta}(k_d) - L(k_d,m) }{C} = \beta\, k_d^2 \, .
\end{equation}
Examples of such fits are shown in \autoref{fig:fitting:kdsqfit}.

To determine the uncertainty in the polarizability, a jackknife error estimate is performed.  A
second-order jackknife is used to obtain uncertainties on the correlation functions and correlation
function ratios.  Uncertainties for fit values such as the magnetic polarizability energy shift for
each field strength are obtained from individual first-order jackknife subensembles.  Finally, the
fit of the energy shifts as a function of field strength is repeated on each jackknife sub-ensemble
and the error on the ensemble average calculated as the jackknife error.

In this particular case, we obtain $\beta_p=2.25(21)\times 10^{-4}\,$fm$^3$ for the proton at
$\kappa=0.13727$, $\mpi=570\,$MeV. This process is repeated for every baryon at every quark
mass.

\section{Lattice Results}\label{sec:latticeresults}

\begin{figure*}
    \subfloat[\label{fig:quarkmodel:uncorrected}]{
        \includegraphics[width=0.45\textwidth]{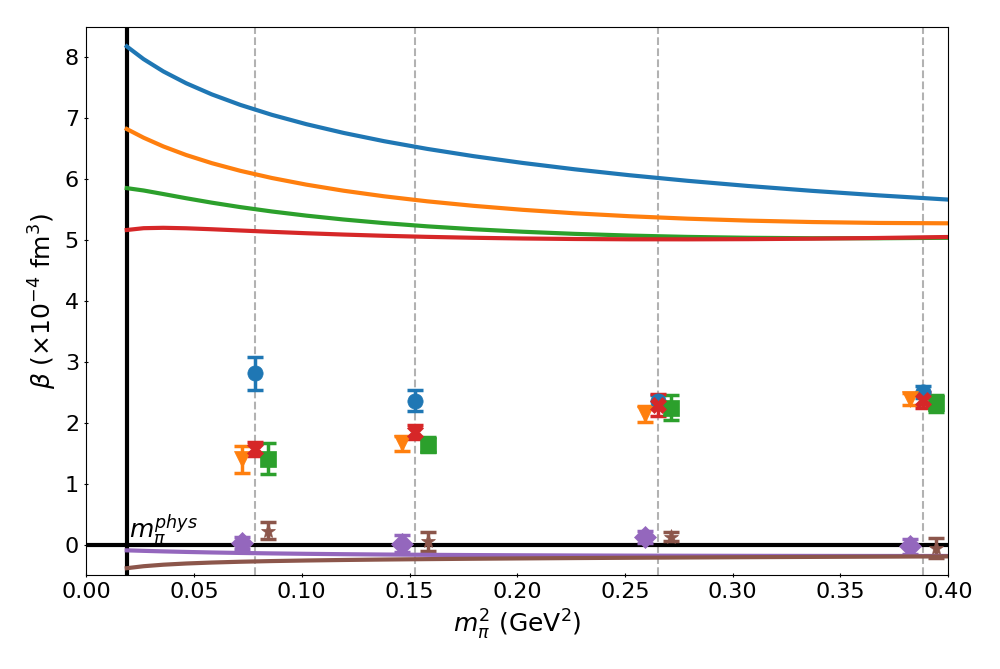}
    }
    \subfloat[\label{fig:quarkmodel:correctedandlattice}]{
        \includegraphics[width=0.45\textwidth]{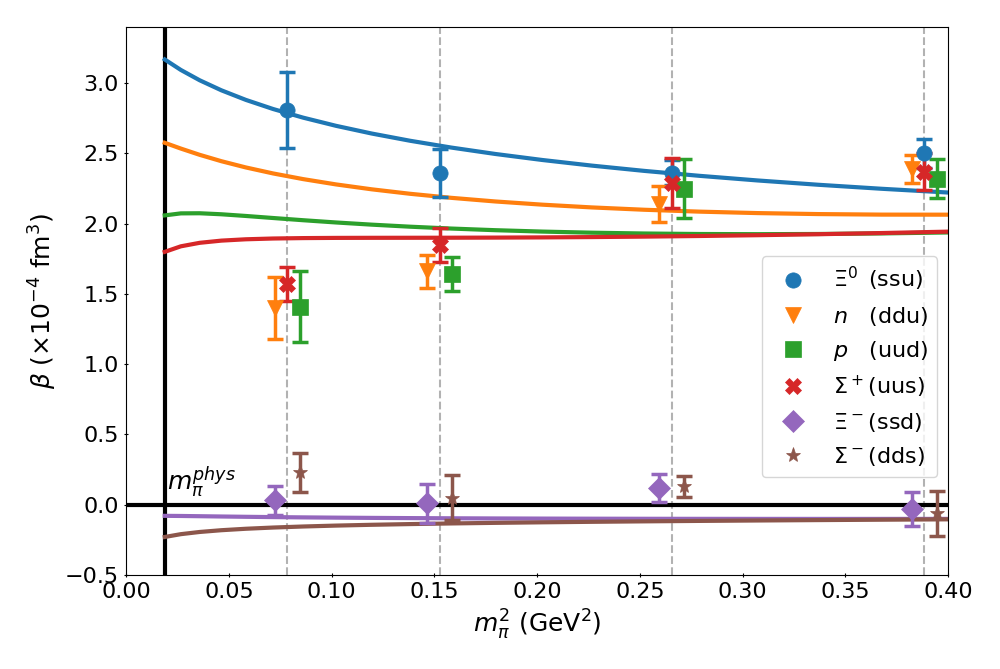}
    }

    \caption{Original (a) and scaled (b) quark model predictions for the magnetic polarizability on the PACS-CS ensembles are displayed as interpolated curves. The lattice data from \autoref{tab:latticeresults:allpolarisability} are plotted as points on both figures. The legend is common to both plots and ordered to match the vertical ordering at the first dashed line. Dashed lines represent the pion masses of the PACS-CS ensembles used in this work.} \label{fig:quarkmodelimplementation:latticepolarisabilitypredictions}

\end{figure*}

Magnetic polarizability values are presented in \autoref{tab:latticeresults:allpolarisability}
and illustrated in \autoref{fig:quarkmodelimplementation:latticepolarisabilitypredictions} in the
context of the constituent quark model of \autoref{sec:quarkmodel}. One observes a discrepancy in
magnitude between the quark model predictions and the lattice results and we will address this
below.

However, a few observations of the lattice results are worthy of note. The uncertainties in the
lattice results are sufficiently small to reveal interesting structure in the octet-baryon
polarizabilities. The $\Xi^0$ stands out in magnitude in accord with quark model expectations.
This is followed by a cluster of $n$, $p$, and $\Sigma^+$ magnetic polarizabilities of similar
magnitude, again in accord with the quark model.  Finally, the negatively charged $\Xi^-$ and
$\Sigma^-$ baryons have very small polarizabilities as the quark model predicts.

\begin{table*}[p]
    \caption{Magnetic polarizability results for the outer octet baryons on the PACS-CS
        ensembles. Polarizability values are in units of $\times10^{-4}\,$fm$^3$, and pion masses are
        in the Sommer scheme as described in \autoref{sec:simulationdetails:gaugeensembles}.} \label{tab:latticeresults:allpolarisability}
    \begin{ruledtabular}
        \begin{tabular}{cccd{1.5}cd{1.6}d{1.6}d{1.5}}
            \noalign{\smallskip}
            $\kappa$ & $m_{\pi}/$MeV               & $p$                            & \multicolumn{1}{c}{$n$}     & $\Sigma^+$
                     & \multicolumn{1}{c}{$\Xi^0$} & \multicolumn{1}{c}{$\Sigma^-$} & \multicolumn{1}{c}{$\Xi^-$}                                                   \\
            \noalign{\smallskip}\hline\noalign{\smallskip}
            0.13700  & 623                         & 2.32(14)                       & 2.390(98)                   & 2.37(13)   & 2.50(10)  & -0.06(16)  & -0.03(12) \\
            0.13727  & 515                         & 2.25(21)                       & 2.14(13)                    & 2.29(18)   & 2.364(89) & -0.130(73) & -0.12(11) \\
            0.13754  & 390                         & 1.64(12)                       & 1.66(12)                    & 1.85(12)   & 2.36(17)  & -0.05(16)  & -0.01(14) \\
            0.13770  & 280                         & 1.41(25)                       & 1.40(22)                    & 1.57(12)   & 2.81(27)  & 0.23(14)   & -0.03(10) \\
            \noalign{\smallskip}
        \end{tabular}
    \end{ruledtabular}
\end{table*}

\subsection{Improved quark model}

To address the magnitude difference between our lattice QCD results and the quark model, we
consider a simple scaling of the quark model. Recalling Eq.~\ref{eqn:quarkmodel:polarisability}
\begin{align}
\label{eqn:QM:summary}
    \beta &=
    \frac{1}{2\pi}\sum_{\mathcal{B}^*}\frac{\abs{\bra{\mathcal{B}}\hat{\mu}_z\ket{\mathcal{B}^*}}^2}{E_{\mathcal{B}^*}-E_{\mathcal{B}}}
    - \sum_{f=1}^3\frac{q_f^2\,\alpha}{6 m_f}\expval{r^2}_f \, , \nonumber \\
    &\equiv \beta_1 - \beta_2 \, ,
\end{align}
there are two contributions to the magnetic polarizability. To retain the negative contribution of
$\beta_2$, we include two positive fit parameters $a_1,\,a_2$
\begin{equation}
    \beta = a_1\,\beta_1 - a_2\,\beta_2 \, ,
\end{equation}
and scale the model by fitting these two parameters to describe the magnetic polarizabilities of
all six octet baryons at each of the four quark masses considered in lattice QCD. A simple least
squares minimization produces the fit parameters
\begin{equation}
    a_1=0.401,\quad a_2=0.532 \, .
\end{equation}
We see an approximate reduction in both terms of a half. 

There are several factors beyond the considerations of the simple quark model model presented here
that can combine to generate corrections of this magnitude.  Consider for example the quark model
calculation of Ref.~\cite{Darewych:1983yw} for the photon-decay amplitudes and widths of low-lying
strange baryons, generalized to the outer members of the baryon octet in
Ref.~\cite{Leinweber:1992pv:emtransitions}. Here additional kinematic factors are taken into
account
\begin{equation}
    \mu_{\mathcal{B}\mathcal{B}^*} = \frac{2\sqrt{2}}{3}\sqrt{\frac{m_{\mathcal{B}}}{m_{\mathcal{B}^*}}}\,e^{-K}\left(\mu_D - \mu_S \right) \, .
\end{equation}
Comparing with Eq.~\ref{eqn:QCD:transitionmoment}, the mass-ratio and $e^{-K}$ 
% \simeq 0.97$ 
factors are new and combine to generate a suppression of $\sim 15$\% in the transition moment at
the physical point, a suppression approaching 30\% upon squaring in Eq.~\ref{eqn:QM:summary} for
the transition term, $\beta_1$.

%  $e^{-K}$ is a kinematic factor which provides a small correction of $\sim3\%$. Using physical
%  masses, the mass ratio provides a correction of $\sim13\%$ for $\mu_{p\Delta}$. As the
%  transition magnetic moment is squared in $\beta_1$, these factors provide a total correction of
%  $\sim28\%$. 

Furthermore, the $\Delta$ baryon is a resonance observed in pion-nucleon scattering. Its structure is
richer than a simple three-quark single-particle state.  Two-particle $\pi N$ components mix with
the single-particle component of the quark model to form the resonance and this physics is not
considered in the simple model presented here.  The lattice results suggest that these components
act to further suppress the transition moment.

With regard to the suppression of the quark distribution term $\beta_2$, we note that a reduction
to 70\% of the rms charge radius provides a 50\% correction in $\beta_2$ as the rms charge radius
is squared in Eq.~\ref{eqn:QM:summary}.  In obtaining values for the quark distribution radii, we
have referred to results from lattice QCD. As such, the radii will include contributions from the
meson cloud that dresses the state and becomes an integral part of its structure.  One could argue
that the radii to be used in $\beta_2$ should be the distribution radii of the constituent quark
degrees of freedom within the confinement potential.  These radii, void of meson cloud
contributions, would be smaller and could account for the suppression factor derived from the
lattice QCD polarizabilities.

The scaled model is illustrated in
\autoref{fig:quarkmodelimplementation:latticepolarisabilitypredictions} beside the original
quark model. We see that the quark model now broadly agrees in a qualitative manner with the
lattice values and provides a detailed understanding of the physics underpinning the magnetic
polarizabilities of octet baryons.  Given the simplicity of the constituent quark model, the
quality of and insight provided by its predictions are significant.

We see excellent agreement for the \cascadez. We also see the clear difference in polarizability of
the \cascadem and \sigmam as predicted by the quark model, though we lack sufficient precision to
make a statement about the sign of those magnetic polarizabilities.

The ordering of baryons in the lattice data is very interesting. Correlated differences between the
proton, neutron and \sigmap cannot order them uniquely at $1\sigma$. The ordering changes for
different quark masses.

We also highlight the quark mass dependence of the proton, neutron, and \sigmap in the lattice
results. We see a clear trend downward as we approach the physical point. However, this trend is
not captured well by the quark model.

We now proceed to examine the implications of the chiral physics which dominate the interactions
near the physical point and connect these lattice results with experiment.

\section{Chiral extrapolation}\label{sec:chiralextrapolation}

\subsection{Formalism}

\begin{figure*}
    \centering
    \subfloat{
        \includegraphics[width=0.4\textwidth]{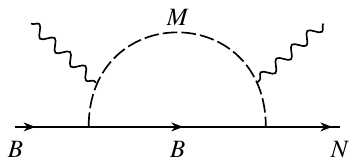}
    }
    \subfloat{
        \includegraphics[width=0.4\textwidth]{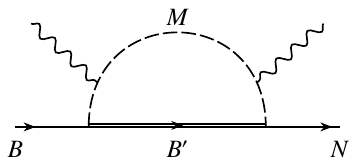}
    }
    \caption{Two-photon interactions relevant to the baryon magnetic polarizability.  In coupling
        the photons to the intermediate meson, these diagrams generate the most important
        leading nonanalytic behavior in the quark mass dependence of the magnetic polarizability of a
        baryon.  Intermediate baryon states include those degenerate with the initial baryon (left)
        and those where the intermediate state has a mass different from the initial baryon (right).
    }\label{fig:chiralextrapolation:loopdiagrambprime}
\end{figure*}

To complete our analysis, we now connect our lattice QCD results to the physical world via chiral
extrapolation.  In doing so, we draw on chiral effective field theory ($\chi$EFT) to inform
corrections for systematic uncertainties associated with the finite volume of the lattice and
electroquenching of the quark sea. This analysis follows the formalism established in
Refs.~\cite{hall2014chiral} and~\cite{bignell2020nucleon} and extends the formalism to address all
six of the outer octet baryons considered herein.

As our lattice QCD results can be described well by a fit linear in $\mpis$, we consider the
following chiral expansion for the magnetic polarizability
\begin{align}
    \beta^B(\mpis)=
    a_0^\Lambda
    &+ a_2^\Lambda \, \mpis \nonumber \\
    &+ \sum_M\,\beta^{MB}(\mpis,\Lambda) \nonumber \\
    &+ \sum_{M,B^{\prime}}\,\beta^{MB^{\prime}}(\mpis,\Lambda)+\order{m_{\pi}^3}
    , \label{eqn:chiralextrapolation:fullexpression}
\end{align}
where $a_0^\Lambda$, and $a_2^\Lambda$ are residual series coefficients which we constrain using
volume-corrected lattice QCD results. $\Lambda$ is the renormalization scale and $\beta^{MB}$ and
$\beta^{MB^{\prime}}$ are the leading-order loop contributions to the magnetic polarizability from
the diagrams of Fig.~\ref{fig:chiralextrapolation:loopdiagrambprime}.  The scale-dependent
residual series coefficients are combined with the analytic scale-dependent terms of the loop
contributions to recover the renormalized series expansion
\cite{Young:2002ib,Hall:2010ai,Hall:2012pk,Hall:2013oga}.

The loop contributions have integral forms in the heavy-baryon approximation given by
\begin{equation}\label{eqn:chiralextrapolation:betab}
    \beta^{MB}(m_{\pi}^2,\Lambda) =
    \frac{e^2}{4\pi}\frac{1}{288\,\pi^3\,f^2_{\pi}}\,\chi_{MB}
    \int d^3k\,\frac{\vec{k}^2\,u^2(k,\Lambda)}{\omega^6_{\vec{k}}}\, ,
\end{equation}
and for $\Delta M = m_{B^{\prime}} - m_B$,
\begin{widetext}
    \begin{equation}
        \beta^{MB^{\prime}}(m_{\pi}^2,\Lambda) =
        \frac{e^2}{4\pi}\frac{1}{288\,\pi^3\,f^2_{\pi}}
        \,\chi_{MB^{\prime}}
        \int d^3k\, {\vec{k}^2\,u^2(k,\Lambda)}
        \, \frac{
        \omega^2_{\vec{k}}\,\Delta M \left(
        3\,\omega_{\vec{k}}+\Delta M
        \right)
        + \vec{k}^2 \left(
        8\,\omega^2_{\vec{k}} + 9\,\omega_{\vec{k}}\, \Delta M + 3\,(\Delta M)^2
        \right)
        }{
        8\,\omega^5_{\vec{k}}\left(
        \omega_{\vec{k}}+\Delta M
        \right)
        } \, ,
        \label{eqn:chiralextrapolation:betabprime}
    \end{equation}
\end{widetext}
where $\omega_{\vec{k}}=\sqrt{\vec{k}^2+m_M^2}$ is the energy carried by the meson $M$ with three-momentum $\vec{k}$, $f_{\pi}=92.4\,$MeV is the pion-decay constant, and
\begin{equation}
    u(k,\Lambda) = \frac{1}{(1+\vec{k}^2/\Lambda^2)^2},
\end{equation}
is a dipole regulator which ensures only soft momenta flow through the effective field theory
degrees of freedom.
Equation~\ref{eqn:chiralextrapolation:betab} generates the leading nonanalytic contribution to the
magnetic polarizability proportional to $1/m_\pi$.
It is through these known contributions that we can correct for the electroquenched nature of the
calculation and estimate finite-volume corrections to the lattice results.

For each of the baryons in this work, $p$, $n$, $\Sigma^{\pm}$, $\Xi^{0,-}$, we consider
transitions to all possible octet and decuplet baryons. These transitions involve the mesons
$\pi^{\pm,0}$, $K^{\pm,0}$, $\overline K^{0}$, $\eta$, and $\eta'$. We note that
all transitions featuring a neutral meson in the intermediate state vanish in the leading-order
terms of full QCD, but not in the electroquenched theory where sea-quark charges are effectively
zero.

\begin{figure*}
    \centering
    \includegraphics[width=0.8\textwidth]{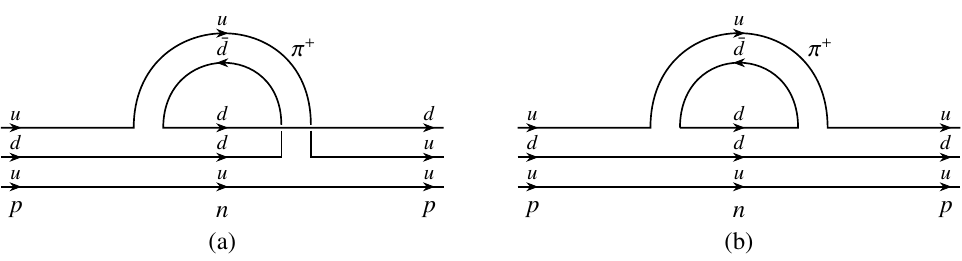}
    \caption{The fully connected quark-flow diagram (a) features only valence quarks. The
        quark-flow disconnected diagram (b) contains a sea-quark-loop contribution to a baryon
        transition to an intermediate meson-baryon state. In this case, we have a proton $p$
        transitioning to an intermediate $\pi^+ n$ intermediate state. It is straightforward to
        extend these quark-flow diagrams to consider any of the outer octet baryons with doubly and
        singly represented quark flavors.}  \label{fig:chiralextrapolation:quarkflow}
\end{figure*}

To correct for the electroquenching present in the lattice results we must first determine the
contributions $\chi_{MB}$ and $\chi_{MB^{\prime}}$ associated with $\beta^{MB^{\prime}}$ and
$\beta^{MB^{\prime}}$, respectively. These contributions are derived from the interactions in
\autoref{fig:chiralextrapolation:loopdiagrambprime} with reference to the quark-flow diagrams of
\autoref{fig:chiralextrapolation:quarkflow}.  There are two cases to examine; one where the quark
flow is fully connected and one with a disconnected sea-quark-loop contribution.

In each case we must attach two photons to the meson that is formed in the interaction. In the
connected case, all quarks must be connected to valence quarks, however in the disconnected case we
have the option of connecting one or two photons to a disconnected sea quark. As such, we have
three types of interactions. We call these "valence-valence", "valence-sea" and
"sea-sea". In equations, these are abbreviated to their initials.

In the lattice results, only the valence-valence contributions are present. The sea quarks
are electrically neutral and diagrams where a photon couples to a sea quark vanish.  As such, some
contributions to the magnetic polarizability are not present and some contributions which should
cancel out in summing the contributions of connected and disconnected quark flows do not do so.

As the chiral coefficients are analytically known, we can proceed by fitting the lattice QCD
results in the electroquenched effective field theory and then explicitly add the missing
disconnected sea-quark contributions.  Each contribution contributes proportionally with the charge
of the interacting quarks, so in the example $p\rightarrow n\,\pi^+$ shown in
\autoref{fig:chiralextrapolation:quarkflow} (right) we have
\begin{align}
    \chi_{v-v}&\propto q_u^2, \\
    \chi_{v-s}&\propto 2q_u\,q_{\bar{d}}, \\
    \chi_{s-s}&\propto q_{\bar{d}}^2,
\end{align}
where the factor of two comes from the two possible orderings of the photon attachment.

The coupling strengths may be obtained in a partially quenched chiral perturbation theory scheme
\cite{Leinweber:2002qb} where the flavor of the disconnected sea quark is relabeled to the missing
SU(3) quark flavor. In this way the quark disconnected contribution can be isolated in terms of
known meson-baryon dressing coefficients.  For the case under consideration here, that is the
strange quark. In this case the intermediate state would become \mbox{$\Lambda / \Sigma^0$ +
    $K^+$}, hence the sea-sea contribution becomes
\begin{align}
    \chi_{s-s} &\propto q_{\bar d}^2\, (\chi_{K^+\Sigma^0}^2 + \chi_{K^+\Lambda}^2), \nonumber\\
    &= q_{\bar d}^2 \, \bigl((D-F)^2 + \frac{1}{3}(D+3F)^2\bigr) \, ,
\end{align}
where $F$ and $D$ are the standard axial coupling constants.
The valence-sea contributions are obtained equivalently by simply adjusting the
charge coefficients and counting the two possible orderings for the photon couplings
\begin{align}
    \chi_{v-s} &\propto 2\, q_u \, q_{\bar d} \, (\chi_{K^+\Sigma^0}^2 + \chi_{K^+\Lambda}^2), \nonumber\\
    &= 2\, q_u \, q_{\bar d} \, \bigl((D-F)^2 + \frac{1}{3}(D+3F)^2\bigr)\, .
\end{align}

With the disconnected quark-flow contributions determined, the valence-valence contribution
is obtained by subtracting the valence-sea and sea-sea contributions from the full
QCD contribution.  In some cases the full QCD contribution may be zero leading to equal and
opposite contributions from the connected and disconnected quark-flow contributions.  However, in
the case under consideration here, the $p\to n\pi^+$ transition makes a contribution proportional
to $\chi_{\pi^+ n}^2$ such that
\begin{align}
    \chi_{v-v} &\propto q_{\pi^+}^2 \, \chi_{\pi^+ n}^2 - (2\, q_u \, q_{\bar d} + q_{\bar d}^2) \, (\chi_{K^+\Sigma^0}^2 + \chi_{K^+\Lambda}^2), \nonumber\\
    &=(2D + F)^2 \nonumber\\
    &\quad- (2\, q_u \, q_{\bar d} + q_{\bar d}^2) \, \bigl((D-F)^2 + \frac{1}{3}(D+3F)^2\bigr)\, .
\end{align}
This process is repeated for all possible transitions for each of the outer octet baryons.  All
relevant chiral coefficients and associated contributions are summarized in
\autoref{appendix:chiralcoeffs}.

We note that we have not included transitions to an intermediate $\bar s s$ pseudoscalar meson.
While it does not contribute to full QCD processes, its consideration can in principle lead to a
contribution in the process of separating valence and sea contributions.  However, the mass of the
$\bar s s$ pseudoscalar meson is large at approximately $\sqrt{2 m_K^2 - m_\pi^2} \simeq 685\,$MeV.  In the
finite-range regularization used here, the contributions from such large-mass mesons are naturally
suppressed, such that the partially quenched correction is negligible.

It is interesting to note that transitions involving $\pi^0$, $\eta$, and $\eta'$ intermediate
mesons do not contribute to this analysis.  In full QCD these mesons are electrically neutral and
do not generate a contribution to the leading loop integrals. In partially quenched QCD, their
composition of $\bar u u$, $\bar d d$, and $\bar s s$ matched flavors means they are not relevant
to the disconnected flow calculations which by definition include a third quark flavor that
differs from the quark flavors present in the outer octet baryons under consideration.  As such,
subtleties associated with SU(3) flavor symmetry breaking in the $\pi^0$, $\eta$, and $\eta'$
masses do not enter \cite{Leinweber:2022guz}.

\subsection{Implementation}

In the process of addressing finite-volume corrections and electroquenching, we proceed as follows.
We first leverage the leading loop integrals to inform the size of finite volume effects. The
finite volume of the lattice requires infinite-volume integrals over momenta such as those in
Eqs.~\ref{eqn:chiralextrapolation:betab} and \ref{eqn:chiralextrapolation:betabprime} to be
calculated as sums. As such, the finite-volume effects associated with the effective field theory
are given by the difference between the continuous infinite-volume integrals and the discrete
sums. Hence, the finite-volume-corrected (FVC) valence-valence lattice polarizability is
\begin{align}\label{eqn:chiralextrapolation:volumecorrecteddata}
    \beta^{\rm FVC}_{v-v}(\mpis) = &\beta^{\rm lattice}_{v-v}(\mpis) \nonumber \\
    + &\sum_{MB}\,\Big(\beta^{MB}_{\rm integ}(\mpis,\Lambda^{FV}) - \beta^{MB}_{\rm sum}(\mpis,\Lambda^{FV})\Big) \nonumber\\
    + &\sum_{MB^{\prime}}\Big(\beta^{MB^{\prime}}_{\rm integ}(\mpis,\Lambda^{FV}) -
    \beta^{MB^{\prime}}_{\rm sum}(\mpis,\Lambda^{FV})\Big),
\end{align}
where the idea is to subtract the finite-volume sum (sum) contained within the lattice results and
replace it with the infinite-volume integral (integ).  Here, the regulator parameter $\Lambda^{FV}$
is selected to be different from that of the main expression as one needs to avoid a collision
between infrared and ultraviolet effects \cite{hall2014chiral}.  We consider, $\Lambda^{FV} =
    2.0\,$GeV as in Ref.~\cite{hall2014chiral}.

The residual series coefficients $a_0(\Lambda)$,$a_2(\Lambda)$ are now obtained by fitting to the
finite volume corrected, valence-valence lattice results according to
\begin{align}
    \beta^{\rm FVC}_{v-v}(\mpis) =
    a_0^\Lambda
    &+ a_2^\Lambda\, \mpis \nonumber \\
    &+ \sum_M\,\beta^{MB}_{v-v}(\mpis,\Lambda) \nonumber \\
    &+ \sum_{M,B^{\prime}}\,\beta^{MB^{\prime}}_{v-v}(\mpis,\Lambda)+\order{m_{\pi}^3},
    \label{eqn:chiralextrapolation:valencevalenceextrapolation}
\end{align}
where the subscripts ${v-v}$ indicate the use of the valence-valence chiral coefficients in
place of the full QCD coefficients $\chi_{MB}$ and $\chi_{MB'}$ in
Eqs.~\ref{eqn:chiralextrapolation:betab} and \ref{eqn:chiralextrapolation:betabprime}.

This time we take the phenomenologically motivated value from the induced pseudoscalar form factor
of the nucleon with $\Lambda=0.8\,$GeV for the regulator parameter
\cite{Wang:2008vb,Young:2002cj,Leinweber:2004tc,Leinweber:2006ug,Wang:2008xpc}. Such a value defines
a pion cloud contribution to masses \cite{Young:2002cj}, magnetic moments \cite{Leinweber:2004tc},
and charge radii \cite{Wang:2008vb} allowing for the correction
of the sea-quark-loop contributions to the pion cloud which play a significant role, especially at
small pion masses. At this regulator mass, the nucleon core contribution governed by the residual
series coefficients is insensitive to sea-quark-loop contributions.

With the values for $a_0^\Lambda$ and $a_2^\Lambda$ determined in the fit of
Eq.~\ref{eqn:chiralextrapolation:valencevalenceextrapolation}, the electroquenching correction
is incorporated by simply replacing the valence-valence coefficients by the full-QCD
coefficients, {\it i.e.}  by simply evaluating the full expression for the magnetic
polarizability in Eq.~\ref{eqn:chiralextrapolation:fullexpression}.  The resulting values for the
octet-baryon magnetic polarizabilities at the physical pion mass $m^{\rm phys}_{\pi} = 140\,$MeV are
given in \autoref{tab:chiralextrapolation:extrapolatedresults}.

Systematic uncertainties associated with the chiral extrapolation are estimated through variation
of the regulator parameter. The parameter is varied over the broad range $0.6 \leq \Lambda \leq 1.0\,$
GeV to allow the estimation of the uncertainty associated with the higher terms
of the chiral expansion. This systematic uncertainty is also included in
\autoref{tab:chiralextrapolation:extrapolatedresults}.

\begin{table}[tb]
    \centering
    \caption{Magnetic polarizability values for the outer octet baryons in full QCD at the
        physical pion mass, $m^{\rm phys}_{\pi} = 0.140\,$GeV, in infinite volume. The values include
        both finite-volume and electroquenching corrections as described in the text.  All
        values are in units of $\times 10^{-4}\,$fm$^3$. Uncertainties include statistical
        simulation uncertainties and systematic uncertainties associated with the chiral
        extrapolation. These are combined in quadrature in the final column.
    } \label{tab:chiralextrapolation:extrapolatedresults}
    \begin{ruledtabular}
        \begin{tabular}{ccccc}
            %        \noalign{\smallskip}
                       &         & \multicolumn{3}{c}{Uncertainties}                         \\
            Baryon     & $\beta$ & Statistical                       & Systematic & Combined \\
            \noalign{\smallskip}\hline\noalign{\smallskip}
            $p$        & 2.10    & 0.17                              & 0.16       & 0.23     \\
            $n$        & 2.11    & 0.15                              & 0.16       & 0.22     \\
            $\Sigma^+$ & 1.83    & 0.12                              & 0.06       & 0.13     \\
            $\Sigma^-$ & 0.51    & 0.15                              & 0.09       & 0.17     \\
            $\Xi^0$    & 2.58    & 0.18                              & 0.04       & 0.19     \\
            $\Xi^-$    & 0.11    & 0.11                              & 0.03       & 0.11     \\
            %        \noalign{\smallskip}
        \end{tabular}
    \end{ruledtabular}
\end{table}
%}

\begin{figure*}
    \subfloat[\label{fig:chiralextrapolation:proton_extrap}]{
        \includegraphics[width=0.49\textwidth]{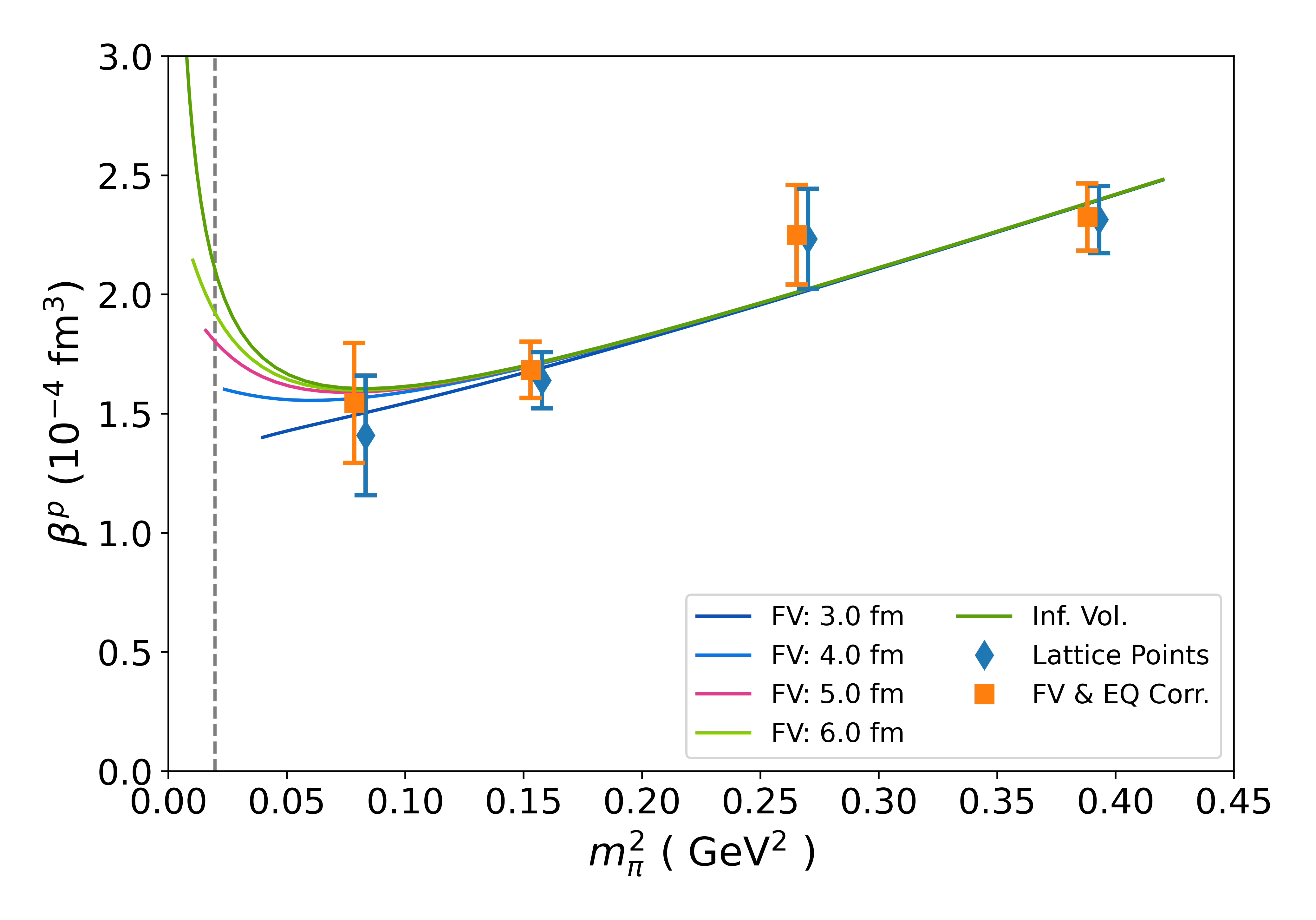}
    }
    \subfloat[\label{fig:chiralextrapolation:neutron_extrap}]{
        \includegraphics[width=0.49\textwidth]{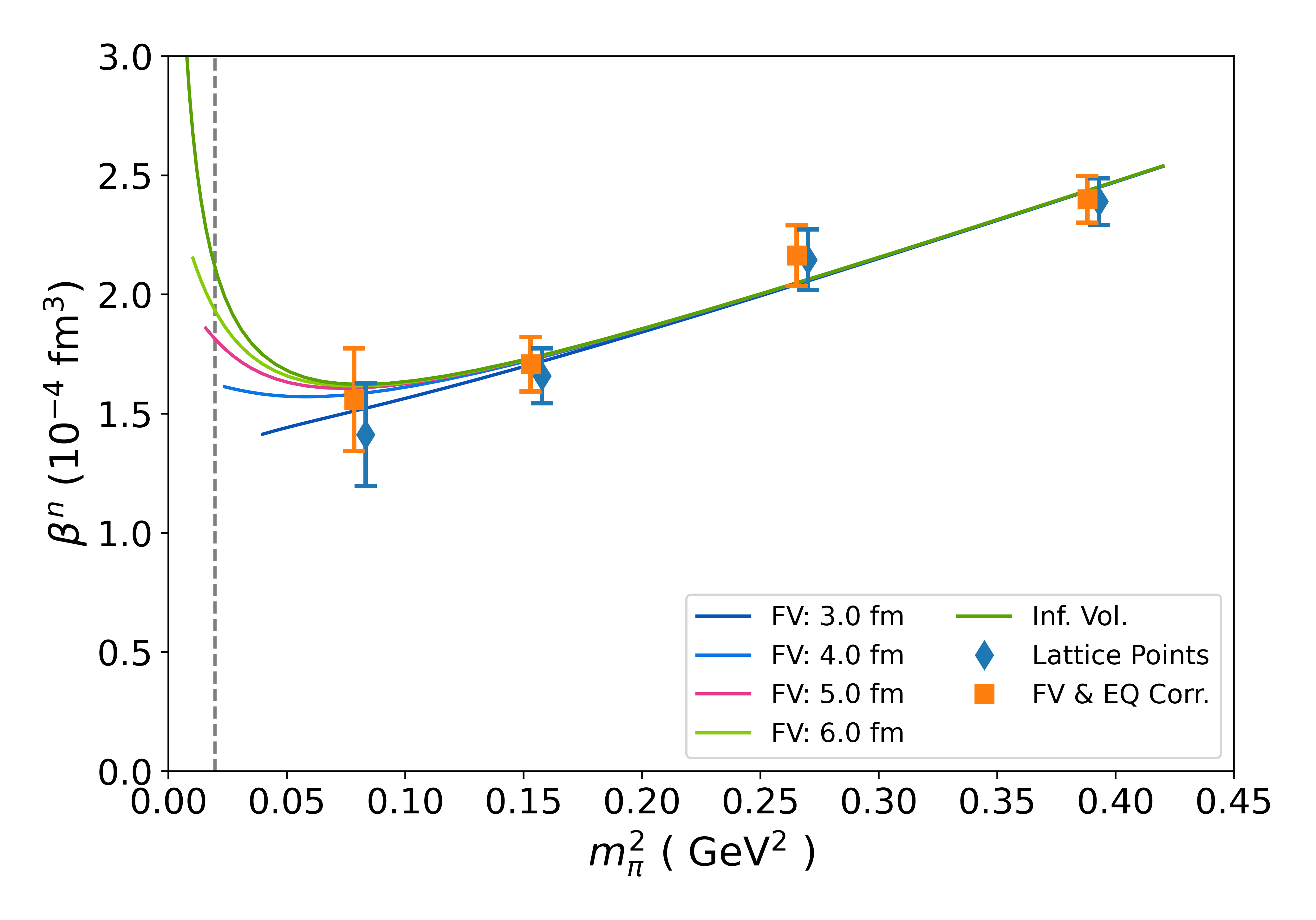}
    }
    \hspace{0mm}
    \subfloat[\label{fig:chiralextrapolation:sigmap_extrap}]{
        \includegraphics[width=0.49\textwidth]{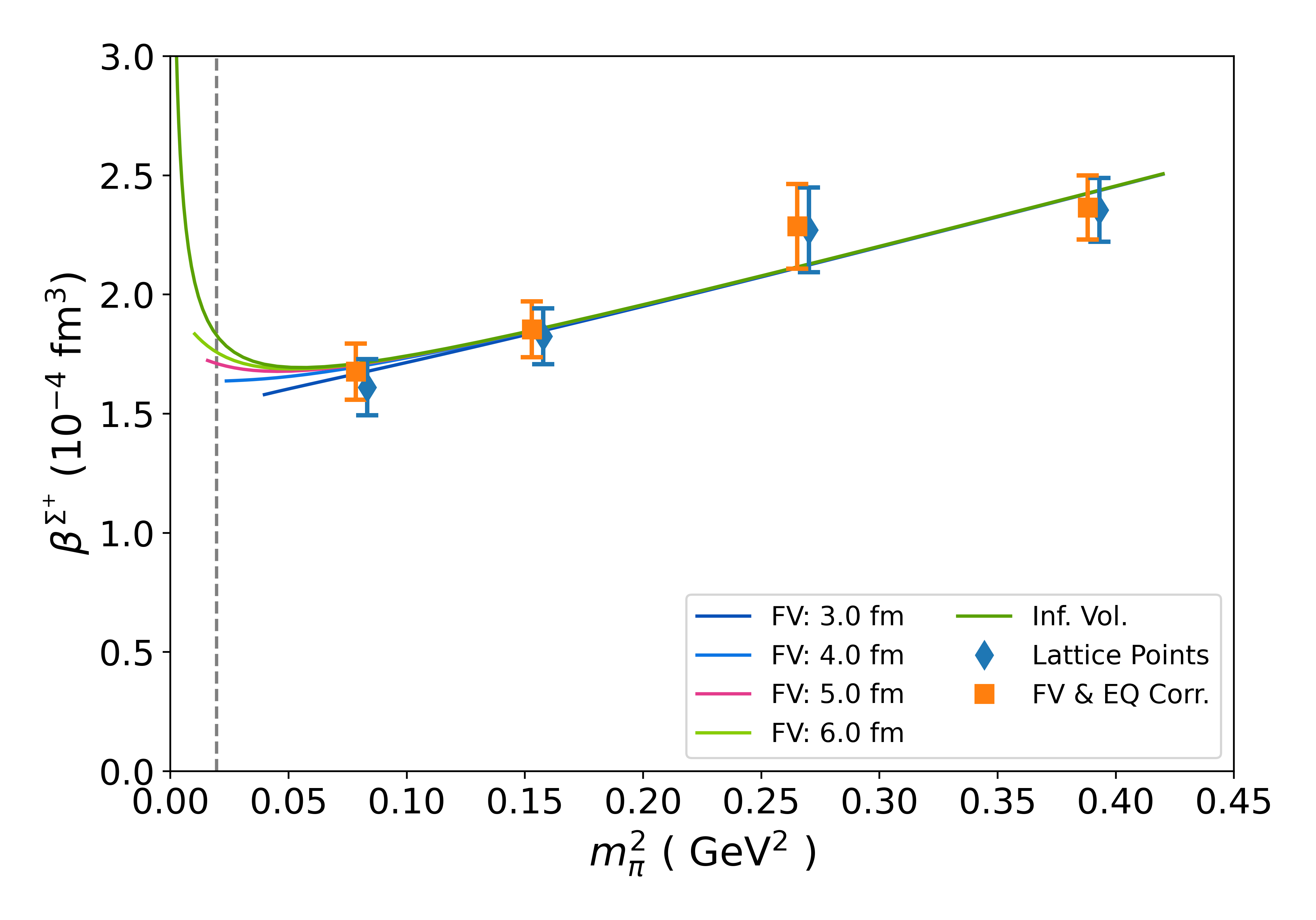}
    }
    \subfloat[\label{fig:chiralextrapolation:sigmam_extrap}]{
        \includegraphics[width=0.49\textwidth]{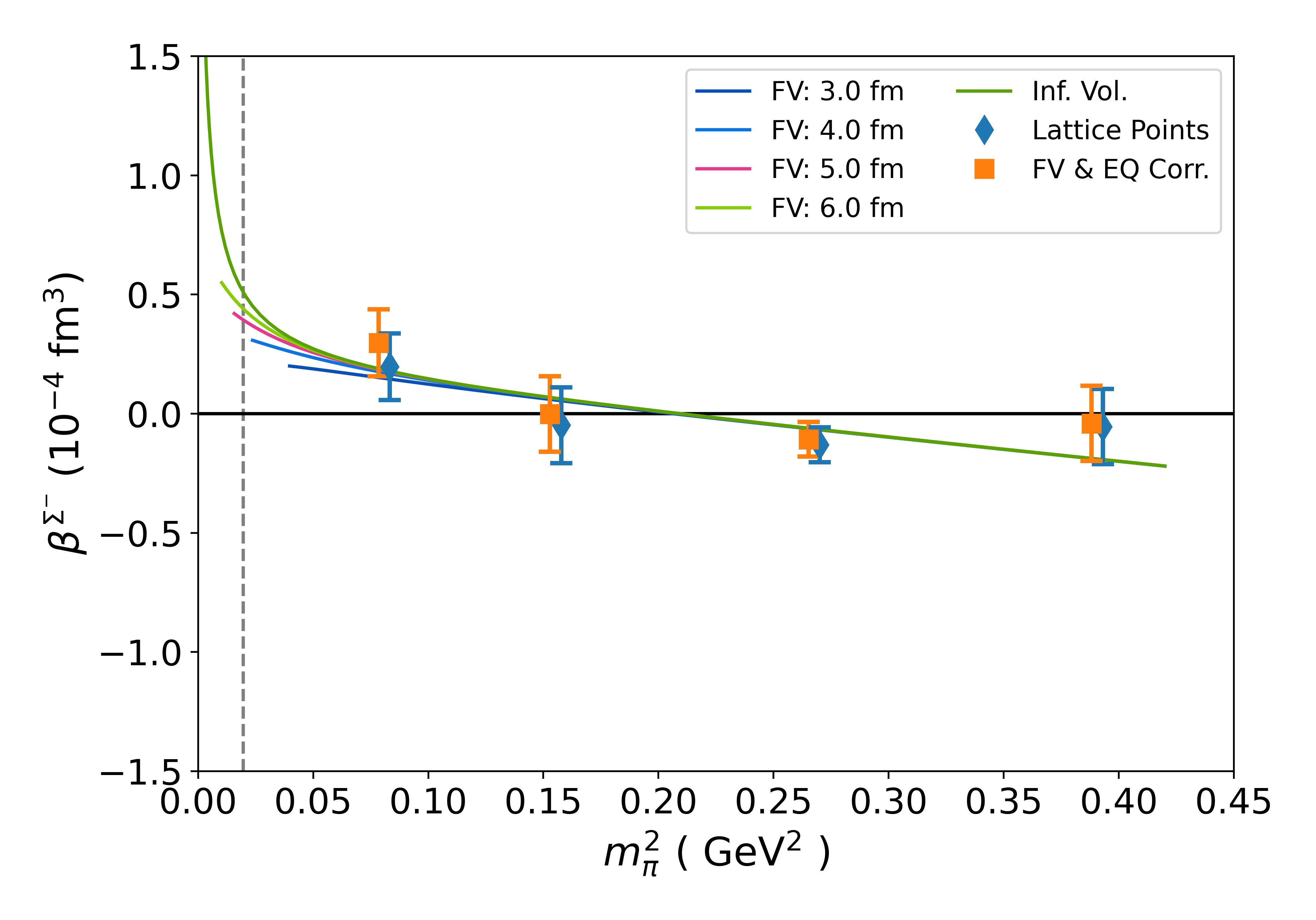}
    }
    \hspace{0mm}
    \subfloat[\label{fig:chiralextrapolation:cascade0_extrap}]{
        \includegraphics[width=0.49\textwidth]{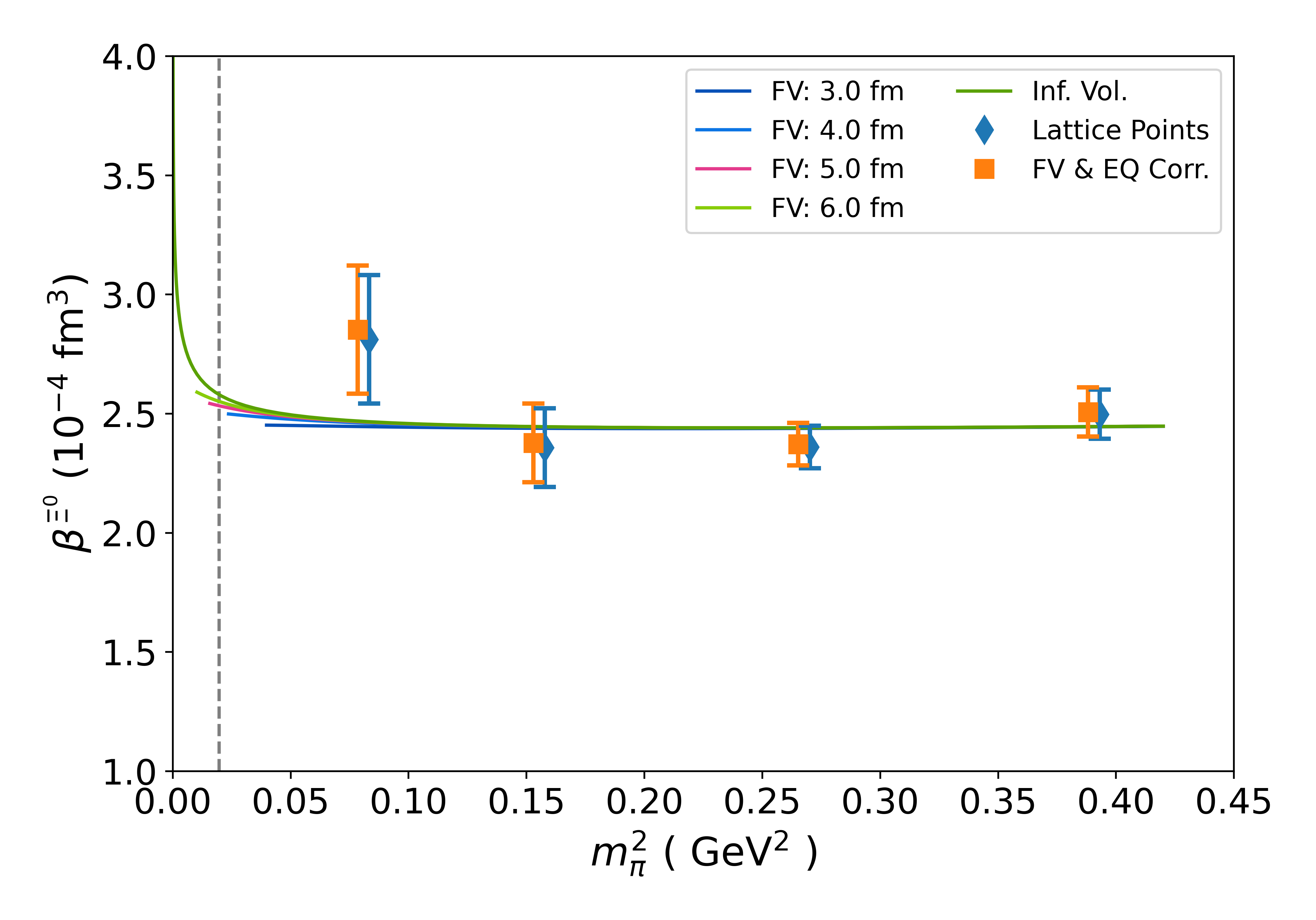}
    }
    \subfloat[\label{fig:chiralextrapolation:cascadem_extrap}]{
        \includegraphics[width=0.49\textwidth]{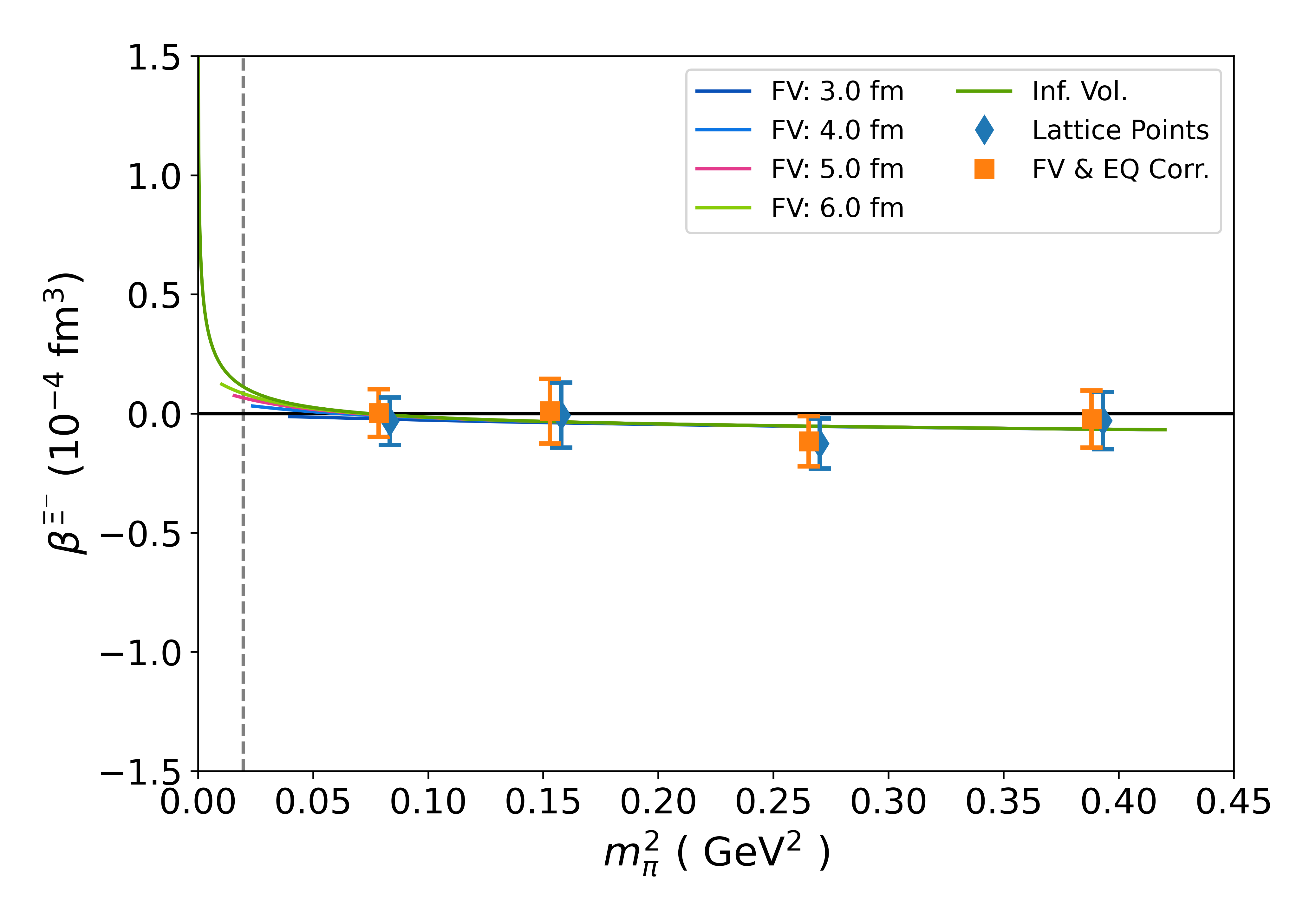}
    }
    \caption{Chiral extrapolation of the finite-volume- and electroquenched-corrected (FV \& EQ
        Corr.) octet-baryon magnetic polarizabilities to the physical pion mass indicated by the
        vertical dashed line. Uncorrected lattice points (Lattice Points) are horizontally offset
        from the corrected points.  Extrapolation curves for future larger finite volumes (FV:
        $\cdots$ fm) are shown to illustrate the requirements for observing chiral curvature in full
        QCD. The curves are shown for $m_\pi\, L >$ 3, where $L$ is the spatial lattice length.
        Finally, the infinite-volume extrapolation (Inf.\ Vol.) relevant to nature is
        illustrated. (a) Proton, (b) neutron, (c) \sigmap, (d) \sigmam, (e) \cascadez, and (f) \cascadem extrapolation.
    } \label{fig:chiralextrapolation:extrapolationplots}
\end{figure*}

It is interesting to note that one can also use
Eq.~\ref{eqn:chiralextrapolation:volumecorrecteddata} to explore the chiral curvature to be seen in
future lattice QCD calculations.  This time, one subtracts the infinite-volume integral and adds a
finite-volume sum appropriate to the future lattice simulation volume.  This idea is explored as we
present the chiral extrapolations.

Extrapolations of the octet-baryon magnetic polarizabilities to the physical point are presented in
\autoref{fig:chiralextrapolation:extrapolationplots}. The $p$, $n$, and $\Sigma^+$ baryons have
very similar profiles as predicted by the simple quark model of \autoref{sec:quarkmodel}.
Recall the $\Xi^0$ baryon is unique due to its small octet-decuplet mass splitting and containing
only one $u$ quark which contributes well to the transition term of the model while minimising the
negative effect in the charge distribution term by only having one $u$ quark. Together, these
properties generate a large magnetic polarizability for the $\Xi^0$ and this is seen in our lattice
QCD results.

The negatively charged baryons are very interesting. We observe that they have very small magnetic
polarizability values as predicted by the quark model. However, with the extrapolation incorporating
divergent chiral physics and including both electroquenching corrections and finite-volume
corrections, both baryons are predicted to have positive values in the infinite-volume world of
full QCD.  Carefully examining the results in
\autoref{tab:chiralextrapolation:extrapolatedresults}, we see there is a clear prediction for the
$\Sigma^-$ to have a positive nonvanishing magnetic polarizability.  On the other hand, the positive
polarizability for the $\Xi^-$ is only a $1\, \sigma$ effect.

\begin{figure*}
    \includegraphics[width=\linewidth]{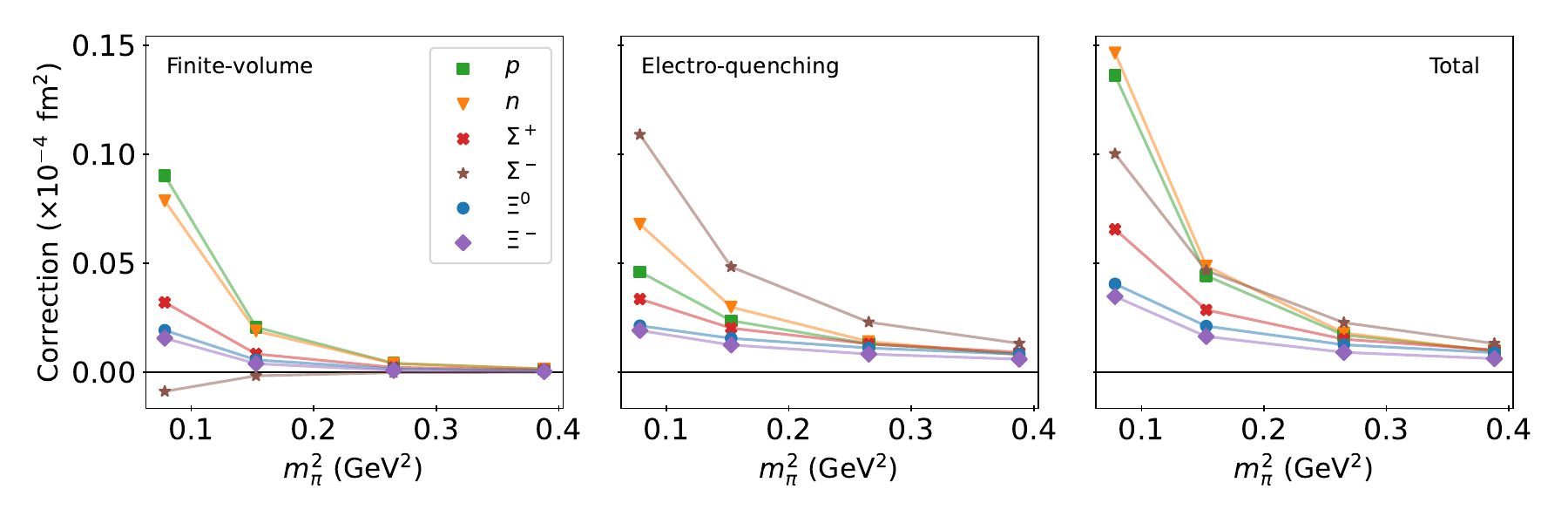}
    \caption{The valence-valence finite-volume (FV), electroquenching (EQ), and total
        (FV+EQ) corrections for each outer octet baryon as a function of pion mass.  These
        corrections are added to the original results from the lattice.}
    \label{fig:chiralextrapolation:chiralcorrections}
\end{figure*}

Also of interest is the very small difference between the values of the proton and neutron. Based
on our discussion of the quark model in \autoref{sec:quarkmodel}, we would expect the neutron to
have a larger magnetic polarizability. The additional up quark in the proton acts to increase the
magnitude of the negative charge distribution term.  Otherwise, the transition term is identical
for the proton and neutron.  We will examine this difference more carefully in
\autoref{sec:proton-neutron}.

We now turn our attention to the subtle electroquenching and finite-volume corrections for the
baryons. The details of these corrections are shown in
\autoref{fig:chiralextrapolation:chiralcorrections}. With the exception of the
\emph{valence-valence} volume correction for the \sigmam, these corrections act to increase the
magnetic polarizability. The \sigmam is unique and is discussed further below.

In all cases, the dominant correction at heavy pion masses is the electroquenching correction.
However, this correction is still small. As the pion mass approaches the physical value, both
volume and electroquenching corrections increase in magnitude, but the volume corrections increase
much more quickly, resulting in corrections of approximately equal magnitude at the lightest pion
mass considered in this work.

\begin{figure}[tb]
    \includegraphics[width=0.99\columnwidth]{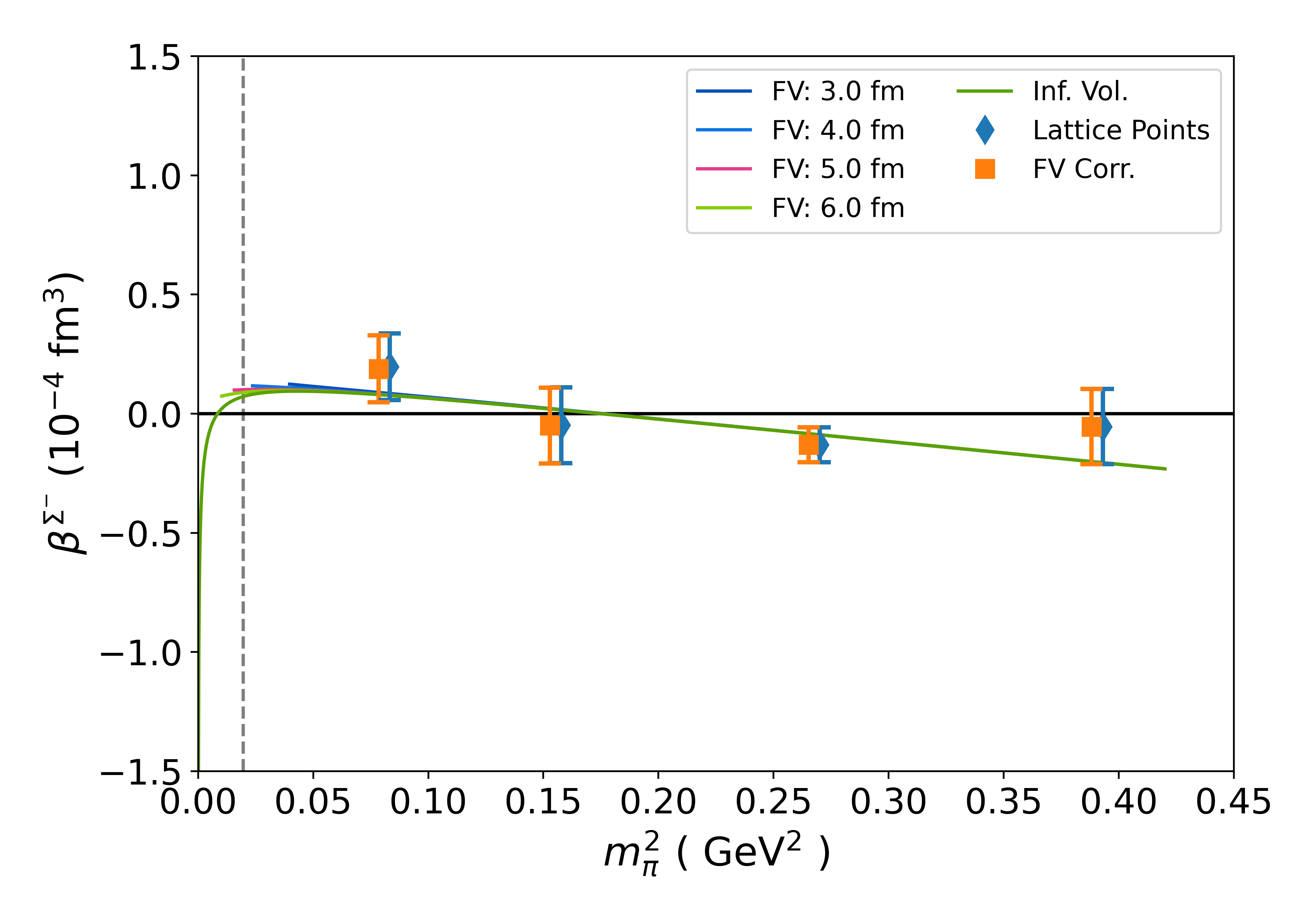}
    \includegraphics[width=0.99\columnwidth]{Figures-Chiral-sigmam-chiEFT-total_plots.png}
    \caption{The chiral extrapolation of the finite-volume-corrected valence-valence lattice
        QCD data (top) compared to the finite-volume and electroquenching-corrected extrapolation
        (bottom) for the \sigmam. The magnitude of the electroquenching correction results in a sign
        change for the nonanalytic chiral curvature, a behavior not observed for any other baryon.}
    \label{fig:chiralextrapolation:sigmamcurvature}
\end{figure}

The \sigmam is interesting as it has large electroquenching corrections.  In the
\emph{valence-valence} sector, the sign of the chiral curvature is negative, such that the
finite-volume correction of the \emph{valence-valence} sector acts to decrease the magnetic
polarizability.  This is shown in the upper plot of
\autoref{fig:chiralextrapolation:sigmamcurvature}. Upon implementing the electroquenching
correction, the sign of the chiral curvature becomes positive as illustrated in the lower plot of
\autoref{fig:chiralextrapolation:sigmamcurvature}.

This change in sign underscores the importance of the order in which the corrections are
implemented.  Finite-volume corrections in the \emph{valence-valence} sector are implemented first
as this is the frame in which the finite-volume lattice results are obtained.  Once volume
corrected the lattice results can be fit to determine the coefficients of the residual series.
Provided one uses a physical regulator parameter, the residual series coefficients are insensitive
to sea-quark-loop contributions and one can model the electroquenching corrections by changing the
coefficients of the loop-integral coefficients. With the full QCD result in hand, one can plot the
finite-volume- and electroquenched-corrected points and draw the infinite volume extrapolation
curve.  From here one can also explore other finite-volumes as done in
Figs.~\ref{fig:chiralextrapolation:extrapolationplots} and
\ref{fig:chiralextrapolation:sigmamcurvature}.

\section{Comparison to other predictions}\label{sec:comparison}

\begin{figure}[tb]
    \includegraphics[width=0.99\columnwidth]{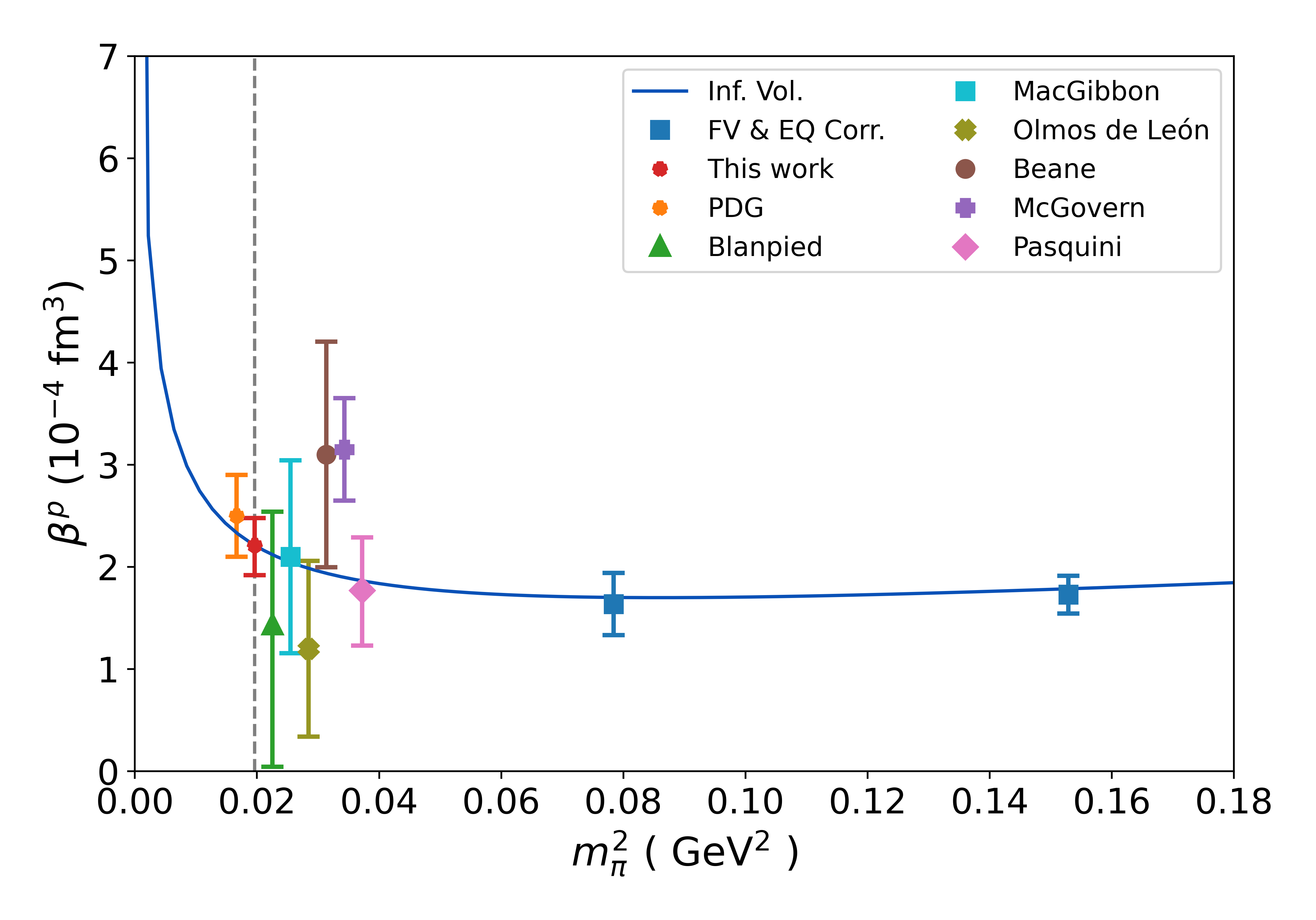}
    \includegraphics[width=0.99\columnwidth]{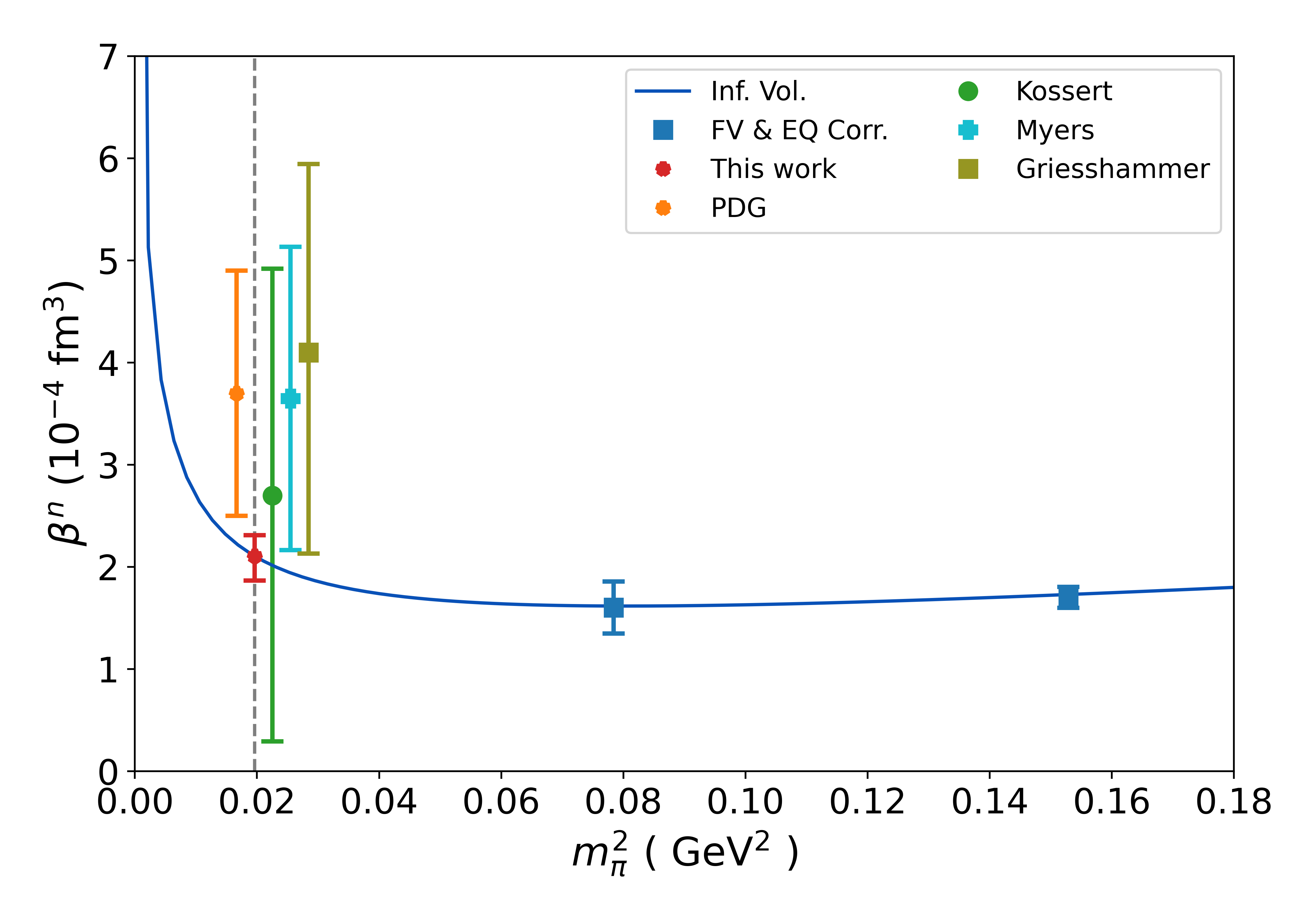}
    \caption{The chiral extrapolation of volume- and electroquenched-corrected lattice QCD
      results for the proton (top) and neutron (bottom) are compared with PDG averages \cite{PDG2022},
      experimental results and theory-based analyses.
% results obtained through re-interpretation of experimental data. 
      Results at the physical point are offset for clarity.  Blanpied \etal~\cite{Blanpied:2001ae},
      MacGibbon \etal~\cite{MacGibbon:1995in}, Olmos de Leon \etal~\cite{deLeon:2001dnx} and
      Kossert \etal~\cite{Kossert:2002ws} report experimental results.  Myers
      \etal~\cite{COMPTONMAX-lab:2014cve} provide a result from deuteron scattering with effective
      field theory input.  McGovern \etal~\cite{McGovern:2012ew}, Beane \etal~\cite{Beane:2002wn}
      and Griesshammer \etal~\cite{Griesshammer:2015ahu} report effective-field-theory based
      analyses.  Pasquini \etal~\cite{Pasquini:2019nnx} utilizes dispersion relations to obtain the
      result presented.}
    \label{fig:nucleon-comp-plot}
\end{figure}

The Particle Data Group~\cite{PDG2022} quotes values for the magnetic polarizability of the proton
and neutron. The values
\begin{align}
    \beta_p^{\rm PDG} = 2.5(4)\times 10^{-4}\,\text{fm}^3, \\
    \beta_n^{\rm PDG} = 3.7(12)\times 10^{-4}\,\text{fm}^3,
\end{align}
are aggregated from a number of Compton scattering experiments.
The quark model prediction that the neutron should have larger polarizability than the neutron is
observed in the central values, though the large uncertainty of the neutron measurement precludes a
definitive statement.
Our final lattice QCD values reported in \autoref{tab:chiralextrapolation:extrapolatedresults}
compare favorably with the PDG values.  Figure \ref{fig:nucleon-comp-plot} shows our chiral
extrapolation in comparison to experimental measurements and theory-based analyses for the
proton~\cite{PDG2022,Pasquini:2019nnx,MacGibbon:1995in,Blanpied:2001ae,McGovern:2012ew,deLeon:2001dnx,Beane:2002wn,Griesshammer:2015ahu}
and
neutron~\cite{PDG2022,Kossert:2002jc,Kossert:2002ws,COMPTONMAX-lab:2014cve,Griesshammer:2012we,Griesshammer:2015ahu}.

\begin{figure*}
    \hspace{0mm}
    \subfloat[\sigmap.\label{fig:chiralextrapolation:sigmap_comp}]{
        \includegraphics[width=0.49\textwidth]{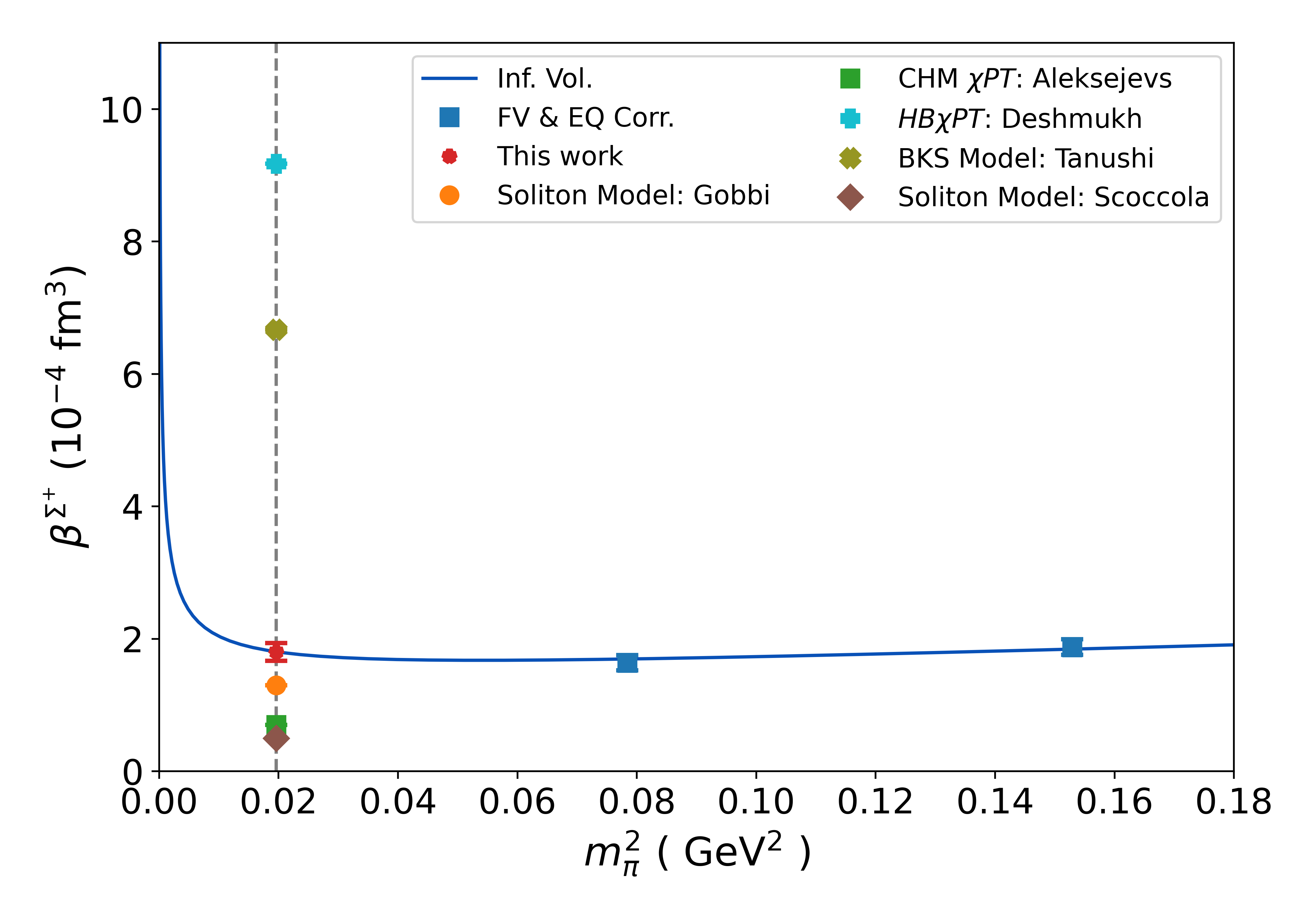}
    }
    \subfloat[\sigmam.\label{fig:chiralextrapolation:sigmam_comp}]{
        \includegraphics[width=0.49\textwidth]{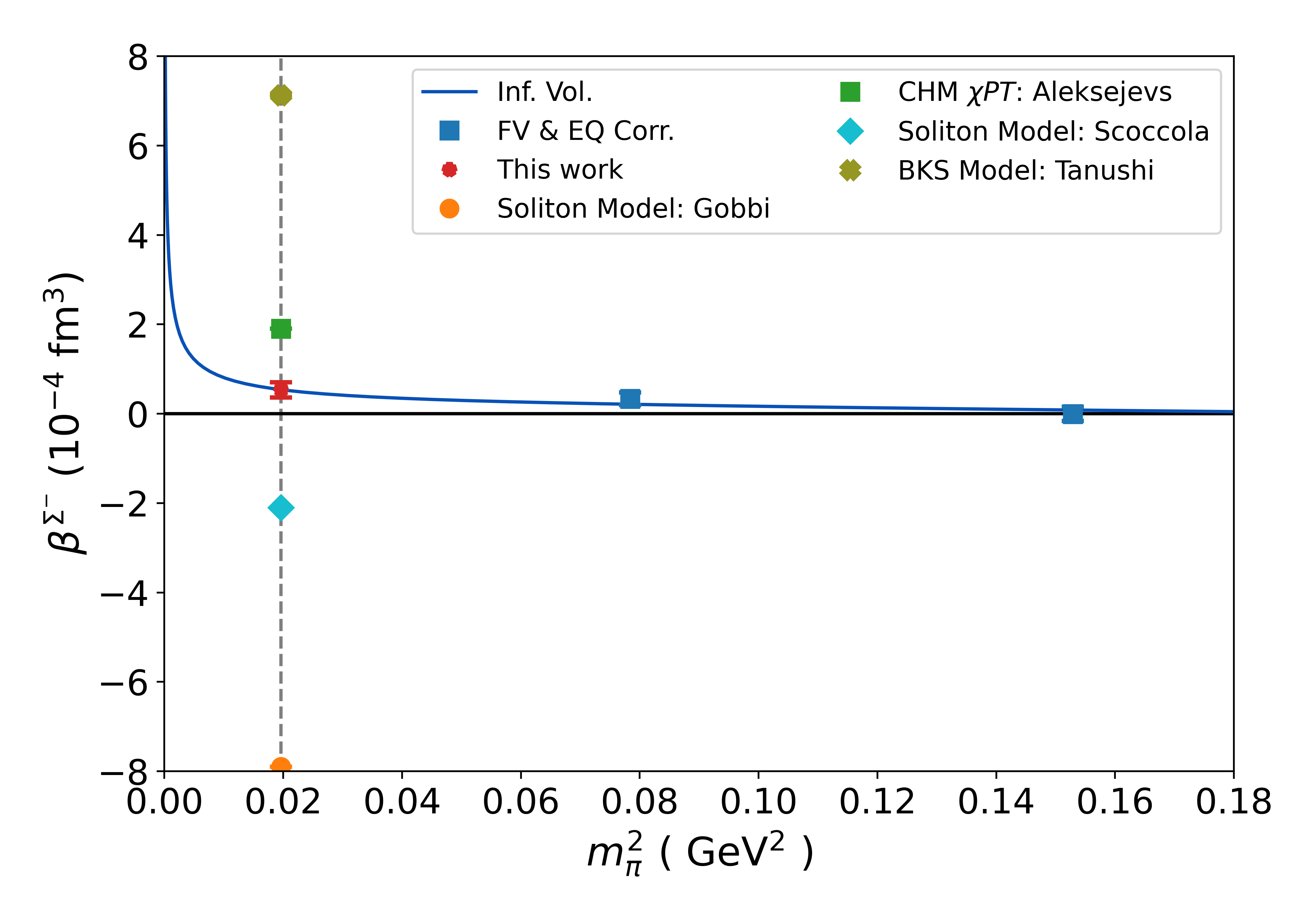}
    }
    \hspace{0mm}
    \subfloat[\cascadez.\label{fig:chiralextrapolation:cascadezcomp}]{
        \includegraphics[width=0.49\textwidth]{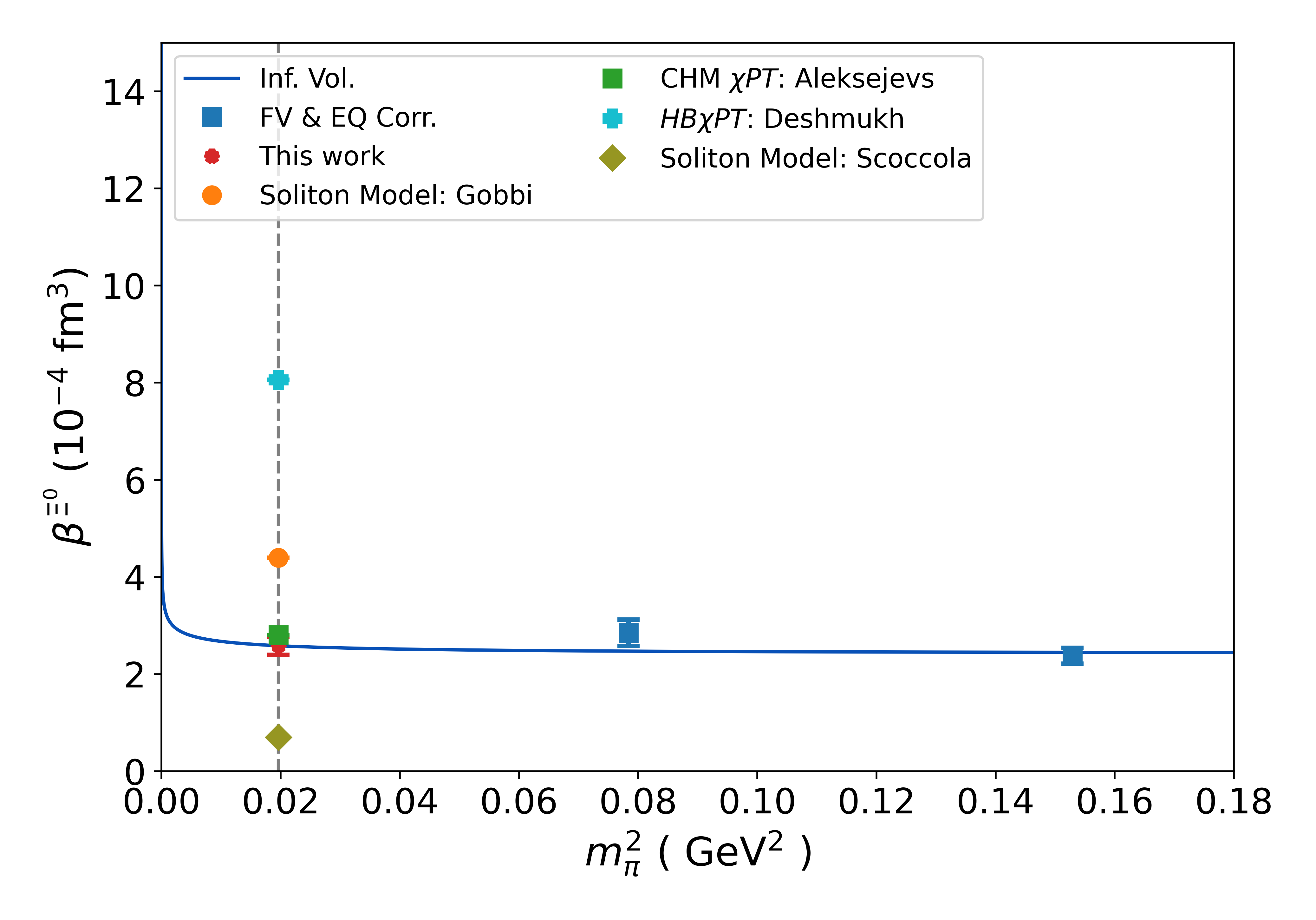}
    }
    \subfloat[\cascadem.\label{fig:chiralextrapolation:cascademcomp}]{
        \includegraphics[width=0.49\textwidth]{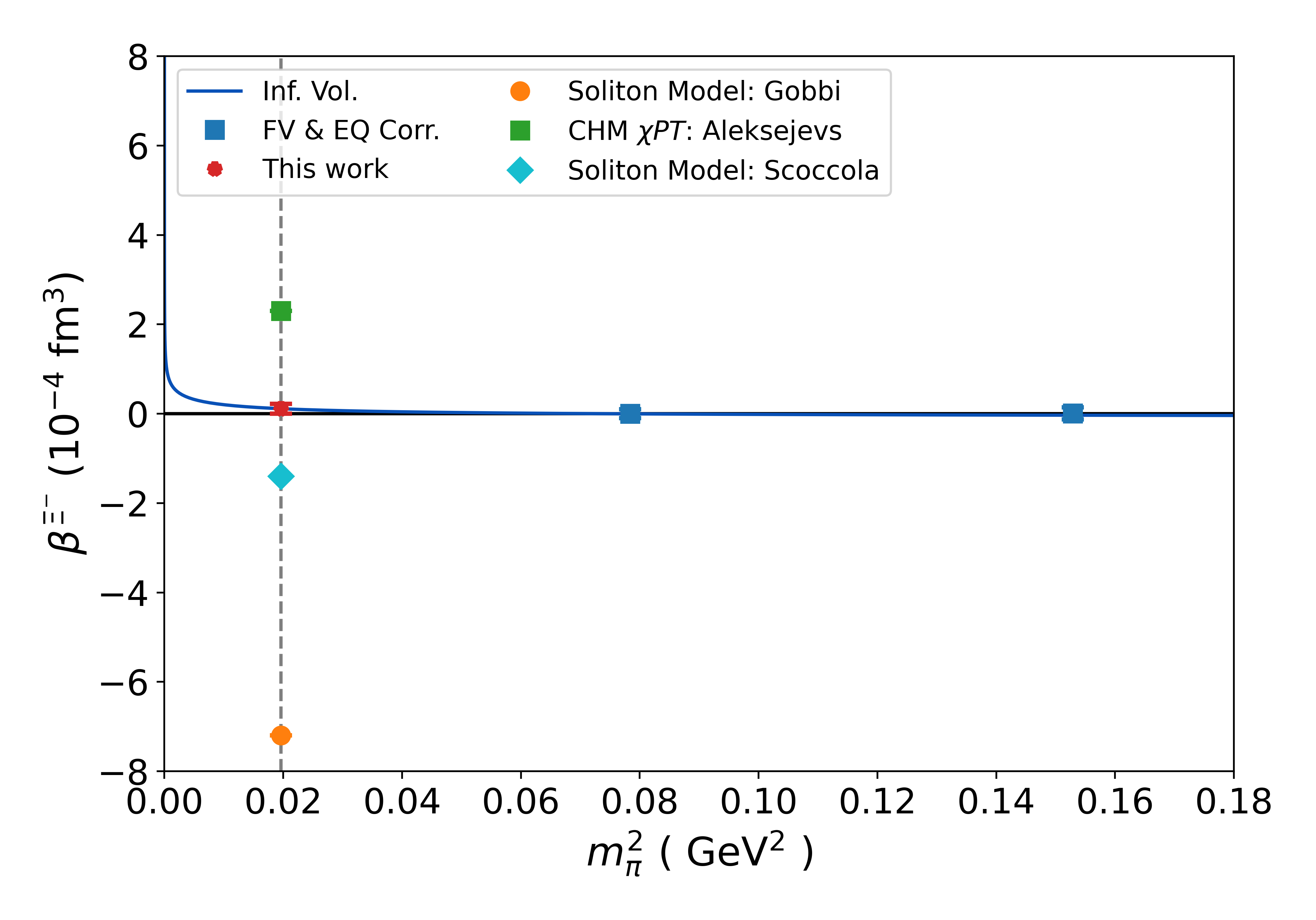}
    }
    \caption{The chiral extrapolation of volume- and electroquenched-corrected lattice QCD results are
        compared with other phenomenological estimates \cite{Aleksejevs:2010zw,Gobbi:1995de,Tanushi:2001ya,Scoccola:1996kh,Deshmukh:2018:octet} for the hyperons.} \label{fig:hyperon-comp-plot}
\end{figure*}

It is also interesting to place our results in the context of previous lattice QCD studies.  Here
we discuss the results of three
works~\cite{Lee:2006:polarisability,Chang:2015:polarisability,bignell2020nucleon}. We first note
that Refs.~\cite{Lee:2006:polarisability} and \cite{Chang:2015:polarisability} were carried out in
the quenched approximation.  In addition, Refs.~\cite{Lee:2006:polarisability} and
\cite{Chang:2015:polarisability} did not use a background-field-corrected clover action
to remove the additive mass renormalization associated with the Wilson term
when a background field is applied~\cite{bignell:2019:cloverpion}.

In the early work of Ref.~\cite{Lee:2006:polarisability} the magnetic polarizabilities of all octet
baryons were determined at a range of relatively heavy pion masses. In that work, the Landau term
was not considered.  However, we find the energy shift associated with the Landau term to be of
similar magnitude to the total energy shift, therefore we do not compare with their results for
charged baryons.  Accounting for their different definition of the magnetic polarizability, we
divide their results by $4\pi$ and compare the \cascadez and neutron magnetic polarizabilities.  We
find our results for \cascadez to be much larger than that reported in
Ref.~\cite{Lee:2006:polarisability}, approaching a factor of 3. Our results for the neutron at the
lightest quark mass considered in Ref.~\cite{Lee:2006:polarisability} are 60\% larger.

The observed discrepancy may be associated with the boundary conditions explored in
Ref.~\cite{Lee:2006:polarisability}. While a fixed spatial boundary condition avoids the uniform
magnetic-field quantization condition of \autoref{eq:qc}, it is difficult to avoid the spatial
boundary in the lattice QCD simulations introducing new systematic errors.

The lattice simulation of Ref.~\cite{Chang:2015:polarisability} determined the polarizability of
the proton and neutron at $m_{\pi}\sim 806\,$MeV. Again, the neutron results presented there are
small at approximately half that presented here. In the case of the proton,
they do attempt to fit the Landau term, using their smaller field strength correlation functions to
identify which Landau level the particle rests in. They fit for the $n$th Landau level, allowing
$n$ to take a positive real value. As $n$ should be an integer, they round the resulting
value. However, due to the magnitude of the Landau term relative to the energy shift, their
approach introduces a large uncertainty into the energy shift. As such, we do not consider their
proton results further.  Perhaps it is worthy to note, our use of Landau-mode projection at the
baryon sink removes all uncertainty surrounding the Landau levels.

Finally, Bignell {\it et al.} \cite{bignell2020nucleon} determined the proton and neutron magnetic
polarizabilities on the PACS-CS ensembles using similar methods which we have extended herein. The
most important extensions include the single-state isolation analysis in all correlators, the
consideration of several fit windows through weighted averaging, and much higher statistics.

It is the latter consideration that has led to a notable change in the magnetic polarizability of
the proton.  The current work has 8 times the statistics and this has enabled the explicit
examination of the excited-state contamination in each of the correlation functions prior to
combining them to obtain the polarizability.  This contrasts the analysis possible in Bignell {\it
  et al.}  \cite{bignell2020nucleon} where excited-state contamination was examined only in the
ratio after combining the correlation functions to access the magnetic polarizability energy shift
$\delta E_{\beta}$.

We have now discovered the proton's correlation function ratio displays an early plateau in $\delta E_{\beta}$
before the underlying correlation functions have reached single-state isolation.  Thus,
the results in Ref.~\cite{bignell2020nucleon} are based on fit windows commencing at earlier Euclidean
times.

For example at $\mpipacs=296\,$MeV, we showed in \autoref{sec:fitting} and
\autoref{fig:fitting:protonheatmaps} that the underlying correlation functions for the proton do
not reach single-state isolation until $t=24$. In Ref.~\cite{bignell2020nucleon}, the fit window
$[t_s,t_f]=[20,24]$ was utilized.  Excited-state contamination gives rise to a difference in the
results reported.

This behavior is shared by the neutron, though to a lesser degree as $\delta E_{\beta}$ for the
neutron tends to show plateaulike behavior only later in Euclidean time, allowing the underlying
correlation functions to have settled further to single-state isolation.  Still, the neutron results
presented here have a larger slope with respect to $m_\pi^2$ but extrapolate to a similar value at
the physical point.

Increased statistics and weighted averaging of acceptable fit windows have resolved smaller
polarizability values for the proton at the two lightest quark masses considered, increasing the
slope with respect to $m_\pi^2$ and producing smaller proton magnetic polarizabilities at the
physical point.  These systematic advances bring the previous value of
$2.79(22)(16) \times 10^{-4}$ to $2.12(17)(16) \times 10^{-4}\,$fm$^3$,
just outside $1\sigma$ agreement.

Looking more broadly, there are several phenomenological studies of the hyperons.  In the
absence of experimental measurements, we compare our results with theoretical models in
\autoref{fig:hyperon-comp-plot}.  It is here that one can observe the value of our lattice QCD
simulation results for the hyperons.  Our uncertainties are very small on the scale of variation in
model predictions.

\section{Proton-Neutron Magnetic Polarizability Difference}\label{sec:proton-neutron}

The difference between the magnetic polarizability of the proton and neutron can provide a test of
Reggeon dominance~\cite{Gasser:2015dwa,Gasser:2020hzn}.  Under the assumption of Reggeon dominance,
chiral perturbation theory and Baldin sum rules can predict the difference of magnetic
polarizabilities.

\begin{figure}[t]
    \includegraphics[width=\columnwidth]{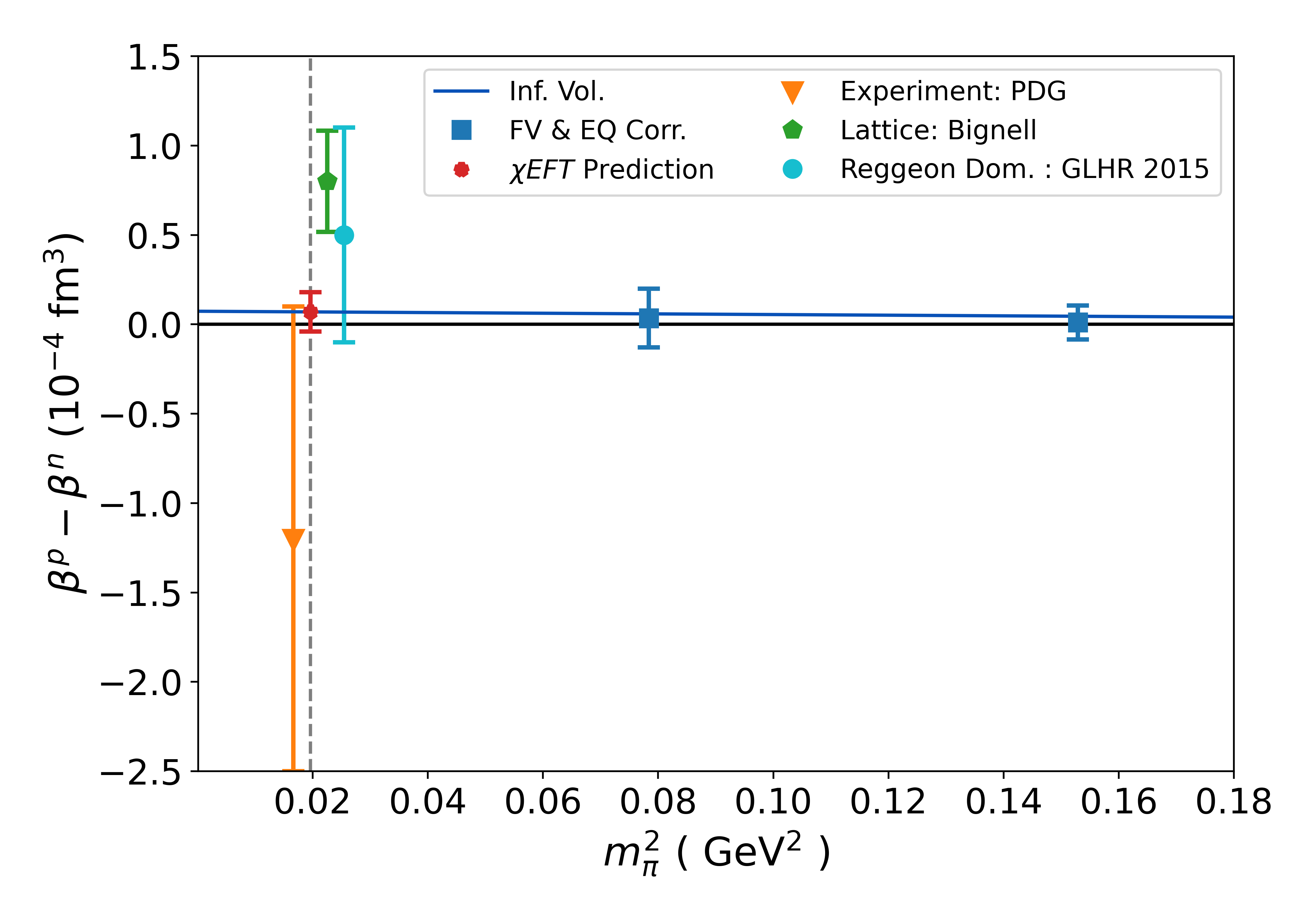}
    \caption{The difference between the proton and neutron magnetic polarizabilities. This work's
        finite-volume- and electroquenched-corrected lattice QCD results are
        extrapolated to the infinite volume physical point through chiral
        extrapolation. These results are compared to the experimental results of the
        PDG~\cite{PDG2022}, a Reggeon dominance prediction~\cite{Gasser:2020hzn}, and a previous
        lattice QCD calculation~\cite{bignell2020nucleon} which are horizontally offset at the
        physical pion mass for clarity. } \label{fig:proton-neutron:plot}
\end{figure}

We calculate the difference between the magnetic polarizability of the proton and neutron by construction of a correlation-function ratio analogous to Eq.~\ref{eqn:fitting:energyshift}, which provides direct access to the polarizability difference
\begin{align}\label{eqn:proton-neutron:energyshift}
  \delta E_{\beta_p - \beta_n}(B, t) &=
  \frac{1}{2}\frac{1}{\delta t}\lim_{t\rightarrow \infty}\log\Bigl(
  \frac{R_p(B,t)}{R_p(B,t+\delta t)}\frac{R_n(B,t+\delta t)}{R_n(B,t)}
  \Bigr), \nonumber \\
  &= \left(\frac{\abs{q_p}}{m_p} - \frac{\abs{q_n}}{m_n}\right)\frac{\abs{e B}}{2}
  - \frac{4\pi}{2}\left(\beta_p - \beta_n\right)|B|^2 \nonumber \\
  &\phantom{=}+ \order*{B^3}.
\end{align}
$\delta E_{\beta_p-\beta_n}$ is fit using the techniques discussed in
\autoref{sec:fitting}. We then extrapolate to the physical regime using the formalism discussed in
\autoref{sec:chiralextrapolation}. When taking the polarizability difference, the $u-d$ symmetry in
the leading loop-integral coefficients of the chiral expansion in full QCD causes the chiral
contributions to cancel. As such, the extrapolation becomes a simple linear extrapolation. The
resulting value at the physical point is
\begin{equation}
    \beta_p - \beta_n = 0.09(11) \times 10^{-4} \, \textrm{fm}^3 \, ,
\end{equation}
a much better estimate for the difference between the two polarizabilities.  Here the statistical
uncertainty is given in parentheses and the systematic uncertainty is negligible.
Figure~\ref{fig:proton-neutron:plot} shows the extrapolation to the physical regime
and includes the PDG value~\cite{PDG2022}, a result derived using Reggeon
dominance~\cite{Gasser:2020hzn} and a previous lattice QCD calculation~\cite{bignell2020nucleon}.

The key observation is that our result is now in $1\sigma$ agreement with both experiment and the
Reggeon dominance prediction, resolving a discrepancy observed in Ref.~\cite{bignell2020nucleon}.
Moreover, our results provide a very precise prediction for $\beta_p - \beta_n$.

\section{Conclusion}\label{sec:conclusion}

A generalized expression for the magnetic polarizability was derived in a simple constituent quark
model.  Once scaled, this simple model provides accurate predictions for the octet-baryon
magnetic polarizabilities examined herein. It also provides the ability to identify key characteristics
of the magnetic polarizability and deep insight into the physics that drives the observed
patterns. In particular, we identified the importance of opposite-charge quark flavors in
generating large magnetic polarizability values.  Here the importance of the up quark is manifest.
We also focused on the octet-decuplet mass splitting and associated hyperfine interactions that
govern the magnitude of the magnetic polarizability transition term.

Turning our attention to lattice simulation techniques, we investigated the behavior of the
underlying correlation functions associated with the correlation-function ratio required to extract
the magnetic polarizability in the background-field method. We saw that in many cases, the
correlation-function ratio exhibits plateaulike behavior in spite of the underlying correlation
functions not yet having reached single-state isolation.  We developed new methods to address this,
ensuring the suppression of excited-state contamination, and ensuring the uncertainties in our
fitted mass shifts were not underestimated.

We also implemented a weighted averaging method to systematically extract the magnetic
polarizability from the associated energy shifts obtained from candidate fit windows.  This is
especially important in cases where different fit windows introduce variation in the fitted values
of the polarizability energy shifts.  These improved analysis techniques allow the extraction of
magnetic polarizability values for the outer octet baryons at a variety of pion masses with
unprecedented precision.

The systematics of the lattice QCD calculation were then addressed.
Drawing on chiral effective field theory and accounting for both electroquenching and finite
volume effects we extrapolated our simulation results to the physical point. %%
This process produces results
that compare favorably with the experimental values for the proton and neutron.  Excellent
agreement with the precise experimental value for the proton is observed and our prediction for the
neutron is much more precise.

Comparison of phenomenological values for the hyperons indicate our results are very precise on the
scale of variation in QCD models. Finally, a new precise calculation of the difference in the
proton and neutron magnetic polarizabilities has been presented.

We have revealed complex dynamics underpinning the magnetic polarizabilities of octet baryons.  It
would be interesting to examine these dynamics at a more microscopic level, where the quark mass
dependence and environment dependence of individual quark sector contributions to the baryon
polarizabilities are exposed.  The techniques of the current presentation are flexible enough to
admit such an analysis and we anticipate reporting on this in the near future.

Another challenge facing the community is gaining access to the light-quark mass regime. Using the
techniques presented here, we are unable to resolve a signal of statistical interest for the
lightest PACS-CS ensemble.  We are currently exploring the possibility that the electroquenching
of the sea-quark sector is creating improbable gauge fields when the light-quark ensemble is used
at finite magnetic-field strength.  Filtering techniques to identify exceptional configurations are
under development and the veracity of the approach is under investigation.

Finally, the $u d s$ members of the baryon octet, the $\Lambda$, and $\Sigma^0$ remain to be examined.
As highlighted in the Introduction, a perturbative calculation
\cite{Wilcox:2021:towards,Lee:2023:pion} can offer some important advantages in obtaining a
clear understanding of the magnetic polarizabilities of these baryons. \\

The results presented herein are founded on the PACS-CS ensembles~\cite{PACS-CS2008ensembles}.
The ensembles are created by the PACS-CS collaboration, Aoki {\it et al.} and are available from~\cite{jldg}. The ensembles in question are part of the April 2009 release. 

\begin{acknowledgments}
    It is a pleasure to acknowledge beneficial communications with Harald Griesshammer.
    We thank the PACS-CS Collaboration for making their $2+1$ flavor configurations available and the
    ongoing support of the International Lattice Data Grid (ILDG).  Baryon correlation functions were
    constructed using the \texttt{COLA} software library, developed at the University of Adelaide~\cite{COLA}.
    WK was supported
    by the Pawsey Supercomputing Centre through the Pawsey Centre for Extreme Scale Readiness (PaCER)
    program. RB is grateful for support via STFC grant ST/T000813/1 and acknowledges support from a
    Science Foundation Ireland Frontiers for the Future Project award with grant number
    SFI-21/FFP-P/10186 (data not available). 
    This work was supported with
    supercomputing resources provided by the Phoenix HPC service at the University of Adelaide. This
    research was undertaken with the assistance of resources from the National Computational
    Infrastructure (NCI), which is supported by the Australian Government. This research is supported
    by Australian Research Council through Grants No.~DP190102215 and No.~DP210103706.
\end{acknowledgments}

\null
\bigskip
\appendix
\section{Chiral coefficients} \label{appendix:chiralcoeffs}

\begin{table*}[t]
    \centering
    \caption{Chiral coefficients $\chi^2_{MB}$ required to calculate quark-flow connected and
      disconnected coefficients for the leading chiral contribution to the magnetic
      polarisability. These coefficients address outer octet baryon transitions to an intermediate
      octet-baryon (column) octet-meson (row) states. Note that $\pi^0$ intermediate states do not
      contribute to the analysis.  In full QCD the $\pi^0$ is electrically neutral and in partially
      quenched QCD, its composition of $\bar u u$ and $\bar d d$ matched flavours means it is not
      relevant to the disconnected flow calculations.}
    \label{tab:appendix:chiralcoeffs:octet}
    \begin{ruledtabular}
    \begin{tabular}{ccccccccc}
        \noalign{\smallskip}
               & $p$           & $n$           & $\Xi^0$       & $\Xi^-$       & $\Sigma^+$    & $\Sigma^-$    & $\Sigma^0$ & $\Lambda$ \\
        \noalign{\smallskip}
        \hline
        \noalign{\smallskip}
        $\pip$ &               & $2\rb{D+F}^2$ &               & $2\rb{D-F}^2$ &               &               & $4F^2$       & $\frac{4}{3}D^2$ \\
        $\pim$ & $2\rb{D+F}^2$ &               & $2\rb{D-F}^2$ &               &               &               & $4F^2$       & $\frac{4}{3}D^2$ \\
        $\Kp$  &               &               & $2\rb{D+F}^2$ &               &               & $2\rb{D-F}^2$ & $\rb{D-F}^2$ & $\frac{1}{3}\rb{D+3F}^2$ \\
        $\Kz$  &               &               &               & $2\rb{D+F}^2$ & $2\rb{D-F}^2$ &               & $\rb{D-F}^2$ & $\frac{1}{3}\rb{D+3F}^2$ \\
        $\Kzb$ & $2\rb{D-F}^2$ &               &               &               &               & $2\rb{D+F}^2$ & $\rb{D+F}^2$ & $\frac{1}{3}\rb{D-3F}^2$ \\
        $\Km$  &               & $2\rb{D-F}^2$ &               &               & $2\rb{D+F}^2$ &               & $\rb{D+F}^2$ & $\frac{1}{3}\rb{D-3F}^2$ \\
        \noalign{\smallskip}
    \end{tabular}
    \end{ruledtabular}
\end{table*}

\begin{table*}[t]
    \centering
    \caption{Chiral coefficients $\chi^2_{MB}$ required to calculate quark-flow connected and
      disconnected coefficients for the leading chiral contribution to the magnetic
      polarisability. These coefficients address outer octet baryon transitions to an intermediate
      decuplet-baryon (column) octet-meson (row) states. Again $\pi^0$ intermediate states do not
      contribute.}
    \label{tab:appendix:chiralcoeffs:decuplet}
    \begin{ruledtabular}
    \begin{tabular}{ccccccccccc}
        \noalign{\smallskip}
               & $\Delta^{++}$       & $\Delta^{+}$        & $\Delta^0$          & $\Delta^-$          & $\Sigma^{*+}$       & $\Sigma^{*0}$       & $\Sigma^{*-}$       & $\Xi^{*0}$          & $\Xi^{*-}$          & $\Omega^-$ \\
        \noalign{\smallskip}
        \hline
        \noalign{\smallskip}
%        $\piz$ &                     & $\frac{8}{3}\mcC^2$ & $\frac{8}{3}\mcC^2$ &                     & $\frac{2}{9}\mcC^2$ & $\frac{2}{3}\mcC^2$ & $\frac{2}{3}\mcC^2$ & $\frac{2}{9}\mcC^2$ & $\frac{2}{3}\mcC^2$ &                     \\
        $\pip$ &                     &                     & $\frac{4}{9}\mcC^2$ & $\frac{4}{3}\mcC^2$ &                     & $\frac{2}{9}\mcC^2$ &                     &                     & $\frac{4}{9}\mcC^2$ &                     \\
        $\pim$ & $\frac{4}{3}\mcC^2$ & $\frac{4}{9}\mcC^2$ &                     &                     &                     & $\frac{2}{9}\mcC^2$ &                     & $\frac{4}{9}\mcC^2$ &                     &                     \\
        $\Kp$  &                     &                     &                     &                     &                     & $\frac{2}{9}\mcC^2$ & $\frac{4}{9}\mcC^2$ & $\frac{4}{9}\mcC^2$ &                     & $\frac{4}{3}\mcC^2$ \\
        $\Kz$  &                     &                     &                     &                     & $\frac{4}{9}\mcC^2$ & $\frac{2}{9}\mcC^2$ &                     &                     & $\frac{4}{9}\mcC^2$ & $\frac{4}{3}\mcC^2$ \\
        $\Kzb$ &                     & $\frac{4}{9}\mcC^2$ & $\frac{8}{9}\mcC^2$ & $\frac{4}{3}\mcC^2$ &                     & $\frac{2}{9}\mcC^2$ & $\frac{4}{9}\mcC^2$ &                     &                     &                     \\
        $\Km$  & $\frac{4}{3}\mcC^2$ & $\frac{8}{9}\mcC^2$ & $\frac{4}{9}\mcC^2$ &                     & $\frac{4}{9}\mcC^2$ & $\frac{2}{9}\mcC^2$ &                     &                     &                     &                     \\
        \noalign{\smallskip}
    \end{tabular}
    \end{ruledtabular}
\end{table*}

\begin{table*}[p]
	\caption{Chiral coefficients for the leading-order loop integral contributions for the
          neutron.} \label{tab:appendix:chipt:neutron}
        \begin{ruledtabular}
	\begin{tabular}{lccc}
	\noalign{\smallskip}
	Process &Total &Valence-sea &Sea-sea \\
	\noalign{\smallskip}
	\hline
	\noalign{\smallskip}
		
        $n\rightarrow N\pi$ & & & \\
        \hline
        \noalign{\smallskip}
        \enspace$n\rightarrow n\pi^0$         
            & $0$ 
            & $2q_dq_{\bar{d}}(\chi^2_{K^0\Sigma^0}+\chi^2_{K^0\Lambda})
               + 2q_uq_{\bar{u}}\,\chi^2_{K^+\Sigma^-}$ 
            & $q^2_{\bar{d}}(\chi^2_{K^0\Sigma^0}+\chi^2_{K^0\Lambda})
               + q^2_{\bar{u}}\,\chi^2_{K^+\Sigma^-}$ \\
        \enspace$n\rightarrow p\pi^-$         
            & $\chi^2_{\pi^-p}$ 
            & $  2q_dq_{\bar{u}}(\chi^2_{K^0\Sigma^0}+\chi^2_{K^0\Lambda})$ 
            & $  q^2_{\bar{u}}(\chi^2_{K^0\Sigma^0}+\chi^2_{K^0\Lambda})$ \\
        \enspace$n\rightarrow n^-\pi^+$       
            & $0$ 
            & $  2q_uq_{\bar{d}}\,\chi^2_{K^+\Sigma^-}$ 
            & $q^2_{\bar{d}}\,\chi^2_{K^+\Sigma^-}$\\
	\noalign{\smallskip}
        \hline
	\noalign{\smallskip}

        $n\rightarrow \Sigma K$ & & & \\
        \hline
        \noalign{\smallskip}
        \enspace$n\rightarrow (\Sigma^0,\,\Lambda)K^0$ 
            & $0$ 
            & $  2q_dq_{\bar{s}}(\chi^2_{K^0\Sigma^0}+\chi^2_{K^0\Lambda})$ 
            & $q^2_{\bar{s}}(\chi^2_{K^0\Sigma^0}+\chi^2_{K^0\Lambda})$\\
        \enspace$n\rightarrow \Sigma^- K^+$            
            & $\chi^2_{K^+\Sigma^-}$ 
            & $  2q_uq_{\bar{s}}\,\chi^2_{K^+\Sigma^-}$ 
            & $q^2_{\bar{s}}\,\chi^2_{K^+\Sigma^-}$ \\
	\noalign{\smallskip}
        \hline
	\noalign{\smallskip}

        $n\rightarrow \Delta\pi$ & & & \\
        \hline
        \noalign{\smallskip}
        \enspace$n\rightarrow \Delta^0\pi^0$
            & $0$ 
            & $2q_dq_{\bar{d}}\,\chi^2_{K^0\Sigma^{*0}}
               + 2q_uq_{\bar{u}}\,\chi^2_{K^+\Sigma^{*-}}$ 
            & $q^2_{\bar{d}}\,\chi^2_{K^0\Sigma^{*0}}
               + q^2_{\bar{u}}\,\chi^2_{K^+\Sigma^{*-}}$\\
        \enspace$n\rightarrow \Delta^+\pi^-$
            & $  \chi^2_{\pi^-\Delta^+}$ 
            & $  2q_dq_{\bar{u}}\,\chi^2_{K^0\Sigma^{*0}}$ 
            & $q^2_{\bar{u}}\,\chi^2_{K^0\Sigma^{*0}}$\\
        \enspace$n\rightarrow \Delta^-\pi^+$
            & $\chi^2_{\pi^+\Delta^-}$ 
            & $  2q_uq_{\bar{d}}\,\chi^2_{K^+\Sigma^{*-}}$ 
            & $q^2_{\bar{d}}\,\chi^2_{K^+\Sigma^{*-}}$\\
	\noalign{\smallskip}
        \hline
        \noalign{\smallskip}

        $n\rightarrow \Sigma^* K$ & & & \\
        \hline
        \noalign{\smallskip}
        \enspace$n\rightarrow \Sigma^{*0}K^0$ 
            & $0$ 
            & $  2q_dq_{\bar{s}}\,\chi^2_{K^0\Sigma^{*0}}$ 
            & $q^2_{\bar{s}}\,\chi^2_{K^0\Sigma^{*0}}$\\
        \enspace$n\rightarrow \Sigma^{*-}K^+$ 
            & $\chi^2_{K^+\Sigma^{*-}}$ 
            & $  2q_uq_{\bar{s}}\,\chi^2_{K^+\Sigma^{*-}}$ 
            & $q^2_{\bar{s}}\,\chi^2_{K^+\Sigma^{*-}}$\\
        \noalign{\smallskip}
    \end{tabular}
    \end{ruledtabular}
\end{table*}

\begin{table*}[p]
	\caption{Chiral coefficients for the leading-order loop integral contributions for the proton.} \label{tab:appendix:chipt:proton}
        \begin{ruledtabular}
	\begin{tabular}{lccc}
	\noalign{\smallskip}
	Process &Total &Valence-sea &Sea-sea \\
	\noalign{\smallskip}
	\hline
	\noalign{\smallskip}
		
        $p\rightarrow N\pi$ & & & \\
        \hline
        \noalign{\smallskip}
        \enspace$p\rightarrow p\pi^0$         
            & $0$ 
            & $2q_uq_{\bar{u}}(\chi^2_{K^+\Sigma^0}+\chi^2_{K^+\Lambda})
               + 2q_dq_{\bar{d}}\,\chi^2_{K^0\Sigma^+}$ 
            & $q^2_{\bar{u}}(\chi^2_{K^+\Sigma^0}+\chi^2_{K^+\Lambda})
               + q^2_{\bar{d}}\,\chi^2_{K^0\Sigma^+}$ \\
        \enspace$p\rightarrow n\pi^+$         
            & $\chi^2_{\pi^+n}$ 
            & $  2q_uq_{\bar{d}}(\chi^2_{K^+\Sigma^0}+\chi^2_{K^+\Lambda})$ 
            & $q^2_{\bar{d}}(\chi^2_{K^+\Sigma^0}+\chi^2_{K^+\Lambda})$\\
        \enspace$p\rightarrow p^{++}\pi^-$    
            & $0$ 
            & $  2q_dq_{\bar{u}}\,\chi^2_{K^0\Sigma^+}$ 
            & $q^2_{\bar{u}}\,\chi^2_{K^0\Sigma^+}$\\
	\noalign{\smallskip}
        \hline
        \noalign{\smallskip}

        $p\rightarrow \Sigma K$ & & & \\
        \hline
        \noalign{\smallskip}
        \enspace$p\rightarrow (\Sigma^0,\,\Lambda)K^+$ 
            & $\chi^2_{K^+\Sigma^0}+\chi^2_{K^+\Lambda}$ 
            & $  2q_uq_{\bar{s}}(\chi^2_{K^+\Sigma^0}+\chi^2_{K^+\Lambda})$ 
            & $q^2_{\bar{s}}(\chi^2_{K^+\Sigma^0}+\chi^2_{K^+\Lambda})$\\
        \enspace$p\rightarrow \Sigma^+ K^0$            
            & $0$ 
            & $2q_dq_{\bar{s}}\,\chi^2_{K^0\Sigma^+}$ 
            & $q^2_{\bar{s}}\,\chi^2_{K^0\Sigma^+}$\\
	\noalign{\smallskip}
        \hline
        \noalign{\smallskip}

        $p\rightarrow \Delta\pi$ & & & \\
        \hline
        \noalign{\smallskip}
        \enspace$p\rightarrow \Delta^+\pi^0$         
            & $0$ 
            & $2q_uq_{\bar{u}}\,\chi^2_{K^+\Sigma^{*0}}
               + 2q_dq_{\bar{d}}\,\chi^2_{K^0\Sigma^{*+}}$ 
            & $q^2_{\bar{u}}\,\chi^2_{K^+\Sigma^{*0}}
               + q^2_{\bar{d}}\,\chi^2_{K^0\Sigma^{*+}}$ \\
        \enspace$p\rightarrow \Delta^0\pi^+$         
            & $\chi^2_{\pi^+\Delta^0}$ 
            & $  2q_uq_{\bar{d}}\,\chi^2_{K^+\Sigma^{*0}}$ 
            & $q^2_{\bar{d}}\,\chi^2_{K^+\Sigma^{*0}}$\\
        \enspace$p\rightarrow \Delta^{++}\pi^-$      
            & $\chi^2_{\pi^-\Delta^{++}}$ 
            & $  2q_dq_{\bar{u}}\,\chi^2_{K^0\Sigma^{*+}}$ 
            & $q^2_{\bar{u}}\,\chi^2_{K^0\Sigma^{*+}}$\\
	\noalign{\smallskip}
        \hline
        \noalign{\smallskip}

        $p\rightarrow \Sigma^* K$ & & & \\
        \hline
        \noalign{\smallskip}
        \enspace$p\rightarrow \Sigma^{*0}K^+$ 
            & $\chi^2_{K^+\Sigma^{*0}}$ 
            & $  2q_uq_{\bar{s}}\,\chi^2_{K^+\Sigma^{*0}}$ 
            & $q^2_{\bar{s}}\,\chi^2_{K^+\Sigma^{*0}}$\\
        \enspace$p\rightarrow \Sigma^{*+}K^0$ 
            & $0$ 
            & $  2q_dq_{\bar{s}}\,\chi^2_{K^0\Sigma^{*+}}$ 
            & $q^2_{\bar{s}}\,\chi^2_{K^0\Sigma^{*+}}$\\
        \noalign{\smallskip}
    \end{tabular}
    \end{ruledtabular}
\end{table*}

\begin{table*}[p]
	\caption{Chiral coefficients for the leading-order loop integral contributions for the $\Sigma^+$.} \label{tab:appendix:chipt:sigmap}
        \begin{ruledtabular}
	\begin{tabular}{lccc}
	\noalign{\smallskip}
	Process &Total &Valence-sea &Sea-sea \\
	\noalign{\smallskip}
	\hline
	\noalign{\smallskip}
		
        $\Sigma^+\rightarrow \Sigma\pi$ & & & \\
        \hline
        \noalign{\smallskip}
        \enspace$\Sigma^+\rightarrow \Sigma^+\pi^0$
            & $0$ 
            & $  2q_uq_{\bar{u}}(\chi^2_{\pi^+\Sigma^0}+\chi^2_{\pi^+\Lambda}) $
            % NEED TO CLARIFY - where does this ssbar contribution come in?
            % It is not present in either of Derek's notes nor Ryan's Thesis.
            % It is unlear why not. At this time I will leave it out in all cases.        
            %        + 2q_sq_{\bar{s}}\,\chi^2_{K^0n} )$ 
            % ANSWER: It's negligible.  See text.
            & $q^2_{\bar{u}}(\chi^2_{\pi^+\Sigma^0}+\chi^2_{\pi^+\Lambda})$\\
        \enspace$\Sigma^+\rightarrow (\Sigma^0,\,\Lambda)\pi^+$ 
            & $\chi^2_{\pi^+\Sigma^0}+\chi^2_{\pi^+\Lambda}$ 
            & $2q_uq_{\bar{d}}(\chi^2_{\pi^+\Sigma^0}+\chi^2_{\pi^+\Lambda})$ 
            & $q^2_{\bar{d}}(\chi^2_{\pi^+\Sigma^0}+\chi^2_{\pi^+\Lambda})$ \\
	\noalign{\smallskip}
        \hline
        \noalign{\smallskip}

        $\Sigma^+\rightarrow NK$ & & & \\
        \hline
        \noalign{\smallskip}
        \enspace$\Sigma^+\rightarrow p^{++}K^-$ 
            & $0$ 
            & $  2q_sq_{\bar{u}}\,\chi^2_{K^0p}$ 
            & $q^2_{\bar{u}}\,\chi^2_{K^0p}$\\
        \enspace$\Sigma^+\rightarrow pK^0$
            & $0$ 
            & $  2q_sq_{\bar{d}}\,\chi^2_{K^0p}$ 
            & $q^2_{\bar{d}}\,\chi^2_{K^0p}$\\
	\noalign{\smallskip}
        \hline
        \noalign{\smallskip}

        $\Sigma^+\rightarrow \Xi K$ & & & \\
        \hline
        \noalign{\smallskip}
        \enspace$\Sigma^+\rightarrow \Xi^0K^+$ 
            & $\chi^2_{K^+\Xi^0}$ 
            & $  2q_uq_{\bar{s}}(\chi^2_{\pi^+\Sigma^0}+\chi^2_{\pi^+\Lambda})$
            & $q^2_{\bar{s}}(\chi^2_{\pi^+\Sigma^0}+\chi^2_{\pi^+\Lambda})$\\
	\noalign{\smallskip}
        \hline
        \noalign{\smallskip}

        $\Sigma^+\rightarrow \Sigma^*\pi$ & & & \\
        \hline
        \noalign{\smallskip}
        \enspace$\Sigma^+\rightarrow \Sigma^{*+}\pi^0$
            & $0$ 
            & $2q_uq_{\bar{u}}\,\chi^2_{\pi^+\Sigma^{*0}}$ 
            & $q^2_{\bar{u}}\,\chi^2_{\pi^+\Sigma^{*0}}$\\
        \enspace$\Sigma^+\rightarrow \Sigma^{*0}\pi^+$
            & $\chi^2_{\pi^+\Sigma^{*0}}$ 
            & $  2q_uq_{\bar{d}}\,\chi^2_{\pi^+\Sigma^{*0}}$ 
            & $q^2_{\bar{d}}\,\chi^2_{\pi^+\Sigma^{*0}}$\\
	\noalign{\smallskip}
        \hline
        \noalign{\smallskip}

        $\Sigma^+\rightarrow \Delta K$ & & & \\
        \hline
        \noalign{\smallskip}
        \enspace$\Sigma^+\rightarrow \Delta^{++}K^-$ 
            & $\chi^2_{K^-\Delta^{++}}$ 
            & $  2q_sq_{\bar{u}}\,\chi^2_{K^0\Delta^+}$ 
            & $q^2_{\bar{u}}\,\chi^2_{K^0\Delta^+}$\\
        \enspace$\Sigma^+\rightarrow \Delta^+K^0$    
            & $0$ 
            & $  2q_sq_{\bar{d}}\,\chi^2_{K^0\Delta^+}$ 
            & $q^2_{\bar{d}}\,\chi^2_{K^0\Delta^+}$\\
	\noalign{\smallskip}
        \hline
        \noalign{\smallskip}

        $\Sigma^+\rightarrow \Xi^* K$ & & & \\
        \hline
        \noalign{\smallskip}
        \enspace$\Sigma^+\rightarrow \Xi^{*0}K^+$ 
            & $\chi^2_{K^+\Xi^{*0}}$ 
            & $  2q_uq_{\bar{s}}\,\chi^2_{\pi^+\Sigma^{*0}}$ 
            & $q^2_{\bar{s}}\,\chi^2_{\pi^+\Sigma^{*0}}$\\

        \noalign{\smallskip}
    \end{tabular}
    \end{ruledtabular}
\end{table*}

\begin{table*}[p]
	\caption{Chiral coefficients for the leading-order loop integral contributions for the $\Xi^0$.} \label{tab:appendix:chipt:cascade0}
        \begin{ruledtabular}
	\begin{tabular}{lccc}
	\noalign{\smallskip}
	Process &Total &Valence-sea &Sea-sea \\
	\noalign{\smallskip}
	\hline
	\noalign{\smallskip}
		
        $\Xi^0\rightarrow \Xi\pi$ & & & \\
        \hline
        \noalign{\smallskip}
        \enspace$\Xi^0\rightarrow \Xi^0\pi^0$
            & $0$ 
            & $  2q_uq_{\bar{u}}\,\chi^2_{\pi^+\Xi^-}$ 
            & $q^2_{\bar{u}}\,\chi^2_{\pi^+\Xi^-}$\\
        \enspace$\Xi^0\rightarrow \Xi^-\pi^+$
            & $\chi^2_{\pi^+\Xi^-}$ 
            & $2q_uq_{\bar{d}}\,\chi^2_{\pi^+\Xi^-}$ 
            & $q^2_{\bar{d}}\,\chi^2_{\pi^+\Xi^-}$\\
	\noalign{\smallskip}
        \hline
        \noalign{\smallskip}

        $\Xi^0\rightarrow \Sigma K$ & & & \\
        \hline
        \noalign{\smallskip}
        \enspace$\Xi^0\rightarrow \Sigma^+K^-$ 
            & $\chi^2_{K^-\Sigma^+}$ 
            & $2q_sq_{\bar{u}}(\chi^2_{K^0\Sigma^0}+\chi^2_{K^0\Lambda})$ 
            & $q^2_{\bar{u}}(\chi^2_{K^0\Sigma^0}+\chi^2_{K^0\Lambda})$\\
        \enspace$\Xi^0\rightarrow (\Sigma^0,\,\Lambda)K^0$
            & $0$ 
            & $2q_sq_{\bar{d}}(\chi^2_{K^0\Sigma^0}+\chi^2_{K^0\Lambda})$
            & $q^2_{\bar{d}}(\chi^2_{K^0\Sigma^0}+\chi^2_{K^0\Lambda})$\\
	\noalign{\smallskip}
        \hline
        \noalign{\smallskip}

        $\Xi^0\rightarrow \Xi K$ & & & \\
        \hline
        \noalign{\smallskip}
        \enspace$\Xi^0\rightarrow \Xi^-_{3s}K^+$ 
            & $0$ 
            & $2q_uq_{\bar{s}}\,\chi^2_{\pi^+\Xi^-}$ 
            & $q^2_{\bar{s}}\,\chi^2_{\pi^+\Xi^-}$\\
	\noalign{\smallskip}
        \hline
        \noalign{\smallskip}

        $\Xi^0\rightarrow \Xi^*\pi$ & & & \\
        \hline
        \noalign{\smallskip}
        \enspace$\Xi^0\rightarrow \Xi^{*0}\pi^0$
            & $0$ 
            & $2q_uq_{\bar{u}}\,\chi^2_{\pi^+\Xi^{*-}}$ 
            & $q^2_{\bar{u}}\,\chi^2_{\pi^+\Xi^{*-}}$\\
        \enspace$\Xi^0\rightarrow \Xi^{*-}\pi^+$
            & $\chi^2_{\pi^+\Xi^{*-}}$ 
            & $2q_uq_{\bar{d}}\,\chi^2_{\pi^+\Xi^{*-}}$ 
            & $q^2_{\bar{d}}\,\chi^2_{\pi^+\Xi^{*-}}$\\
	\noalign{\smallskip}
        \hline
        \noalign{\smallskip}

        $\Xi^0\rightarrow \Sigma^* K$ & & & \\
        \hline
        \noalign{\smallskip}
        \enspace$\Xi^0\rightarrow \Sigma^{*+}K^-$ 
            & $\chi^2_{K^-\Sigma^{*+}}$ 
            & $2q_sq_{\bar{u}}\,\chi^2_{K^0\Sigma^{*0}}$ 
            & $q^2_{\bar{u}}\,\chi^2_{K^0\Sigma^{*0}}$\\
        \enspace$\Xi^0\rightarrow \Sigma^{*0}K^0$ 
            & $0$ 
            & $2q_sq_{\bar{d}}\,\chi^2_{K^0\Sigma^{*0}}$ 
            & $q^2_{\bar{d}}\,\chi^2_{K^0\Sigma^{*0}}$\\
	\noalign{\smallskip}
        \hline
        \noalign{\smallskip}

        $\Xi^0\rightarrow \Omega K$ & & & \\
        \hline
        \noalign{\smallskip}
        \enspace$\Xi^0\rightarrow \Omega^-K^+$ 
            & $\chi^2_{K^+\Omega^-}$ 
            & $2q_uq_{\bar{s}}\,\chi^2_{\pi^+\Xi^{*-}}$ 
            & $q^2_{\bar{s}}\,\chi^2_{\pi^+\Xi^{*-}}$\\

        \noalign{\smallskip}
    \end{tabular}
    \end{ruledtabular}
\end{table*}

\begin{table*}[p]
	\caption{Chiral coefficients for the leading-order loop integral contributions for the $\Sigma^-$.} \label{tab:appendix:chipt:sigmam}
        \begin{ruledtabular}
	\begin{tabular}{lccc}
	\noalign{\smallskip}
	Process &Total &Valence-sea &Sea-sea \\
	\noalign{\smallskip}
	\hline
	\noalign{\smallskip}
		
        $\Sigma^-\rightarrow \Sigma\pi$ & & & \\
        \hline
        \noalign{\smallskip}
        \enspace$\Sigma^-\rightarrow \Sigma^-\pi^0$
            & $0$ 
            & $2q_dq_{\bar{d}}(\chi^2_{\pi^-\Sigma^0}+\chi^2_{\pi^-\Lambda})$ 
            & $q^2_{\bar{d}}(\chi^2_{\pi^-\Sigma^0}+\chi^2_{\pi^-\Lambda})$\\
        \enspace$\Sigma^-\rightarrow (\Sigma^0,\,\Lambda)\pi^-$ 
            & $\chi^2_{\pi^-\Sigma^0}+\chi^2_{\pi^-\Lambda}$ 
            & $2q_dq_{\bar{u}}(\chi^2_{\pi^-\Sigma^0}+\chi^2_{\pi^-\Lambda})$ 
            & $q^2_{\bar{u}}(\chi^2_{\pi^-\Sigma^0}+\chi^2_{\pi^-\Lambda})$\\
	\noalign{\smallskip}
        \hline
        \noalign{\smallskip}

        $\Sigma^-\rightarrow NK$ & & & \\
        \hline
        \noalign{\smallskip}
        \enspace$\Sigma^-\rightarrow nK^-$ 
            & $\chi^2_{K^-n}$ 
            & $2q_sq_{\bar{u}}\,\chi^2_{K^-n}$ 
            & $q^2_{\bar{u}}\,\chi^2_{K^-n}$\\
        \enspace$\Sigma^-\rightarrow n^-K^0$
            & $0$ 
            & $2q_sq_{\bar{d}}\,\chi^2_{K^-n}$
            & $q^2_{\bar{d}}\,\chi^2_{K^-n}$\\
	\noalign{\smallskip}
        \hline
        \noalign{\smallskip}

        $\Sigma^-\rightarrow \Xi K$ & & & \\
        \hline
        \noalign{\smallskip}
        \enspace$\Sigma^-\rightarrow \Xi^-K^0$ 
            & $0$ 
            & $2q_dq_{\bar{s}}(\chi^2_{\pi^-\Sigma^0}+\chi^2_{\pi^-\Lambda})$ 
            & $q^2_{\bar{s}}(\chi^2_{\pi^-\Sigma^0}+\chi^2_{\pi^-\Lambda})$\\
	\noalign{\smallskip}
        \hline
        \noalign{\smallskip}

        $\Sigma^-\rightarrow \Sigma^*\pi$ & & & \\
        \hline
        \noalign{\smallskip}
        \enspace$\Sigma^-\rightarrow \Sigma^{*-}\pi^0$
            & $0$ 
            & $2q_dq_{\bar{d}}\,\chi^2_{\pi^-\Sigma^{*0}}$ 
            & $q^2_{\bar{d}}\,\chi^2_{\pi^-\Sigma^{*0}}$\\
        \enspace$\Sigma^-\rightarrow \Sigma^{*0}\pi^-$
            & $\chi^2_{\pi^-\Sigma^{*0}}$ 
            & $2q_dq_{\bar{u}}\,\chi^2_{\pi^-\Sigma^{*0}}$ 
            & $q^2_{\bar{u}}\,\chi^2_{\pi^-\Sigma^{*0}}$\\
	\noalign{\smallskip}
        \hline
        \noalign{\smallskip}

        $\Sigma^-\rightarrow \Delta K$ & & & \\
        \hline
        \noalign{\smallskip}
        \enspace$\Sigma^-\rightarrow \Delta^0K^-$ 
            & $\chi^2_{K^-\Delta^0}$ 
            & $2q_sq_{\bar{u}}\,\chi^2_{K^-\Delta^0}$ 
            & $q^2_{\bar{u}}\,\chi^2_{K^-\Delta^0}$\\
        \enspace$\Sigma^-\rightarrow \Delta^-K^0$
            & $0$ 
            & $2q_sq_{\bar{d}}\,\chi^2_{K^-\Delta^0}$ 
            & $q^2_{\bar{d}}\,\chi^2_{K^-\Delta^0}$\\
	\noalign{\smallskip}
        \hline
        \noalign{\smallskip}

        $\Sigma^-\rightarrow \Xi^* K$ & & & \\
        \hline
        \noalign{\smallskip}
        \enspace$\Sigma^-\rightarrow \Xi^{*-}K^0$ 
            & $0$ 
            & $2q_dq_{\bar{s}}\,\chi^2_{\pi^-\Sigma^{*0}}$
            & $q^2_{\bar{s}}\,\chi^2_{\pi^-\Sigma^{*0}}$\\

        \noalign{\smallskip}
    \end{tabular}
    \end{ruledtabular}
\end{table*}

\begin{table*}[p]
	\caption{Chiral coefficients for the leading-order loop integral contributions for the $\Xi^-$.} \label{tab:appendix:chipt:cascadem}
        \begin{ruledtabular}
	\begin{tabular}{lccc}
	\noalign{\smallskip}
	Process &Total &Valence-sea &Sea-sea \\
	\noalign{\smallskip}
	\hline
	\noalign{\smallskip}
		
        $\Xi^-\rightarrow \Xi\pi$ & & & \\
        \hline
        \noalign{\smallskip}
        \enspace$\Xi^-\rightarrow \Xi^-\pi^0$
            & $0$ 
            & $2q_dq_{\bar{d}}\,\chi^2_{\pi^-\Xi^0}$ 
            & $q^2_{\bar{d}}\,\chi^2_{\pi^-\Xi^0}$\\
        \enspace$\Xi^-\rightarrow \Xi^0\pi^-$
            & $\chi^2_{\pi^-\Xi^0}$ 
            & $2q_dq_{\bar{u}}\,\chi^2_{\pi^-\Xi^0}$ 
            & $q^2_{\bar{u}}\,\chi^2_{\pi^-\Xi^0}$\\
	\noalign{\smallskip}
        \hline
        \noalign{\smallskip}

        $\Xi^-\rightarrow \Sigma K$ & & & \\
        \hline
        \noalign{\smallskip}
        \enspace$\Xi^-\rightarrow \Sigma^-K^0$
            & $0$ 
            & $2q_sq_{\bar{d}}(\chi^2_{K^-\Sigma^0}+\chi^2_{K^-\Lambda})$ 
            & $q^2_{\bar{d}}(\chi^2_{K^-\Sigma^0}+\chi^2_{K^-\Lambda})$\\
        \enspace$\Xi^-\rightarrow (\Sigma^0,\,\Lambda)K^-$ 
            & $\chi^2_{K^-\Sigma^0}+\chi^2_{K^-\Lambda}$ 
            & $2q_sq_{\bar{u}}(\chi^2_{K^-\Sigma^0}+\chi^2_{K^-\Lambda})$  
            & $q^2_{\bar{u}}(\chi^2_{K^-\Sigma^0}+\chi^2_{K^-\Lambda})$\\
	\noalign{\smallskip}
        \hline
        \noalign{\smallskip}

        $\Xi^-\rightarrow \Xi K$ & & & \\
        \hline
        \noalign{\smallskip}
        \enspace$\Xi^-\rightarrow \Xi^-_{3s}K^0$ 
            & $0$ 
            & $2q_dq_{\bar{s}}\,\chi^2_{\pi^-\Xi^0}$ 
            & $q^2_{\bar{s}}\,\chi^2_{\pi^-\Xi^0}$\\
	\noalign{\smallskip}
        \hline
        \noalign{\smallskip}

        $\Xi^-\rightarrow \Xi^*\pi$ & & & \\
        \hline
        \noalign{\smallskip}
        \enspace$\Xi^-\rightarrow \Xi^{*-}\pi^0$
            & $0$ 
            & $2q_dq_{\bar{d}}\,\chi^2_{\pi^-\Xi^{*0}}$  
            & $q^2_{\bar{d}}\,\chi^2_{\pi^-\Xi^{*0}}$\\
        \enspace$\Xi^-\rightarrow \Xi^{*0}\pi^-$
            & $\chi^2_{\pi^-\Xi^{*0}}$ 
            & $2q_dq_{\bar{u}}\,\chi^2_{\pi^-\Xi^{*0}}$ 
            & $q^2_{\bar{u}}\,\chi^2_{\pi^-\Xi^{*0}}$\\
	\noalign{\smallskip}
        \hline
        \noalign{\smallskip}

        $\Xi^-\rightarrow \Sigma^* K$ & & & \\
        \hline
        \noalign{\smallskip}
        \enspace$\Xi^-\rightarrow \Sigma^{*-}K^0$ 
            & $0$ 
            & $2q_sq_{\bar{d}}\,\chi^2_{K^-\Sigma^{*0}}$ 
            & $q^2_{\bar{d}}\,\chi^2_{K^-\Sigma^{*0}}$\\
        \enspace$\Xi^-\rightarrow \Sigma^{*0}K^-$ 
            & $\chi^2_{K^-\Sigma^{*0}}$ 
            & $2q_sq_{\bar{u}}\,\chi^2_{K^-\Sigma^{*0}}$  
            & $q^2_{\bar{u}}\,\chi^2_{K^-\Sigma^{*0}}$\\
	\noalign{\smallskip}
        \hline
        \noalign{\smallskip}

        $\Xi^-\rightarrow \Omega K$ & & & \\
        \hline
        \noalign{\smallskip}
        \enspace$\Xi^-\rightarrow \Omega^-K^0$ 
            & $0$ 
            & $2q_dq_{\bar{s}}\,\chi^2_{\pi^-\Xi^{*0}}$  
            & $q^2_{\bar{s}}\,\chi^2_{\pi^-\Xi^{*0}}$\\

        \noalign{\smallskip}
    \end{tabular}
    \end{ruledtabular}
\end{table*}

As discussed in \autoref{sec:chiralextrapolation}, calculating a correction for electroquenching
requires identification of the various valence-valence, valence-sea and sea-sea contributions to
the magnetic polarizability. These contributions may be written in terms of the SU(3) chiral
coefficients. The coefficients $\chi^2_{MB}$ are derived by considering an appropriate chiral
Lagrangian. The reader is directed to Ref.~\cite{shanahan:2012:chiral} for a convenient catalogue
of these coefficients. The coefficients are written in terms of the SU(6) values $D,F,\mcC$. We
take \mbox{$D+F=g_A=1.267$}, \mbox{$F=\frac{2}{3}D$},
\mbox{$\mcC=-1.52$}~\cite{bignell2020nucleon}.

The chiral coefficients for standard SU(3) axial transitions \cite{shanahan:2012:chiral} are listed
in \autoref{tab:appendix:chiralcoeffs:octet} and
\autoref{tab:appendix:chiralcoeffs:decuplet}. Note, our definition of the coefficients in the loop
integrals differ from Ref.~\cite{shanahan:2012:chiral} such that their octet coefficients have been
squared and divided by 2 and their decuplet coefficients have been squared and multiplied by $4/3$.

The contributions for each possible process are derived in the manner described in
\autoref{sec:chiralextrapolation}. All contributions are listed in Tables
\ref{tab:appendix:chipt:neutron} through \ref{tab:appendix:chipt:cascadem}.

\FloatBarrier

% Bibliography file is main.bib
\bibliography{main}

\end{document}